\documentclass[iop, apj]{emulateapj}
\usepackage{amsmath}
\usepackage{amssymb}
\usepackage{amstext}
\usepackage{apjfonts}
\usepackage{graphicx}
\usepackage{epsfig}
\usepackage{comment}
\usepackage{aas_macros}

\begin{document}

\shorttitle{LBV SN progenitors from binary mergers}
\shortauthors{Justham, Podsiadlowski \& Vink}
\submitted{Original version submitted to ApJ Letters on 21 December 2012. This
version accepted on 7 October 2014.}
\title{Luminous Blue Variables and superluminous supernovae from binary
  mergers}

\author{Stephen Justham}   

\affil{The Key Laboratory of Optical Astronomy, National Astronomical
Observatories, The Chinese Academy of Sciences, Datun Road, Beijing
100012, China
\\ previously at: \\ 
The Kavli Institute for Astronomy and Astrophysics, Peking University, Beijing 100871, China}
\email{SJ: sjustham@bao.ac.cn; PhP: podsi@astro.ox.ac.uk}

\author{Philipp Podsiadlowski} \affil{Astrophysics Sub-Department,
  University of Oxford, Oxford OX1 3RH, United Kingdom}

\author{Jorick S. Vink}
\affil{Armagh Observatory, College Hill, Armagh BT61 9DG, Northern Ireland}

\begin{abstract}
Evidence suggests that the direct progenitor stars of some
core-collapse supernovae (CCSNe) are luminous blue variables (LBVs),
perhaps including some `superluminous supernovae' (SLSNe). We examine
models in which massive stars gain mass soon after the end of core
hydrogen burning. These are mainly intended to represent mergers
following a brief contact phase during early Case B mass transfer, but
may also represent stars which gain mass in the Hertzsprung Gap or
extremely late during the main-sequence phase for other reasons.  The
post-accretion stars spend their core helium-burning phase as blue
supergiants (BSGs), and many examples are consistent with being LBVs
at the time of core collapse. Other examples are yellow supergiants at
explosion. We also investigate whether such post-accretion stars may
explode successfully after core collapse.  The final core properties
of post-accretion models are broadly similar to those of single stars
with the same initial mass as the pre-merger primary star. More
surprisingly, when early Case B accretion does affect the final core
properties, the effect appears likely to favour a successful SN
explosion, i.e., to make the core properties more like those of a
\emph{lower}-mass single star. However, the detailed structures of
these cores sometimes display qualitative differences to any
single-star model we have calculated. The rate of appropriate binary
mergers may match the rate of SNe with immediate LBV progenitors; for
moderately optimistic assumptions we estimate that the progenitor
birthrate is $\sim$1\% of the CCSN rate.
\end{abstract}


\section{Introduction}
\label{sec:intro}

Supernovae (SNe) -- explosions of stars -- have long been studied
as complex physical systems which play a vital role in shaping the
composition and structure of the Universe. 
Despite the extensive history of the field,
recent discoveries have challenged some strongly-held expectations.
One major surprise was that some massive stars appear to explode
during a phase where they appear as ``luminous blue variables''
(LBVs):  observations of radio modulations
from some SNe indicated that the SN ejecta interacted
with circumstellar material (CSM) similar to that found
around LBVs undergoing S Doradus cycles, during which 
envelope material is ejected episodically \citep{Kotak+Vink2006}.
This interpretation was strengthened by observations of
multiple P Cygni absorption profiles in the spectrum of 
the interacting SN~2005gj
\citep{Trundle+2008}, which provided evidence for multiple shells
with characteristic LBV wind velocities, possibly associated with 
multiple LBV outbursts; \citet{Kiewe+2012} subsequently found P Cygni
absorption features in four additional SN IIn.
Further support for this inference was provided when
the immediate progenitor of the type IIn SN~2005gl was identified 
as having been a very luminous star, consistent with an LBV
\citep{Gal-Yam+2007,Gal-Yam+Leonard2009}. 

Perhaps the strongest evidence yet known in favour of
LBV-like outbursts from SN progenitors comes from the systems which produced
SN~2009ip and SN~2010mc, both of which were classified as Type IIn events.
\citet{Ofek+2013Nature} found that the progenitor of SN~2010mc ejected
$\sim 10^{-2}~M_{\odot}$ during an outburst only 40 days before it
exploded as a SN. The system which produced SN~2009ip has
displayed several outbursts since an outbursting LBV was first identified at that location
\citep{Maza+2009CBET, Miller+2009ATel, Berger+2009ATel, Li+2009ATel}. 
The most recent outburst from SN~2009ip may well mark the final explosion of the star
\citep{Mauerhan+2013,Smith+2014}, although there is some
uncertainty over whether the outburst was terminal
\citep{Fraser+2013,Margutti+2014}. 

One complication which arises when interpreting this observational
evidence in terms of understanding SN progenitors is that the
class of LBVs is poorly understood and inhomogeneous. 
In broad terms, LBVs are massive, hot stars located 
near the Eddington limit, and are subject to
occasional outbursts accompanied by mass ejection 
\citep[see, e.g.,][]{VinkReview2009}.  However, the standard S
Doradus-type LBV outbursts are dissimilar from events such 
as the Great Eruption of Eta Carinae. Hence, even though Eta Carinae is referred to as an LBV,
it and other objects which produce similar rare giant eruptions may arise
from a different mechanism than canonical S\,Dor LBVs. 
Furthermore, even though the phenomenology of typical LBV outbursts is fairly well
established \citep{HumphreysDavidson1994}, the specific physical
mechanism responsible for even those S\,Dor LBV mass-loss events is
unclear.  This uncertainty remains despite a great deal of
  theoretical attention, although there is broad agreement that high
  stellar luminosities -- near to the Eddington limit -- could enable
  S\,Dor-type instabilities  \citep[see, e.g.,][]{Joss+1973,Glatzel+Kiriakidis1993,Langer1998,
    Shaviv2001,Vink+deKoter2002,Smith+Owocki2006,Graefener+2012,
    Guzik+Lovekin2014,Owocki2014}. As a result,
we later compare our models to the empirical position
of the S\,Dor instability strip on the HR diagram 
 \citep[see, e.g.,][]{Groh+2009}, not to theoretical models for LBV instabilities.

The term ``LBV'' thus 
defines a broad phenomenology rather than an evolutionary
stage, and the evolutionary nature of LBVs is as yet unknown. 
Nonetheless, canonical LBVs were
not generally expected to be immediate progenitors of core-collapse SNe.
Standard stellar evolution theory predicts that single massive stars
which become LBVs do so near the end or
after the completion of core hydrogen (H) burning, then typically lose their
H-rich envelopes in the LBV phase and become H-deficient
Wolf-Rayet stars, where they spend several $10^5$\,yr burning helium (He)
in the core before they explode in a core-collapse SN (CCSN), long
after they have passed through the LBV phase \citep[for a review see, e.g.,][]{LangerARAA}.

A further challenge when trying to understand the population of these
SNe with apparent LBV progenitors is that the LBV-SNe are unlikely to
correspond with all members of one of the phenomenological SN
types. SNe with presumed LBV progenitors are of type IIn or IIb, and
some previous work has argued that all type IIn SNe may be generated
by LBV-like progenitors \citep[see, e.g.,
][]{Gal-Yam+2007,Kiewe+2012}.  However, the type IIn SN phenomenon can
potentially be produced by a heterogeneous set of circumstances
\citep{Kotak+2004}, as demonstrated by events such as SN~2002ic and
SN~2005gj \citep{Hamuy+2003,Aldering+2006}. Those SNe showed strong
circumstellar interaction like that seen in type IIn SNe, but each is
thought to have been powered by a type Ia SN, not a CCSN.  SN PTF11kx
showed similar, but somewhat weaker interaction \citep{Dilday+2012},
which suggests that there may be a continuum of H emission line
strengths arising from type Ia SN progenitors.  Many other SNe which
were classified as IIn could easily have been disguised SNe Ia.  This
suspicion was strengthened by \citet{Anderson+2012}, who found that
their sample of SNe exhibiting type IIn phenomenology shows less clear
association with star formation than any other SN sub-type generally
attributed to core collapse.  Based on their data, they suggested that
the majority of IIn SNe arise from relatively low-mass progenitors,
i.e., are not associated with LBVs \citep[see also][]{Habergham+2014}.
There may be tension between this conclusion and the inference that
the majority of SN IIn display pre-explosion outbursts
\citep{Ofek+2014}, if those pre-explosion events are shown to be LBV
outbursts from the star which explodes. Nonetheless, the evidence that
most SN which could be classified as belonging to ``Type IIn'' may not
arise from LBVs does not affect the evidence that \emph{some} SNe have
direct LBV progenitors.  However, this heterogeneity is a complication
when trying to determine the rate at which LBVs are formed and
explode.

\subsection{``Superluminous'' supernovae from LBV progenitors?}

The interest surrounding LBVs as immediate SN progenitors
increased further following the discovery of the extraordinarily
luminous type IIn SN~2006gy \citep{Ofek+2007,Smith+2007}, which
was also suggested to have been produced by an LBV star.
Numerous similarly outstanding events have since been identified
\citep{Quimby+2007,Smith+2008,Gezari+2009,Miller+2009,Drake+2010,Gal-Yam2012}.
In this work, we mainly consider a subset of these
`superluminous' SNe (SLSNe), the H-rich `SLSN-II' \citep[for a review, see][]{Gal-Yam2012}.
One popular explanation for the high luminosity
of these SNe is that a standard amount of CCSN energy input is radiated away far
more efficiently than in a canonical CCSN. This is believed to be due to interaction
of the SN ejecta with a dense CSM, which causes
rapid deceleration of the SN shock, thereby converting kinetic energy into radiation
(see, e.g., \citealt{Smith+2007,Ofek+2007,Smith+McCray2007,vanMarle+2010,Gal-Yam2012};
although \citealt{Moriya+2013} found it challening to 
reproduce the SN~2006gy light-curve using CSM interaction). 
Prior to the interaction model for SN~2006gy, several other SNe had shown 
evidence for large amounts of mass ejection soon before the explosion
\citep[][]{Dopita+1984,Chugai+Danziger1994,Chugai+2004}.
As LBVs are a class of single stars which experience
phases of mass loss drastic enough to account for 
the required CSM densities, they have widely been regarded as
potential progenitors of SLSNe.
The properties of LBV ejections before collapse might 
control the CSM densities at the time of explosion, and thereby the 
radiative efficiency of the SN. 
For example, the timing of the last pre-SN LBV outburst may need to
be sufficiently close to the SN to lead to a SLSN, or the amount of
mass ejected may need to be unusually large (at one extreme, a giant
eruption may be required).  Hence similar stellar systems might explain
both the LBV-SNe with normal luminosity and the relevant subset
of SLSNe, separated only by random variations in the pre-SN outburst 
properties.

Nevertheless, a key issue with single-star LBVs as SN
progenitors -- both with normal and exceptionally high luminosity -- is that
stars in the appropriate mass range are typically expected to
produce \emph{faint} SNe (if core collapse leads to any SN explosion at all).  
The reason is that they are predicted to quietly form black holes, 
without the strong outward-moving shock
required for a typical SN explosion energy
(see, e.g., \citealt{Fryer1999,Heger+2003}; for an observational
perspective see, e.g., \citealt{Kochanek+2008,Kochanek2014}).  
There are likely exceptions to this statement, e.g., the
``collapsars'' which are a consequence of rapid-rotation in the core at core collapse
\citep{Woosley1993,MacFadyen+Woosley1999}, but those are thought to be
extremely rare events.

\subsection{Non-LBV models for unusually luminous SNe and for 
  pre-core-collapse mass ejection}

LBV outbursts are not the only way to eject substantial amount of 
mass from the stellar envelope. 
The type of binary interaction known as common-envelope (CE)
evolution may also do the trick. Hence an alternative
possibility for the presence of a massive CSM close to an exploding star
is the recent ejection of a CE in a massive
binary system \citep[][]{Ofek+2007,Chevalier2012}. 
At least some CCSNe are predicted to
occur during such phases \citep{Podsiadlowski+1990}, although it is
unclear whether the empirical event rates could be matched without
fine-tuning. Note that canonical CE ejection may not be
necessary to eject sufficient mass to produce SLSNe, as mergers
may also eject significant amounts of mass
\citep[see,
e.g.,][]{Podsiadlowski+1991,Podsiadlowski1992,Morris+Podsiadlowski2007,
  Morris+Podsiadlowski2009,Ivanova+2013Science}.
A related model is that of \citet{Soker+Kashi2013}, which aims to explain
both the pre-SN outburst of SN~2009ip and the CSM through a particular
binary merger scenario.  In addition, \citet{Mackey+2014} have
  suggested that external photoionization may be able to trap
  the normal winds of red
supergiant stars sufficiently well to explain SNe which
show evidence of interaction with CSM.

Extremely massive stars are predicted to produce
pair-instability SNe (PISNe). The mass limit is generally considered to be in
excess of $\approx 150 M_{\odot}$ (somewhat
dependent on metallicity and other assumptions;
see, e.g., \citealt{Barkat+1967,RakavyShaviv1967,Heger+2003}),
although mixing due to rapid rotation might significantly lower that
limit \citep{Chatz+Wheeler2012ppsn}.  The likely appearance of
such SNe has recently been studied by, e.g., \citet{Kasen+2011}
and \citet{Kozyreva+2014}.
However, there is no reason to believe that such events have been 
confused with the cases which we seek to explain. 
Strong candidates for PISNe have been
identified \citep{Gal-Yam+2009,Gal-Yam2012}, although other models 
have also been proposed \citep[see, e.g.,][]{Moriya+2010ppsn,Nicholl+2013}.  
Slightly less massive stars are expected to produce \emph{pulsational}
PISNe, which provide yet another plausible explanation 
for the CSM around these SN progenitors at the time of
explosion \citep[for which,
see][]{Woosley+2007,Waldman2008,Chatz+Wheeler2012pulsational,Chen+2014}.
Models have also predicted that the envelopes of some luminous red
supergiant (RSG) SN progenitors could
produce pulsation-driven superwinds towards the end of their life
\citep[see, e.g.,][]{Heger+1997,Yoon+Cantiello2010}. Arguments
have even been made in favour of mass-loss driven by the very late
nuclear evolution of the stellar core
\citep[][]{Quataert+Shiode2012,Shiode+Quataert2014,Moriya2014}.
For the specific case of SN~2009ip, \citet{Ofek+2013} argue that their
measurements of the CSM density are more consistent
with a model like that of \citet{Quataert+Shiode2012} than ejections from
pulsational pair-instability. All of these late mass-loss mechanisms
could naturally explain why some core-collapse SNe occur in
denser-than-otherwise-expected CSM. Nonetheless, the
models which require RSGs could clearly not explain SNe
with BSG progenitors (and models which require normal core collapse
from BSGs are incomplete without an explanation for why the stars
reach core collapse as BSGs).

A further class of models for some SLSNe invoke the rapid
spin-down of a magnetar to power the luminosity 
\citep[see, e.g.,][]{Kasen+Bildsten2010,Woosley2010}, which have been
particularly successful in fitting the lightcurves of some extremely luminous
H-poor SNe \citep[][]{Inserra+2013,Howell+2013}.

Obviously more than one of the proposed models may work, 
and even the subset of ``SLSN-II'' might have heterogenous origins.

\subsection{Models for blue supergiant SN progenitors}

Models for direct SN progenitors which are blue supergiants (BSGs) at
core collapse have existed for many years, but only for cases where
the BSG is not an LBV. The most well-known example of a BSG progenitor
is that of SN 1987A.  This paper examines whether a binary merger
model, a variation of a previous model for SN 1987A
\citep{Podsiadlowski1992} is able to explain LBV-SN progenitors. That
merger model is not only able to explain the BSG progenitor of SN
1987A, but it also provides a natural explanation for the distinctive
circumstellar structures seen in the remnant of SN 1987A
\citep{Podsiadlowski+1991,Podsiadlowski1992,Morris+Podsiadlowski2007,Morris+Podsiadlowski2009}.
Here we extend that work and demonstrate that more massive mergers are
capable of producing SN progenitors which are plausibly luminous
enough to be LBVs at the time of explosion.

We suggest that LBV-SN progenitors can form from
massive binary systems which merge soon after the more
massive star has finished core H burning, i.e., as it is
\emph{expanding} across the Hertzsprung Gap (HG). 
The fact that such a star could gain mass is somewhat counter-intuitive, but 
it has been previously studied and accepted as at least plausible; we explain the merger
mechanism below.  Our calculations could also apply to other situations in which massive
stars gain mass soon after the end of the main sequence (i.e.``early Case
B'' accretion; see also \citealt{Podsiadlowski+1992}).
\citet{Podsiadlowski+Joss1989} and \citet{Braun+Langer1995} previously studied
how accretion onto massive main-sequence stars might produce BSG SN
progenitors; in this respect this paper also
extends that work, although stable mass-transfer onto a HG star is
probably far less common than accretion via mergers (as we discuss in \S \ref{sec:stableRLOFrates}).
Roughly coincident with submission of the original version of this
work another paper was submitted which addresses similar possibilities 
\citep{Vanbeveren+2013}. \citet{Glebbeek+2013} also 
published work with similar aspects. 

Binary interactions are expected to have a significant effect on the lives
of a large fraction of massive stars, probably the majority of them, with observational 
studies of massive stars concluding that most massive stars 
occur in interacting binary systems \citep[see,
e.g.,][]{Abt+Levy1976,Abt+Levy1978,Kobulnicky+Fryer2007,Eggleton+Tokovinin2008,Sana+2012,Sana+2013}. 
Close binary systems have long been argued to be responsible for much
of the diversity of observed SN types \citep{Podsiadlowski+1992},
and stellar mergers may well explain all of the B[e] supergiants \citep{Podsiadlowski+2006}.
Indeed, mergers are expected to be so common that \citet{deMink+2014}
predicted that $8^{+9}_{-4}$\% of observed early-type stars are merger products.

In contrast to models for the production of BSG and LBV SNe via binary interactions or
stellar collisions, \citet{Groh+2013} have argued that the larger He cores produced
by stellar rotation allow some single stars to appear as LBVs at the time of
explosion \citep[based on evolutionary calculations
from][]{Ekstrom+2012}. Their proposed LBV-SN progenitors have initial
masses of 20 and 25 $M_{\odot}$, with respective final masses of 7.1
and 9.6 $M_{\odot}$ (and surface He mass fractions of 0.74 and
0.9) following significant mass loss in an RSG phase.
At explosion these stellar models retain little H; Groh et
al.\ state that the He-rich core
accounts for 94\% and 100\% of the stellar mass respectively, which allows
minimal room for additional mass loss via LBV outbursts before
explosion without significantly affecting their surface properties.

Evidence in favour of the idea that some LBVs are produced following
binary mergers arises from the observations that LBVs tend to be
rapidly rotating \citep{Groh+2009} and that they also have a lower
binary fraction than otherwise similar massive stars 
(see, e.g., \citealt{VinkReview2009}; however, whilst
\citealt{Smith+Tombleson2014} agree that LBVs are likely to have
gained mass from a companion, they argue in favour of stable mass
transfer rather than mergers).
Rapid rotation may well promote LBV instabilities \citep{Langer1997,Langer1998}.
However, although observed LBVs are
rapidly rotating, it does not automatically follow that any additional
mixing due to rapid rotation is important for explaining their
properties. For the model which we present, the merger
(or accretion) occurs during the HG, i.e., after strong
molecular-weight gradients have been generated within the
star. It has previously been shown that rotational mixing
is extremely unlikely to occur across those gradients, and therefore
unlikely to significantly affect the evolution of stars
produced by our proposed scenario \citep[see,
e.g.,][]{Mestel1953,Mestel1957,Mestel+Moss1986}.  
Previous authors have argued that Case A mergers of massive stars
could produce sufficient rotationally-driven mixing to cause
homogeneous evolution and thereby produce the progenitors of
long-duration gamma-ray bursts (see, e.g., 
\citealt{WoosleyHeger2006,deMink+2013}, along with related work on
rapidly-rotating single stars by
\citealt{Yoon+Langer2005,Yoon+2006}). 
It may be that early Case A mergers, which occur before large composition gradients
have been generated, can produce stars in which rotational mixing
dominates their future evolution, but because of the molecular weight
gradients such an outcome is less likely for early Case B mergers. 
Late Case A mergers, or mergers of Case A contact
binaries which occur after the primary evolves off the main sequence,
may also have sufficiently well-developed composition profiles to
stabilise them against rotational mixing. 
 
\begin{figure}
\includegraphics[width=8cm]{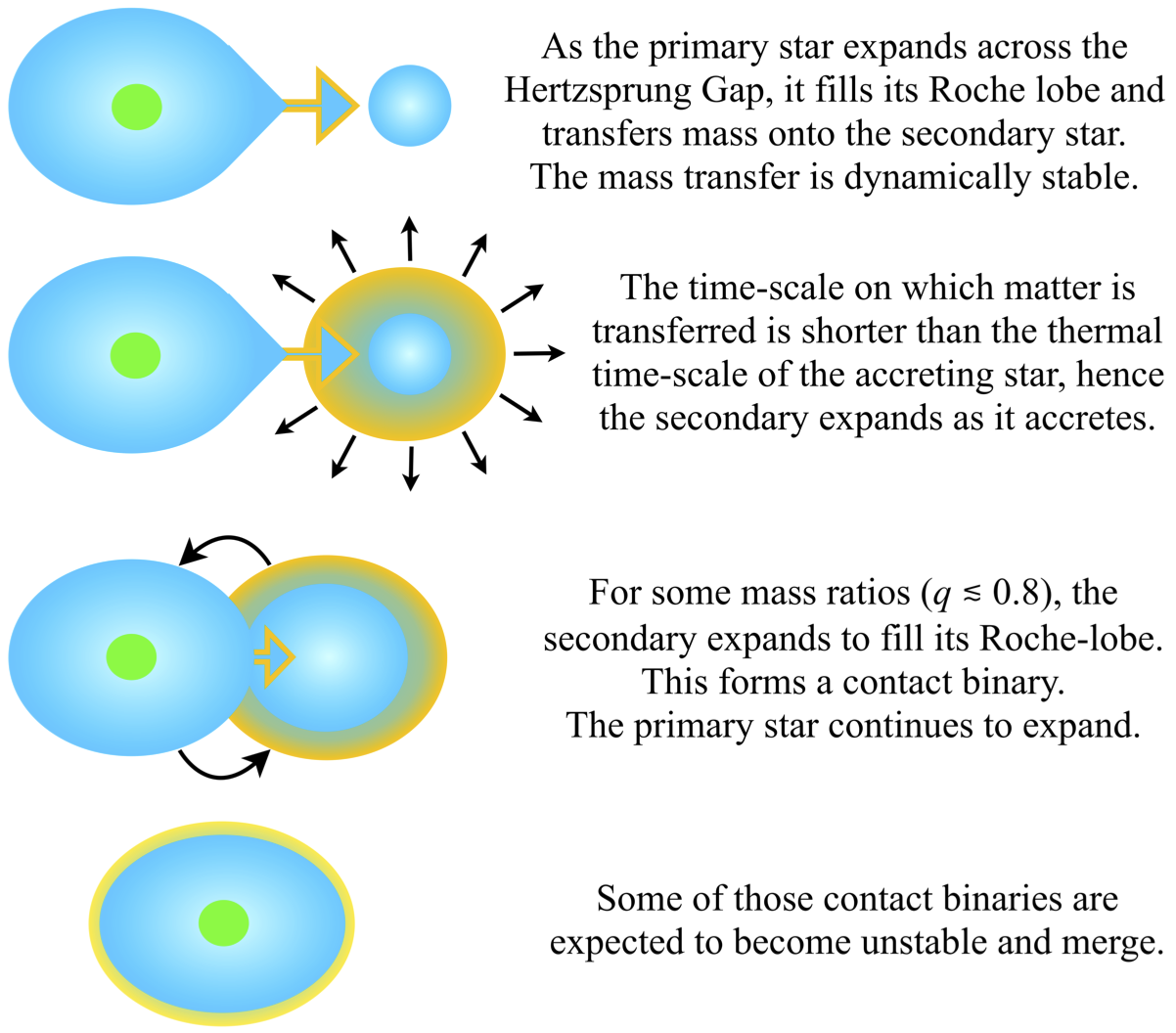}
\includegraphics[width=8cm]{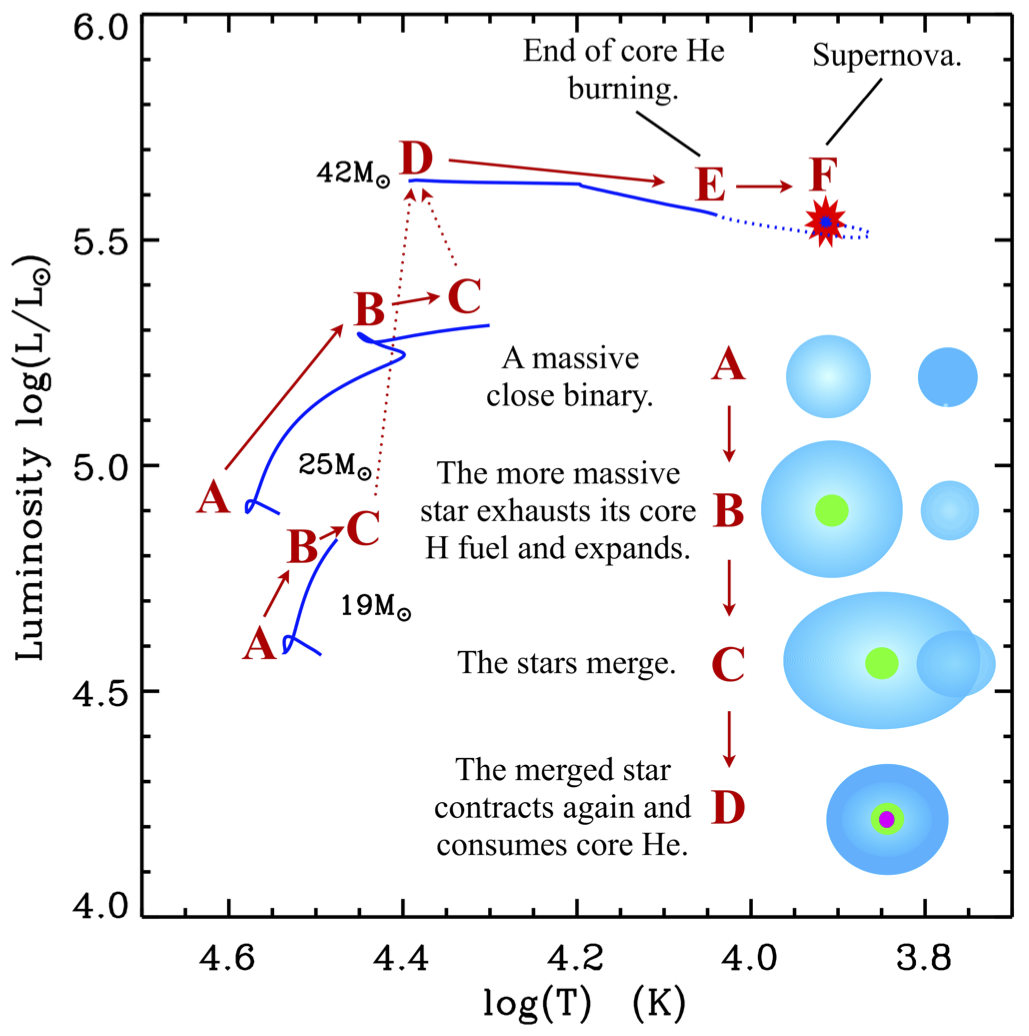}
\caption{\label{fig:schematic}
Schematic outline of the merger scenario which can lead to
LBV-SN progenitors. The top panel
illustrates a mechanism through which dynamically stable mass
transfer leads to a merger via a contact phase (see text for details
and uncertainties); in addition to this early Case B mechanism, some
Case A contact binaries may also become unstable and merge as the
primary leaves the main sequence. 
The lower panel shows an example in which a binary initially composed of
a 25 $M_{\odot}$ star and a 19 $M_{\odot}$ star merges to form a 42
$M_{\odot}$ single star. That merger product becomes a BSG during
core He burning, then explodes soon after entering the region
where LBV outbursts are expected. The dotted part of the curve shows
the evolution after the end of core He-burning, although
uncertainties in our understanding of such stellar envelopes mean that
the precise shape of that part of the curve is very unlikely to be
accurate. \emph{(This figure has a reduced resolution for astro-ph.)}}
\end{figure}

\subsection{The diversity of merger mechanisms}

There are multiple potential causes for stellar mergers. Perhaps the
best-known binary merger scenario occurs for binary systems which
are unstable to mass transfer on a dynamical timescale.  For stars with
radiative envelopes, this is thought to occur only when the donor star is a factor
of a few more massive than the accretor
\citep[see, e.g.,][]{HjellmingWebbink1987}. 
Those high-mass-ratio mergers
would only allow primary stars to increase their mass by a
correspondingly small fraction. 
Mergers following such an instability
would thus not allow stars which are typically expected to produce a
neutron star -- i.e., those with an initial main-sequence mass of
$\approx 25~M_{\odot}$ or less -- to become massive enough to
produce LBVs (unless the mass limit for LBVs is significantly below
$35~M_{\odot}$). 
 
However, there is an alternative merger mechanism that can
produce the LBV-SN progenitors which
this paper aims to explain. 
These mergers would occur as the primary star is
expanding away from the main sequence.
It is well established that some massive binary systems which
transfer mass early in the HG can enter into a contact phase. The
reason for this is that the accreting star is forced to swell up,
since the accretion timescale is much shorter than the thermal timescale of
its envelope.  This is expected to occur when the donor is at least
25\% more massive than the accretor \citep{Pols1994}, although there
is some subtlety in how the precise timing of the mass-transfer phase
affects the formation of contact \citep{Wellstein+2001}. 
Many of the systems which enter contact should then
merge, though it is difficult to be precise about how large a
fraction will do so \citep{Podsiadlowski2010}. We
discuss the likely rates in \S \ref{sec:rates}. Even though the 
occurrence rate of that merger mechanism is not precisely known, we
consider its existence to be relatively
robust. Fig.\,\ref{fig:schematic} presents a schematic of the scenario,
along with a labelled example of a potential binary evolution.

In addition to that early Case B binary merger mechanism, some Case A
massive contact binaries likely become unstable and merge after the
primary starts to expand at the end of the main sequence.  The
structure and evolution of contact binaries is one of the areas of
stellar evolution which is extremely poorly understood, and therefore
we only consider them briefly in the rest of the paper. Nonetheless,
some of these systems might increase the rate of LBV-SN
progenitors from a formation channel that is similar to the one which
we explore in detail.

As yet another possibility, massive stars in dense young clusters do
not need to experience a standard binary instability
to be involved in a merger: for example, stellar dynamics can directly
lead to collisional mergers. 
One previous suggestion for the progenitor of SN 2006gy involved the formation of 
a very massive star by runaway collisions in a dense young star cluster
(\citealt{SPZ+vdH2007}; \textbf{see also \citealt{vdH+SPZ2013}}). 
However, this did not explicitly account for the
inferred properties of the progenitor star at explosion, nor explain
why that merger product could produce a strong SN. 
Our scenario would apply in this case if the multiple mergers occurred soon after the primary had left
the main sequence.  We will not consider whether the timing of these 
mergers based on stellar dynamics is likely to occur often enough to be significant.  
However we will argue that, should early Case B primary
stars gain mass from multiple mergers in stellar clusters, this would
enable the formation of extremely luminous pre-SN BSGs
with cores which still seem likely to avoid direct collapse to a black hole. 
 
\begin{figure*}
\centering
\epsfig{file=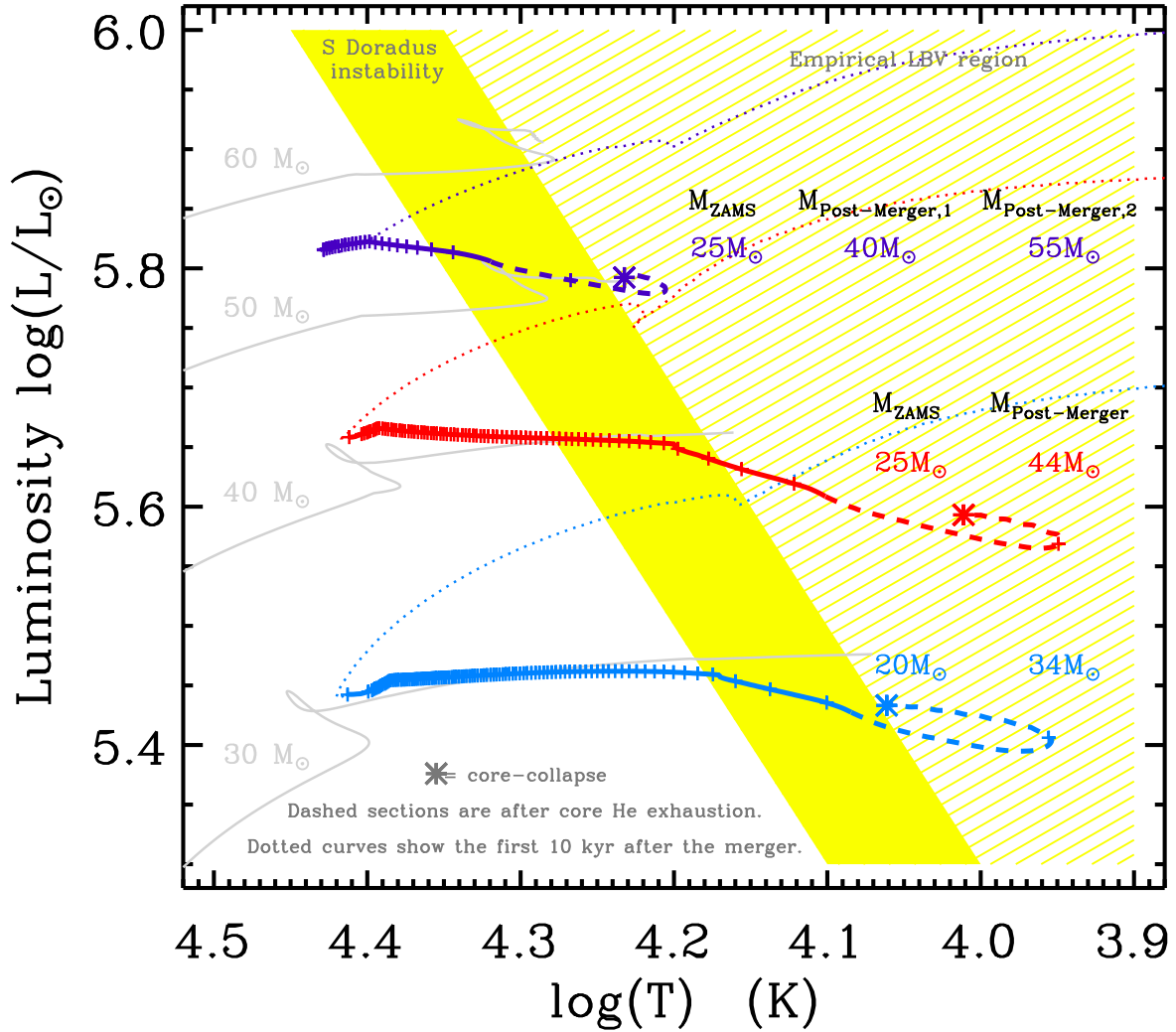, width=17cm}
\caption{\label{fig:HR}
Evolutionary tracks in the Hertzsprung-Russell
(H-R) diagram for three post-merger (or post-accretion) stars,
represented by the three thicker curves, for which the
  initial masses ($M_{\rm ZAMS}$) and post-merger masses are
as indicated.  Of those, the lower two curves
represent the evolution of stars that merged soon after the
completion of core H burning (these are blue and red in the
  online version). 
Most of the post-merger lifetime is spent undergoing
core He burning, shown by the solid section of each curve; the `+'
signs mark points in the evolution separated by $10^{4}$ years.  The
dashed section shows the post-core-He-burning phase, whilst the
dotted section represents the brief post-merger contraction.  The
first merger in each case occurs when the primary star had a surface
temperature of $10^{4.3}$ K. The uppermost of the post-merger curves shows an evolutionary
track which may be produced by a triple star interaction (this
is purple in the online version). In that
case, the second merger or accretion event was triggered at a surface temperature of
$10^{4.25}$ K (see \S \ref{sec:triples} for a discussion of the
likelihood of similar events).
The early evolution of four representative massive single stars is shown as thin
grey curves. The empirical S Doradus instability strip and the region
where these stars are likely to experience LBV outbursts are also
indicated. Note that, especially for the blue and purple curves, only
a few tens of thousands of years are spent in a region which is
typically considered to be potentially unstable to LBV outbursts.}
\end{figure*}

\begin{figure*}
\centering
\includegraphics[width=17cm]{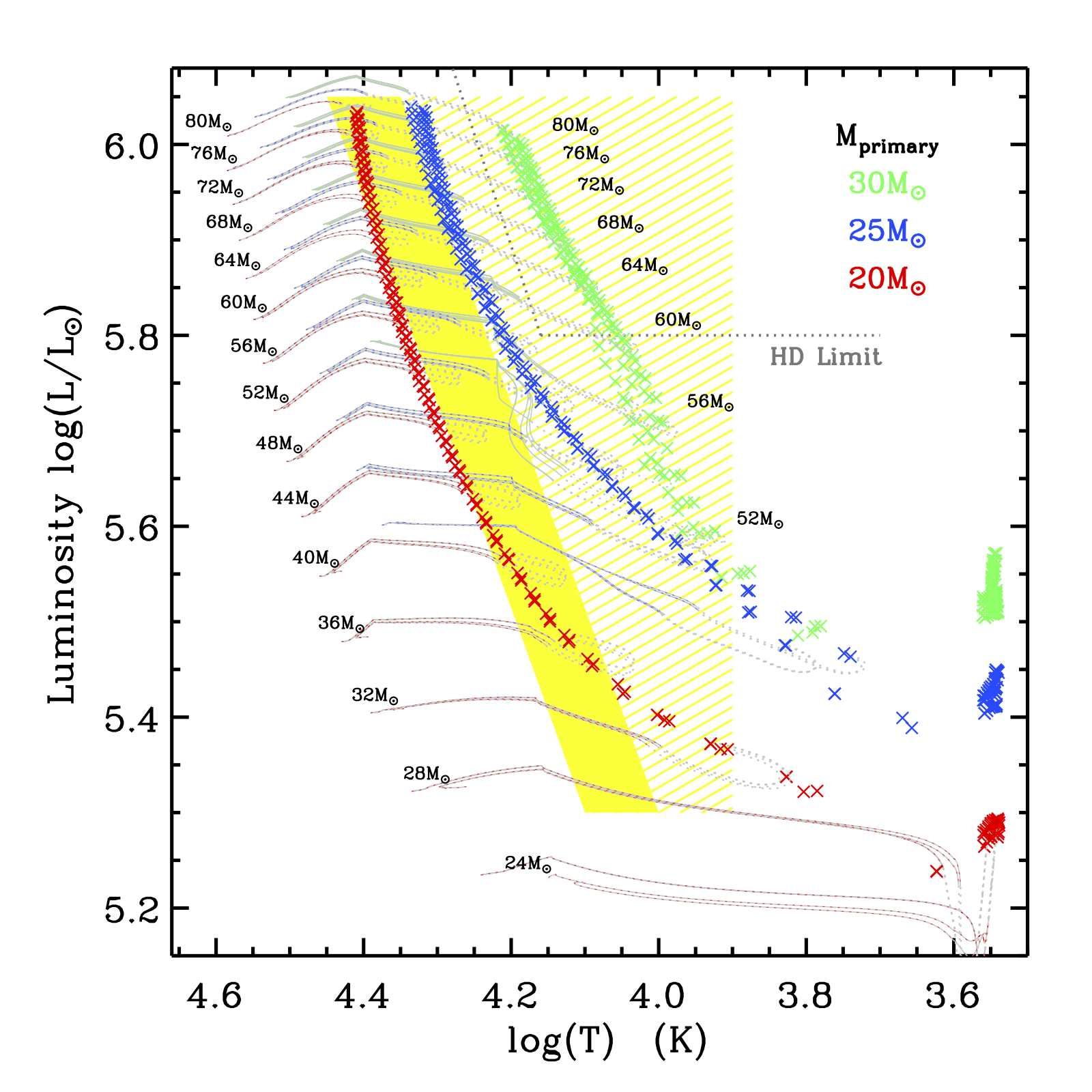}
\caption{\label{fig:HRmulti}
A systematic exploration of the final location
in the H-R diagram for a range of stars which gain mass during their
Hertzsprung Gap for three initial
primary masses (20, 25 and 30 $M_{\odot}$, which respectively produce the
  three sequences from left to right, and which are respectively colored red,
blue and green in the online version).  Not all of these models represent states
accessible to the merger scenario which we consider through most of
this work, but the most massive ones indicate what might happen
after, e.g., multiple mergers.
Models were evolved in increments of 1 $M_{\odot}$ in post-accretion
mass between the initial primary mass and 80 $M_{\odot}$, with the
endpoints marked as crosses. For the marked masses, the evolutionary
tracks are also shown (the first 10 kyr are omitted, and the dashed
segment again represents the post-core-He-burning stage).  For each
post-merger mass, the different tracks represent different points in
the HG at which the merger occurred (see text).  Increasing mass gain allows a star to
reach core collapse as a yellow supergiant or LBV. \emph{(This figure has a reduced resolution for astro-ph.)}}
\end{figure*}

\begin{figure}
\centering
\epsfig{file=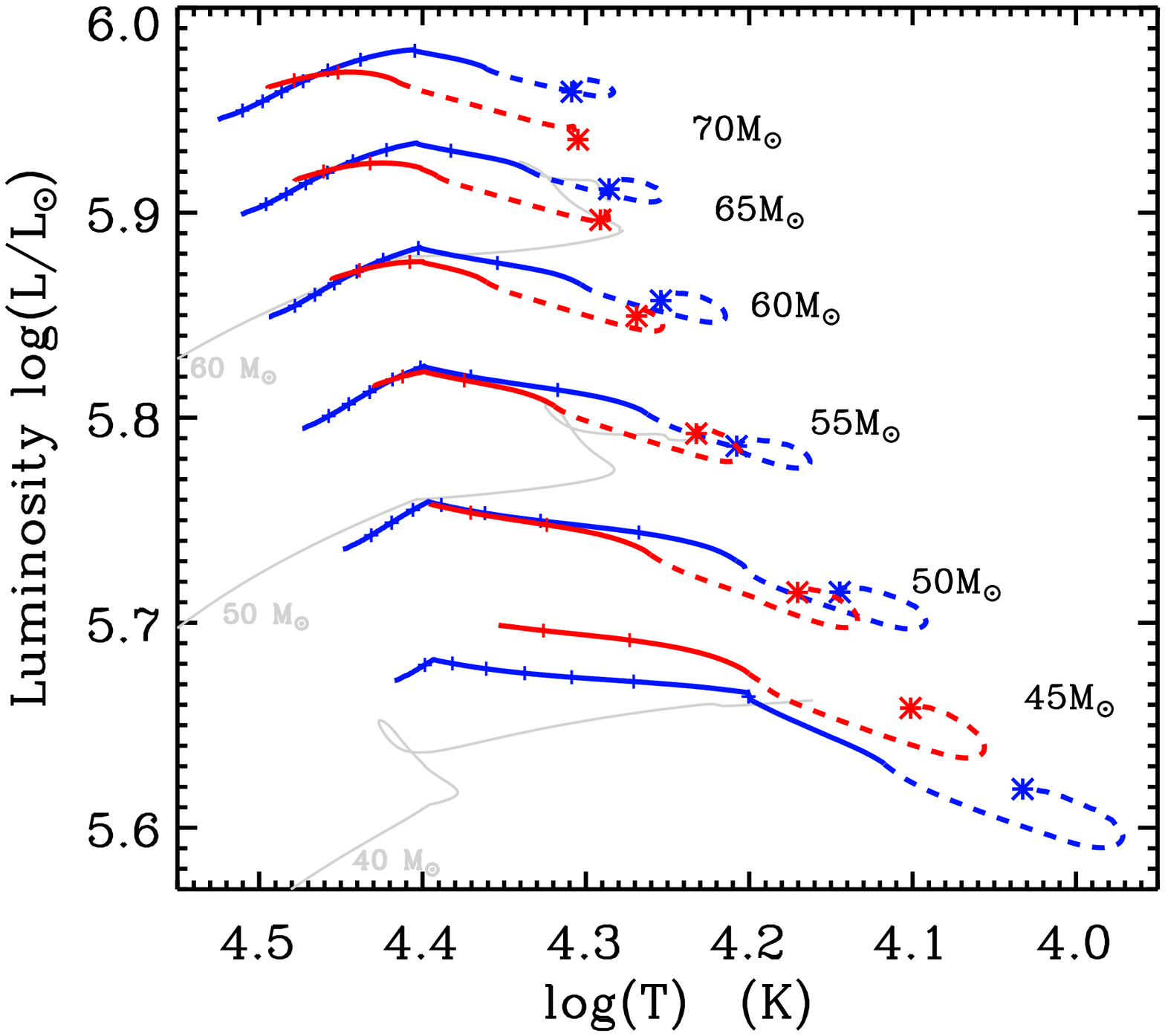, width=9cm}
\caption{\label{fig:HRtriples}
We demonstrate that the final appearance of post-accretion stars are
broadly unaffected by whether the accretion occurs in one phase or
two, for a representative set of models.  
As in Fig.\,\ref{fig:HR}, the dashed segments mark
post-core-He-burning evolution, and we show
the early evolution of some massive single stars with \textbf{thin} grey
curves. Here the `+' mark points separated by $10^{5}$ years. 
As in Fig.\,\ref{fig:HRmulti}, we omit the first 10 kyr of the
post-accretion evolution. The initial primary mass for all cases was
$25~M_{\odot}$.  The darker thick curves (blue in the online
version) represent models in which there was only
one phase of accretion, which began when the primary had a surface
temperature of $10^{4.3}$ K. Further accretion onto the $40~M_{\odot}$
post-accretion model from that set was used to create the models
represented by the lighter thick curves curves (red in the
  online version). That second phase of accretion began
when the first post-merger star had a surface temperature of
$10^{4.25}$ K.  We compare pairs of models with the same
post-accretion mass (as marked on the plot). }
\end{figure}

\subsection{Aims and structure of this work}

This work examines the proposition that LBV-SN progenitors 
may form from massive binary
systems in which the components merge soon after the more
massive star has finished core H burning, i.e., as it is
expanding across the Hertzsprung Gap (HG).  
As described above, we nominally consider the situation in which the
stellar merger follows mass transfer from the early HG
primary to the secondary and then a brief period as a contact
binary. However, most of the calculations we present would also apply
to other situations in which a massive star might gain mass during that
evolutionary phase. Therefore this manuscript will describe
the stellar models we create as both ``post-merger'' and ``post-accretion''. 

In \S \ref{sec:PPE} we study the appearance and evolution of relevant merger
products (or post-accretion stars) using Eggleton's stellar evolution
code. Partly to demonstrate that
our conclusions are robust to reasonable variations in stellar
physics we also perform similar calculations using the MESA stellar
evolution code, presented in \S \ref{sec:MESA}. 

In \S \ref{sec:BHvsNS} we investigate whether these merger products are
likely to produce neutron
star (NS) or black hole (BH) after core collapse.  

Section \ref{sec:rates} then estimates event rates for some of
the routes through which early Case B accretion might occur,
including comparing the potential birth-rates of SN progenitors to 
current observational constraints. Finally, \S \ref{sec:rates} briefly
discusses how often one of these early Case B merger products might accrete even more
mass from a potential tertiary companion before reaching core collapse.

\section{Post-accretion stellar evolution calculations I: Using Eggleton's Code}
\label{sec:PPE}

\subsection{Assumptions}

We created and evolved a set of post-accretion models using
the Eggleton evolutionary code \citep[see,
e.g.,][]{PPE1971,PPE1972,PPE1973,Pols1994,Pols+1995}, and adopting a metallicity of $Z= 0.02$.  The
code assumes the Schwarzschild criterion for convection, and
semi-convection is automatically produced by the code's treatment of
convective mixing. The equation of state follows the treatment of \citet{PPE+1973eos}.
We adopted the overshooting calibration from
\citet{Pols+1998}, which was performed for intermediate-mass stars but is
commonly also assumed for massive-star calculations when using the
Eggleton code \citep[see,
e.g.,][]{EldridgeTout2004,EldridgeVink2006}.  This overshooting is
parametrised in a non-standard way, which corresponds to different
numbers of pressure-scale-heights for different stellar structures,
but the calibrated value is very roughly consistent with 0.25
pressure-scale-heights of overshooting.

Our mass-loss prescription follows the work of 
\citet{Vink+1999},\citet{Vink+2000} and \citet{Vink+2001} for hot stars.\footnote{We adapted the code
publicly available from: http://www.arm.ac.uk/$\sim$jsv/Mdot.pro.}
For stars cooler than $10^{4}$ K, we adopted a mass-loss rate based on
\citet{N+dJ1990}, multiplied by a factor of 0.3 to allow for
updated estimates of the effect of wind clumping \citep[see,
e.g.,][]{PulsVinkNajarro2008}. 
In practice, few of the post-accretion models spend much time with
surface temperatures cooler than $10^{4}$ K, and those are the stars in
which we are least interested (i.e., they do not end their lives as a BSG).

We first evolved a set of single massive-star models to the HG, and
saved snapshots at a range of stages across the HG.
To those models we rapidly added
mass to their envelopes. 
These post-accretion models were then evolved to
give the sequences shown in Figs.\ \ref{fig:HR} and
\ref{fig:HRmulti}.  To some of the evolved post-accretion models we
followed the same procedure again, adding more mass as the star
expanded during or after He-burning. One example of such a
model is shown in Fig.\ \ref{fig:HR}; in Fig.\ \ref{fig:HRtriples} we
show that, for the assumptions we have used,
the final appearance of stars for which the accretion
occurred in one or two phases is relatively small.

We assume that matter is accreted onto the primary star with the
surface entropy and composition of the accretor rather than with,
e.g., the mean composition and entropy of the secondary star.   This
assumption deserves further study in
future work.   However, the thermal structure of the star will
recover after a few thermal timescales, i.e., much less than the remaining
evolutionary time of the merger product.  This suggests to us that our
broad conclusions are unlikely to be affected by this assumption,
which is supported by the fact that we added mass to primaries with a range of surface
temperatures in the HG (and hence studied a range of accretion
entropies, as well as slightly different stellar structures at the
onset of accretion), but the future evolution of these different post-accretion
stars shows only minimal variations, as we will discuss later. 

Any systematic bias introduced by not accreting slightly He-rich
matter is harder to quantitatively estimate. However, an increase in envelope He abundance
increases the parameter range where stars explode as blue supergiants
\citep{BarkatWheeler1988,BarkatWheeler1989corridor,HillebrandtMeyer1989}. 
Hence slight He-enrichment might be
expected to be favourable to the production of BSGs at explosion, and
also to delay the point at which the post-accretion star expands to
become unstable to LBV outbursts. 

As our simulation of the merger process is very simplified, and partly
for reasons of numerical stability, we did
not include wind mass loss when the merger product is contracting
immediately after the merger. 
The amount of mass lost during this very
short time ($\lessapprox10^{4}$ yr) is unlikely to be significant, as
demonstrated in \S \ref{sec:MESA}, where we perform similar
calculations without switching off stellar winds during this phase. 
We also decided not to apply an ad-hoc LBV mass-loss
prescription. Our general conclusions should be unaffected, since the
time spent in the LBV phase is extremely short for most of
the models presented in this study.  Although the final locations of
our stellar models in the H-R diagram are not precise, this would
have been true whichever LBV mass-loss treatment had been applied. 

These calculations include no treatment of stellar rotation. However,
since the accretion occurs after strong composition gradients have already
been generated by nuclear burning on the main sequence, even rapid
rotation caused by accretion of angular momentum is highly unlikely to
lead to additional mixing across those gradients
\citep[see, e.g.,][]{Mestel1953,Mestel1957,Mestel+Moss1986}, and
therefore we consider that the internal evolution will not be
qualitatively altered by rotation.  If the cores of the
merger products were greatly spun-up during the merger then their
future evolution could be affected. This may sometimes occur, but we
expect that in the majority of cases the additional angular momentum will be
gained by the envelope rather than the core.   
(Angular momentum transport between the He core and H
envelope was found to be small in the stellar models calculated by
\citet{Yoon+Langer2005}, though those examples were for stars in which
the core was rotating more rapidly then the envelope, i.e., the
reverse of the seemingly-likely case for these stars.)  On the other hand,
rapid rotation may well increase the likelihood that LBV outbursts
occur, and increase the mass-loss rates from the surface
\citep{Langer1997,Langer1998}. There is some suggestion that the
angular-momentum loss from less massive merger
products during their brief giant phase may be relatively rapid, but
the constraints are far from definitive \citep[][]{Eggleton2010}.
Our MESA calculations -- presented in \S \ref{sec:MESA} -- include comparisons
between non-rotating and rotating post-merger models.

\begin{figure}
\centering
\epsfig{file=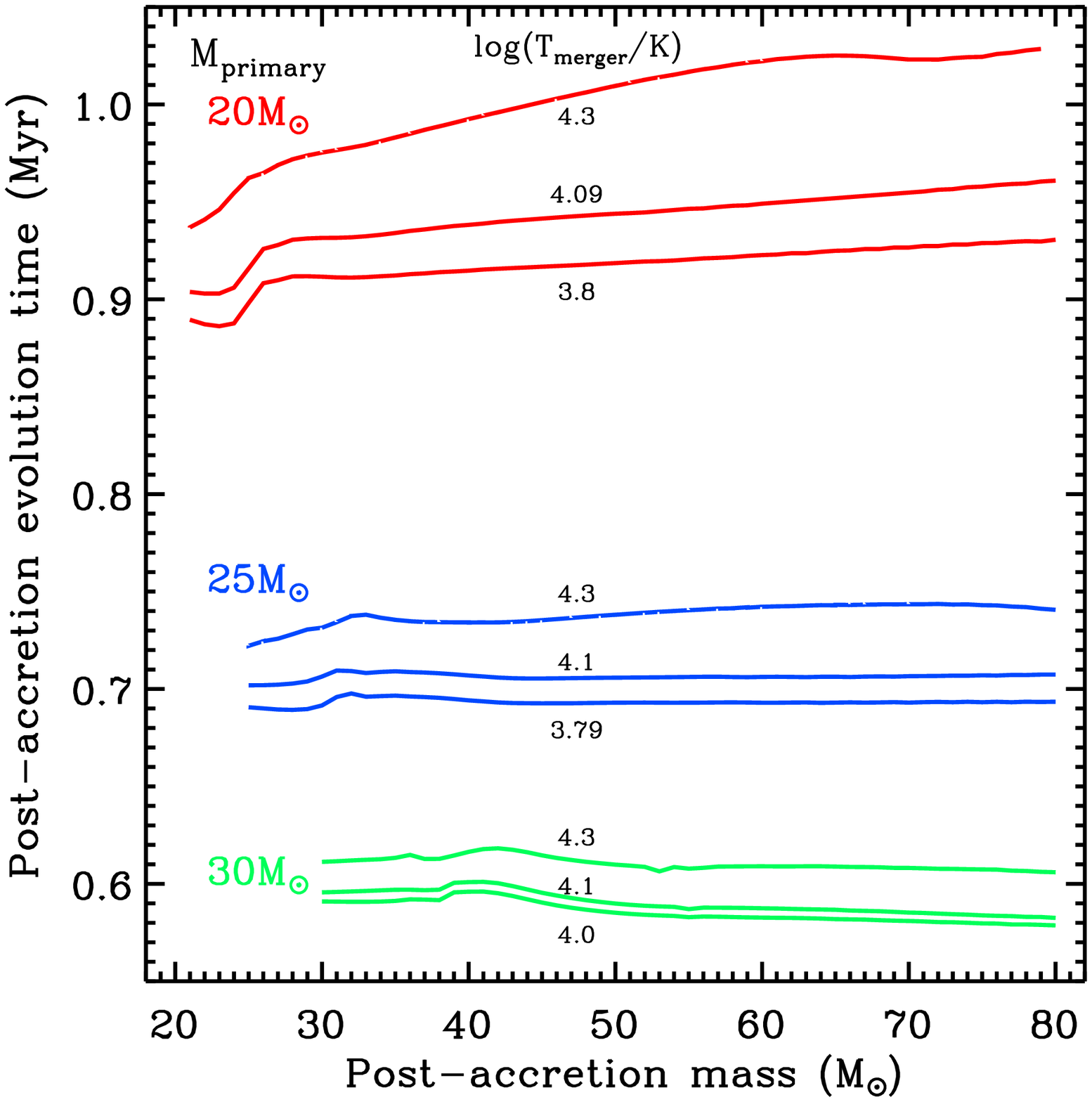,
  width=9cm}
\caption{\label{fig:lifetimes} For our set of calculations using
  Eggleton's code, we show the duration of the post-merger evolution
  as a function of the post-accretion mass (i.e., nominally until the
  end of core carbon burning; carbon burning is numerically unstable,
  and not all models complete that phase successfully, but that would
  only produce a negligible error in this duration). The lifetimes of
  the merger products are governed by the the primary mass (as
  marked), and in most cases are only weakly affected by the amount of
  mass accreted.  The surface temperature of the primary at the start
  of accretion is marked by each curve; this factor does have some
  effect on the lifetime of the merger product, with smaller
  (i.e., hotter) pre-accretion stars living somewhat longer. }
\end{figure}

\begin{figure}
\centering
\epsfig{file=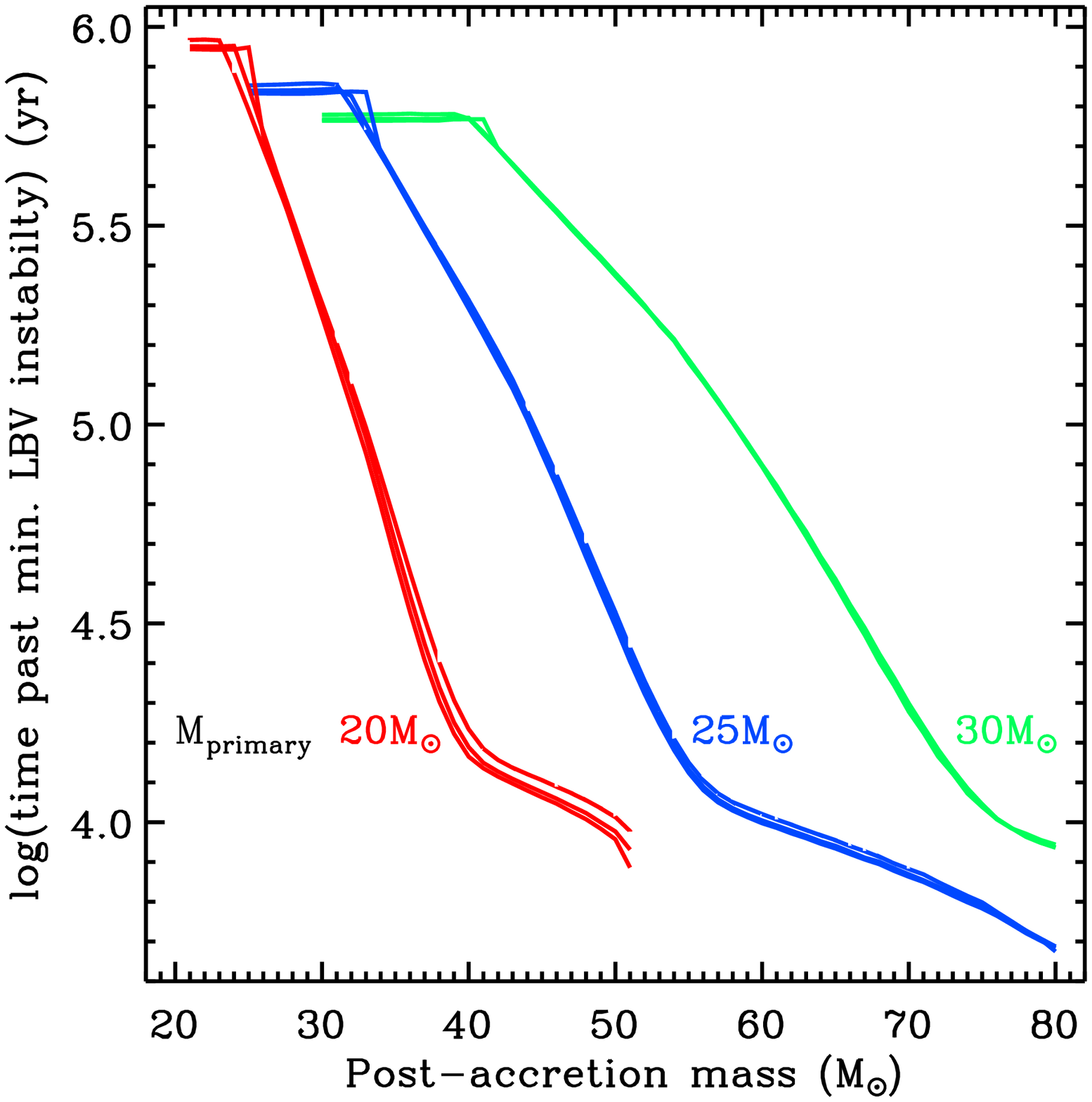,
  width=9cm}
\epsfig{file=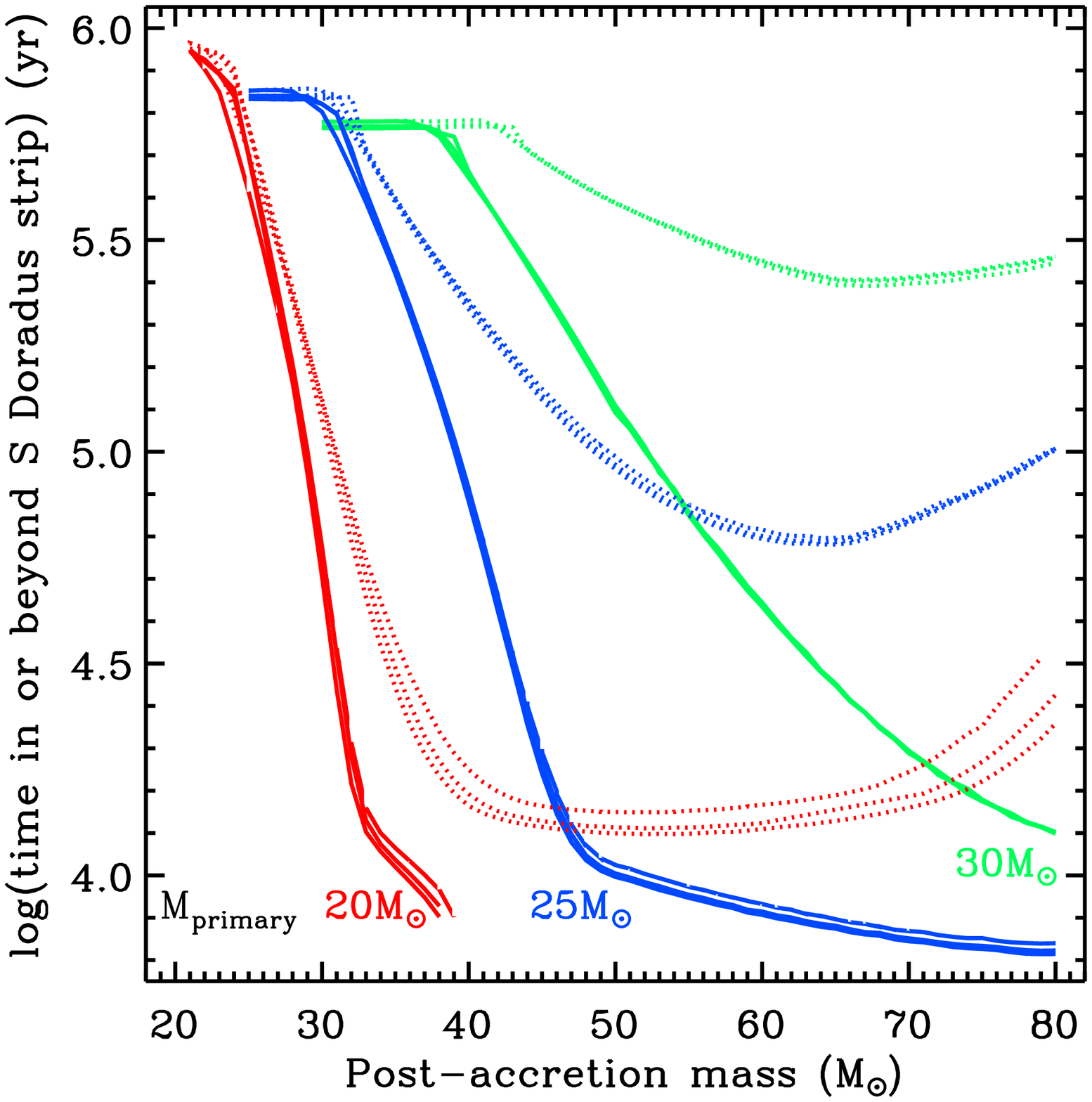,
  width=9cm}
\caption{\label{fig:LBVtimes} 
Based on the calculations in \S \ref{sec:PPE} and different inferred empirical
LBV instability regions, we estimate the length of time for which our
post-merger models might be subject to LBV outbursts.  Again, for each
primary mass (as marked) we show results for different timings of the merger
within the HG; in this case, this almost always leads to small
differences in outcome.  The upper plot
shows the length of time for which the post-merger models exceed the
minimum LBV instability criterion given in \citet{Groh+2009}. The
lower panel uses the S Doradus instability strip plotted in
Figs. \ref{fig:HR} \& \ref{fig:HRmulti} as the instability criterion,
with the broken curves using the left edge of the strip, and the solid
curves the right edge of the strip.  We stress that these are only
intended to be rough estimates (especially since the criteria for LBV
instability are not well understood), and also that these criteria
adopt no minimum luminosity cut-off (i.e., it is extremely unlikely
that a 25 $M_{\odot}$ post-merger star would display LBV
outbursts).}
\end{figure}

\subsection{Results}

As the example sequences in Fig.\,\ref{fig:HR} demonstrate, the
post-merger stars spend most of their remaining lives
burning He in their cores whilst appearing in the
H-R diagram to the left of the regions where LBV
outbursts are thought to occur. Only late in their nuclear
evolution do they start to expand and become potentially unstable to LBV
outbursts.   For the LBV instability region marked in Fig.\,\ref{fig:HR}, 
both the $34~M_{\odot}$ and $44~M_{\odot}$
examples would only spend their last few tens of thousands of years
unstable to outbursts.

One uncertainty which seems to have little qualitative effect on the later
evolution is the timing of the accretion (or merger) within the early HG. 
For the calculations displayed in Fig.\,\ref{fig:HRmulti}, the
ranges of surface temperature at the time of the mergers was
$\rm 10^{4.3}~K$--$\rm 10^{3.8}~K$ (for the $20~M_{\odot}$ and
$25~M_{\odot}$ primary stars) and $\rm 10^{4.3}~K$--$\rm 10^{4.0}~K$
(for the mergers involving a $30~M_{\odot}$ primary).   These temperatures
correspond to ranges in mass at the onset of accretion of
19.49--19.45$~M_{\odot}$, 23.80--23.45$~M_{\odot}$ and
27.86--27.75$~M_{\odot}$ for ZAMS masses of 20, 25 and 30$~M_{\odot}$,
respectively.  Since this change in the timing leads to
relatively minor changes in the post-merger evolution, we conclude that the
precise timing of the onset of accretion within the early HG is not significant for our
current study.   Fig.\,\ref{fig:lifetimes} shows that the post-merger
lifetime is affected by the timing of the merger, but in most cases
the fractional difference is small. 
The range of difference introduced by changing the timing of the
merger could easily be smaller than the other uncertainties in the stellar physics. 

This logarithmic decrease of 0.5 dex in effective temperature 
corresponds to a radius increase by a factor of 10 (assuming constant
luminosity, as is roughly appropriate for the early HG). Our
population estimates later adopt a factor of 10
radius increase within the HG as the range of parameter space during
which mergers can lead to suitable SN progenitors. Based on these
calculations, we suggest that this factor of 10 in stellar expansion
may well be conservative.

Since the location in the HR diagram where LBV outbursts
occur is not precisely known -- and, as noted above, may well be influenced
by rotation -- the S Doradus instability strip
marked in Figs.\,\ref{fig:HR} and \ref{fig:HRmulti} can only be
indicative. \citet{Groh+2009} argued that 
the `minimum LBV instability strip' is steeper in the HR diagram than in previous work,
located at an effective temperature of $\approx \rm 10^{4.2}~K$ for a luminosity of
$10^{5.414}~L_{\odot}$ (and $\approx \rm 10^{4.3}~K$ for a luminosity of
$10^{5.9}~L_{\odot}$).\footnote{\citet{Groh+2009} give ${\rm log} (L/L_{\odot}) = 4.54
  ~{\rm log}(T_{\rm eff}/{\rm K}) - 13.61$.}  This is typically
slightly hotter than the left edge of the instability strip
marked in Figs.\,\ref{fig:HR}  and \ref{fig:HRmulti}. Figure
\ref{fig:LBVtimes} presents the length of time for which our models
stars would be unstable to LBV outbursts for these different
assumptions. Note that the plots do not include any arbitrary assumptions about
a minimum absolute luminosity for LBV outbursts.
Nonetheless from Figs.\,\ref{fig:HRmulti} and \ref{fig:LBVtimes} it
is clear that -- wherever the real minimum
instability strip lies -- models can be created which would only become
unstable to LBV outbursts very late in their evolution,
potentially even after the end of core He burning. As the
instability strip moves to higher temperatures then the primary mass
required to achieve this becomes lower.  This systematic effect
suggests that as the real instability strip becomes hotter then higher
post-merger luminosities will be harder to achieve for canonical binary merger
channels (and so would perhaps require, e.g., multiple mergers in dense
stellar systems or triple-star scenarios). Cooler instability strips
would enable merger products resulting from increasingly massive
primaries to become unstable later in their post-accretion evolution.

The criteria adopted in Fig.\ \ref{fig:LBVtimes} are very uncertain,
and the results shown therein can only be rough
estimates. Nonetheless, we note that the time spent subject to LBV
outbursts is predicted to be a fairly sensitive function of the
post-accretion mass, which is in strong contrast to the post-merger
lifetimes shown in Fig. \ref{fig:lifetimes}.

\subsection{Yellow supergiants}

Figure \ref{fig:HRmulti} also indicates that there is a region of
parameter space where merger products finish their nuclear burning as
yellow supergiants (YSGs). Such SN progenitors were again not predicted in
canonical single-star models but have been suggested observationally
\citep[see, e.g.,][]{Fraser+yellow2010}. 
The models which lead to core collapse as YSGs are those systems in
which less mass was added to the primary
than necessary to produce BSG or LBV SN progenitors. For example, the
$40~M_{\odot}$ post-merger star produced from a $25~M_{\odot}$ primary
finishes its evolution with a surface temperature of $\rm 10^{3.8}~K$.

\begin{figure}
\centering
\epsfig{file=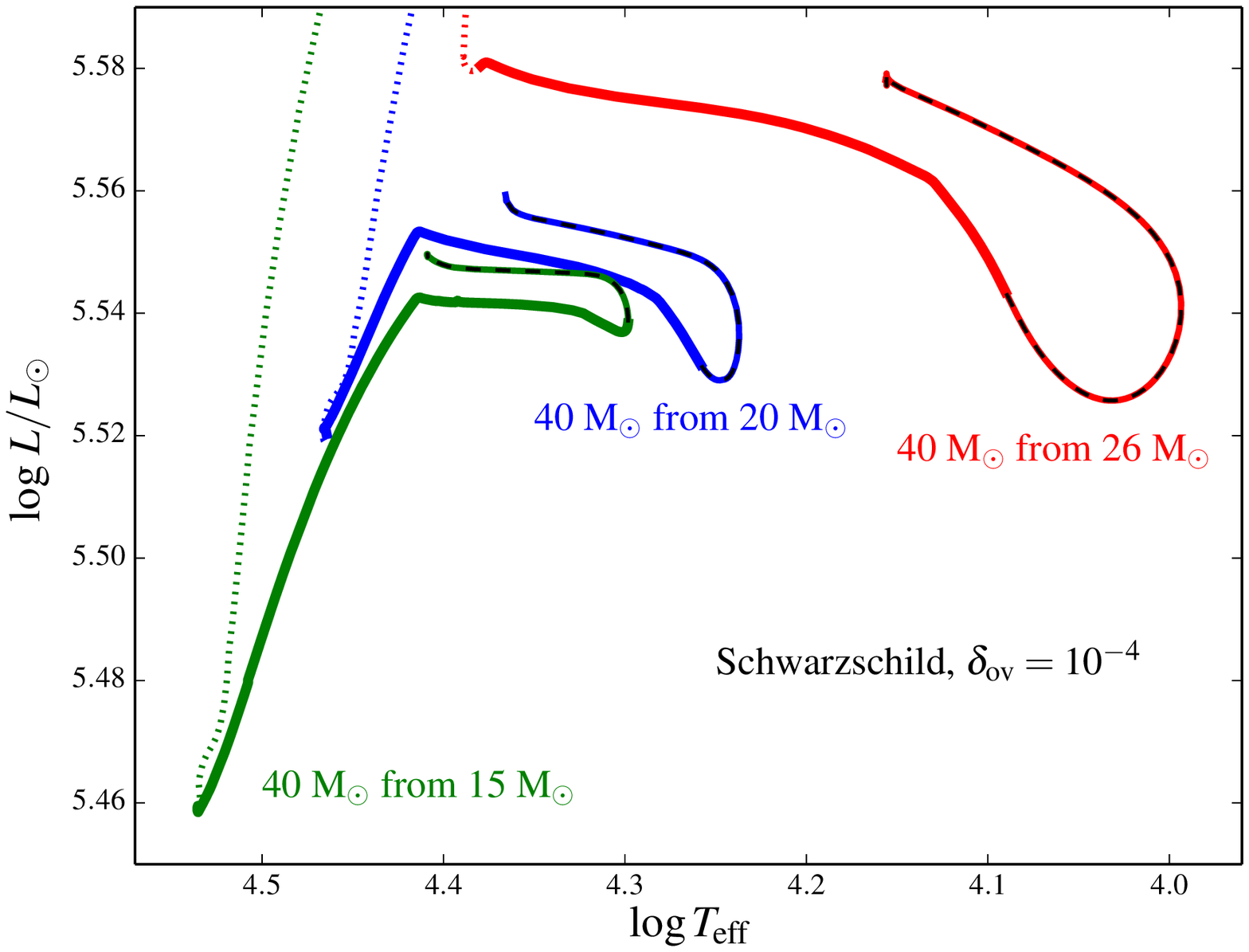,
  width=9cm}
\epsfig{file=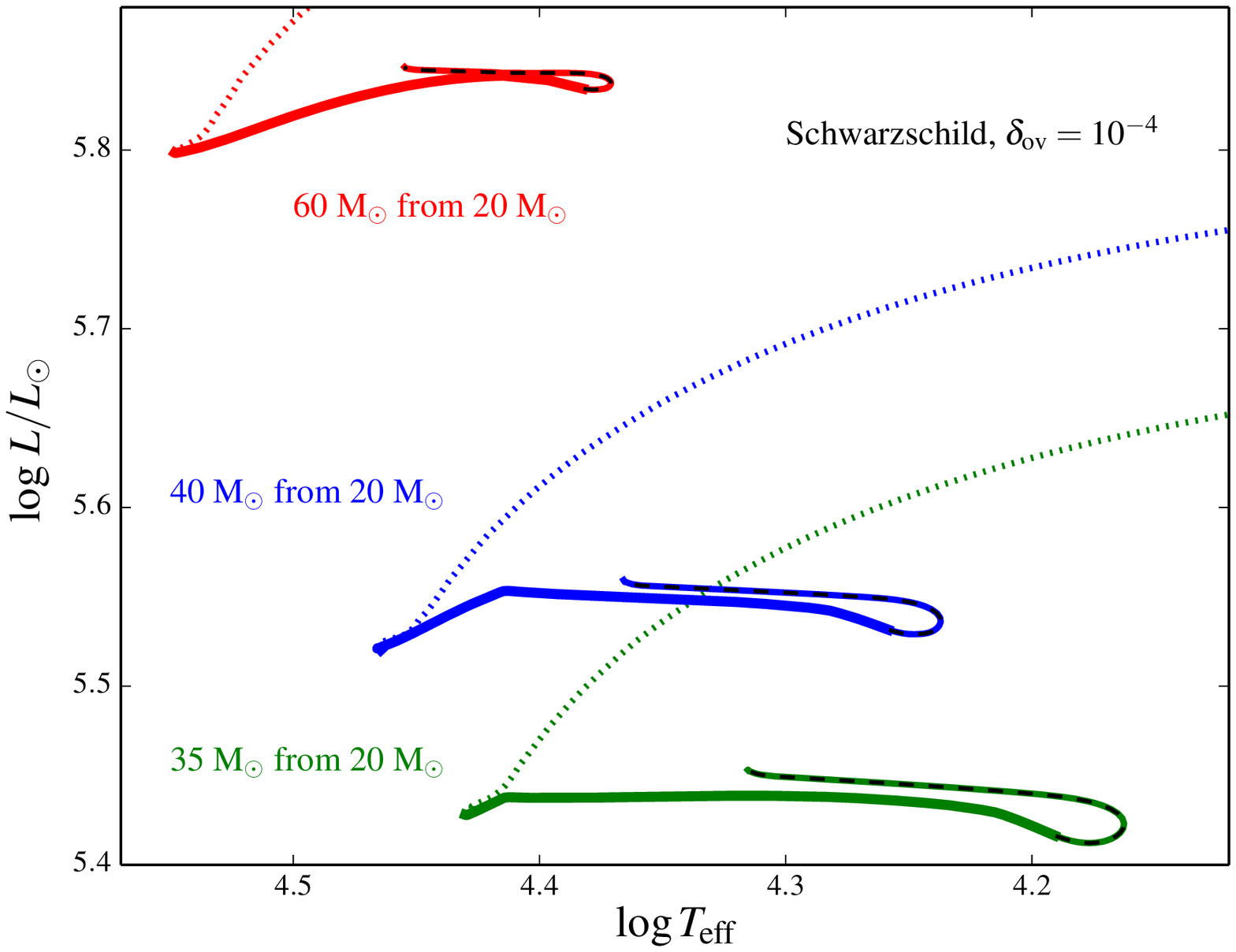,
  width=9cm}
\caption{\label{fig:MESA_HRSchwovs}
The features of the post-accretion evolution for
our MESA calculations are similar to those shown
in \S \ref{sec:PPE}.  These MESA models are of non-rotating
stars and adopt the Schwarzschild criterion with overshooting (see text). The dotted
curves represent the post-accretion thermal contraction phase, whilst
the black-and-color dashed sections of the curves represent the
post-core-He-burning phase. For all of these models, the accretion 
began when the radius of the primary was $30~R_{\odot}$.
\textbf{Upper panel:} we compare the evolution of a 40
$M_{\odot}$ star formed from three different primary masses (as
marked) by early
Case B accretion. \textbf{Lower panel:} we compare the appearance of
three stars each formed by early Case B accretion onto a 20 $M_{\odot}$ primary, with three
different post-accretion masses (as marked).}
\end{figure}

\section{Post-accretion stellar evolution calculations II: Using MESA}
\label{sec:MESA}

We have no reason to doubt the calculations presented in \S
\ref{sec:PPE}. Nonetheless, we recognise that the evidence in favour
of our model would be strengthened by presenting further
calculations using an alternative code, especially since models for
the BSG progenitor of SN 1987A were dependent on the assumed physics
\citep[see, e.g.,][]{Weiss1989,LangerEidBaraffe1989,Podsiadlowski1992}. For this purpose we 
use MESA \citep[as presented by][]{Paxton+2011,Paxton+2013}. 

The MESA calculations presented in this paper were performed
using version 5232.  We chose to use the ``Dutch'' wind-loss option,
based on the choices made in \citet{Glebbeek+2009},
which is similar to the wind-loss rates adopted in \S \ref{sec:PPE}. We again adopted a
metallicity of $Z=0.02$.  For the calculations presented here we used
the streamlined nuclear reaction network ``approx21'', which allows us
to follow approximate nuclear burning to the formation of an iron core.

The majority of the results presented were calculated using the
Schwarzschild criterion for convective instability. We later compare
calculations using the pure Schwarzschild criterion (with no
overshooting), and the Schwarzschild criterion with significant overshooting.
For overshooting we adopt the standard treatment in MESA,
which follows the exponential overshooting treatment of
\citet{Herwig2000}; the set of calculations with overshooting takes
all of those exponential overshooting parameters, $\delta_{\rm ov}$, to be
$10^{-4}$. We will also present a set of calculations which use the
Ledoux test for convective instability, in which we adopt very
efficient semi-convection ($\alpha_{\rm SC} = 0.1$). Uncertainties in
stellar mixing physics are still substantial, and we certainly have not
tested all possible options. 
However, since our parameter choices with respect to overshooting
and semi-convection are on the larger side of the plausible range,
if they are in error, they seem likely to produce cores which
are more massive than might be the case in reality. More massive cores
(i.e., higher fractional core masses) are less favourable when trying
to produce BSG structures. Therefore -- for the scenario
we are testing in this paper -- we consider these parameter choices to
be conservative. We expect that less substantial overshooting, or less efficient
semi-convection, would be more favourable for the production of
massive BSGs. 

We adopted a similar procedure and assumptions to produce the merger
products as when using the Eggleton code except that when using
MESA we did not temporarily switch off wind loss during the brief
post-merger contraction.  The pre-merger stellar models, and the
accretion phase, were typically calculated with higher resolution than
the MESA default (setting the mesh-spacing parameter $C=0.1$, and with the
maximum allowed number of mesh-points increased to 40,000). 
Our post-accretion calculations typically adopted the
default resolution (with $C=1$).  For all of the models presented here,
the assumed accretion rate during the merger phase was $10^{-2} ~M_{\odot}\,
{\rm yr}^{-1}$.

\begin{figure}
\centering
\epsfig{file=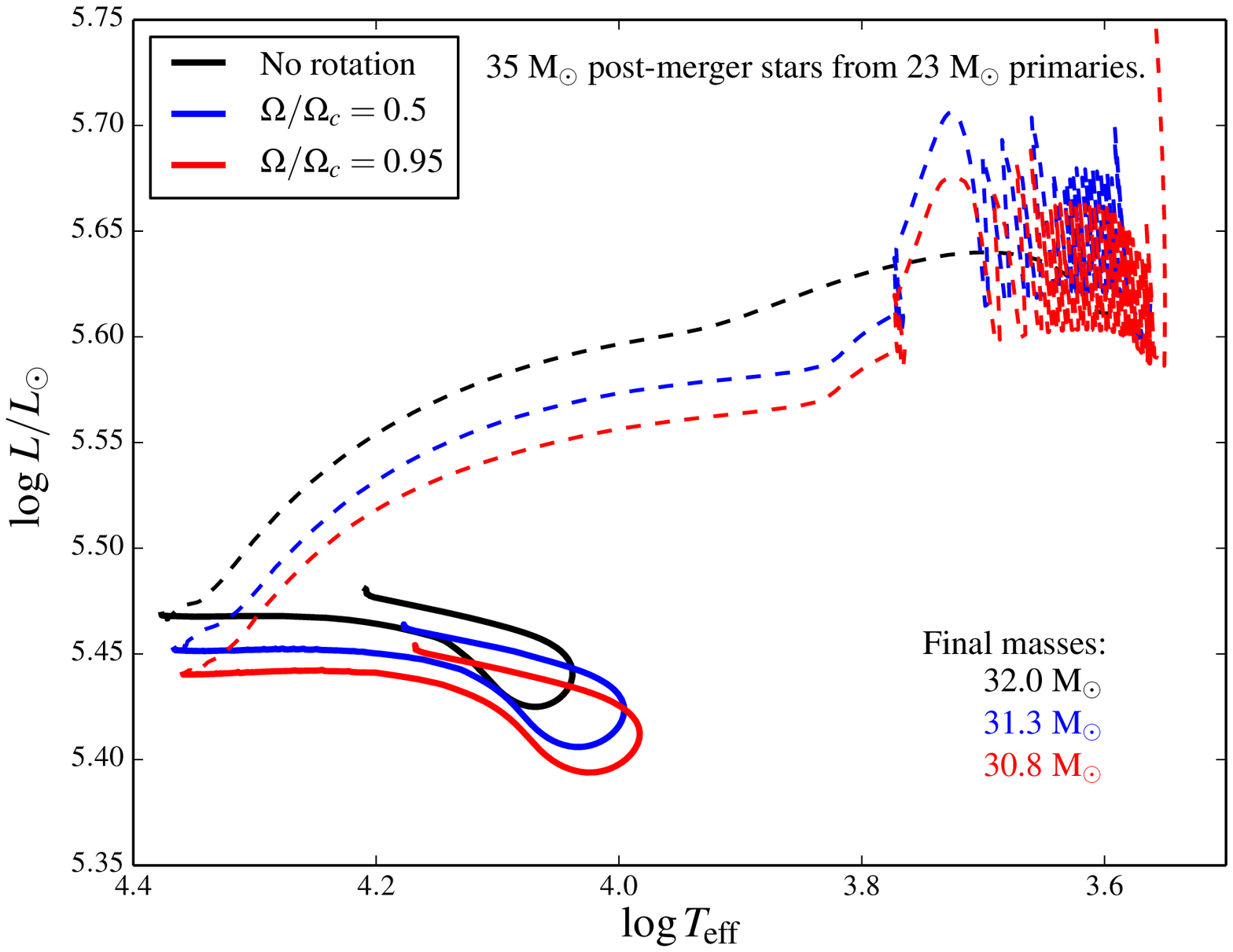,
  width=10cm}
\epsfig{file=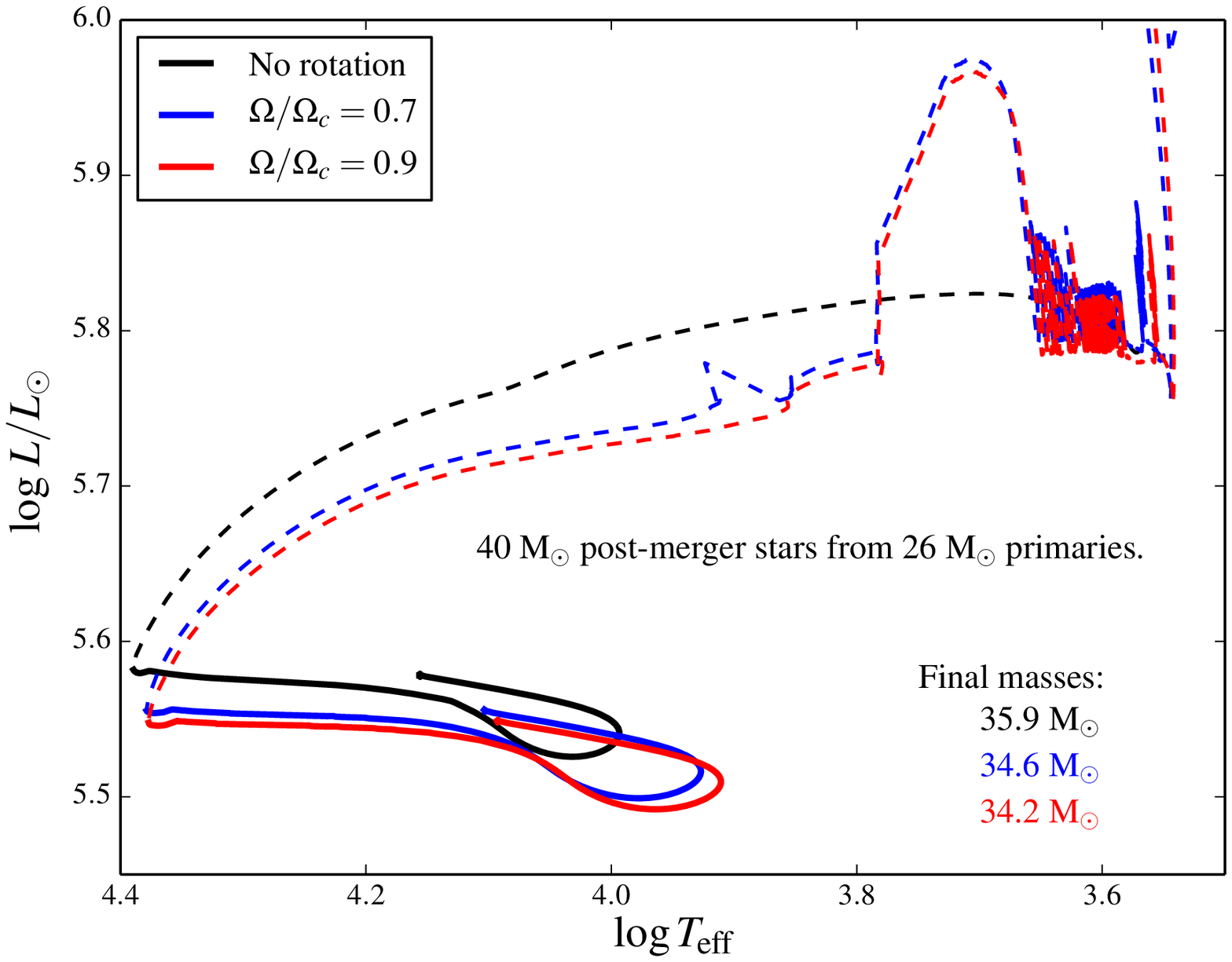,
  width=10cm}
\caption{\label{fig:MESA_rotation}
We show the effect of changing the assumed post-merger spin on the
stellar appearance for two different merger models ($35~M_{\odot}$
from a $23~M_{\odot}$ primary in the upper panel, and $40~M_{\odot}$
from a $26~M_{\odot}$ primary in the lower panel; all models adopt the
Schwarzschild criterion for convection with overshooting).  The
rotation rates
are as marked, given as a fraction of the critical rotation rate. 
As in Fig. \ref{fig:HR}, broken curves plot the first
$10^{4} \rm yr$ after the merger.  Whilst there are differences
during the initial contraction phase, which are associated with mass loss and
lower final stellar masses, the later qualitative evolution is broadly
unaffected by even large rotational velocities.}
\end{figure}

\subsection{The effect of post-merger rotation}

As noted in \S \ref{sec:PPE}, there is good reason to expect
that additional mixing effects due to rotation are minimal across strong
molecular-weight gradients \citep[see
especially][]{Mestel1953,Mestel1957,Mestel+Moss1986}. Since our
pre-merger stars already have H-exhausted cores, we therefore
considered that the internal evolution of the post-accretion stars is unlikely
to be qualitatively altered by rotationally-driven mixing.  
Here we attempt to test that assumption, by adopting the default set
of rotational-mixing physics in the version of 
MESA used for these calculations. This treatment is based
on \citet{Heger+2000mixing,Heger+2005mixing}, with the
main parameter -- the ratio of the turbulent viscosity to the diffusion
coefficient -- set to $1/30$.
Since rotationally-driven mixing is poorly-understood, this test cannot be considered
exhaustive. In addition, we cannot exclude the possibility that a
different post-merger angular momentum distribution would have
produced qualitatively different results. We simply assumed solid-body
rotation at the end of the accretion phase, which seems the most
plausible assumption for the deep convective envelopes possessed by the
post-accretion stellar models. However, it is possible that stellar
cores may be spun-up as a result of mergers.  

Despite those caveats, we consider that our results support our
assumption that rotational mixing does not make a significant qualitative
difference to the evolution of these post-merger stars, at least after
the initial contraction phase. Two representative examples are presented in
Fig.\,\ref{fig:MESA_rotation}. The only significant effect of
even extreme post-accretion rotational velocities is mass shedding as
the merger product contracts and spins-up \citep[see,
e.g.,][]{Heger+Langer1998}. Fig.\,\ref{fig:MESA_rotation} also displays the
final masses of the model stars; more rapidly-rotating post-merger
models do indeed lose more mass before core collapse.

\begin{figure}
\centering
\epsfig{file=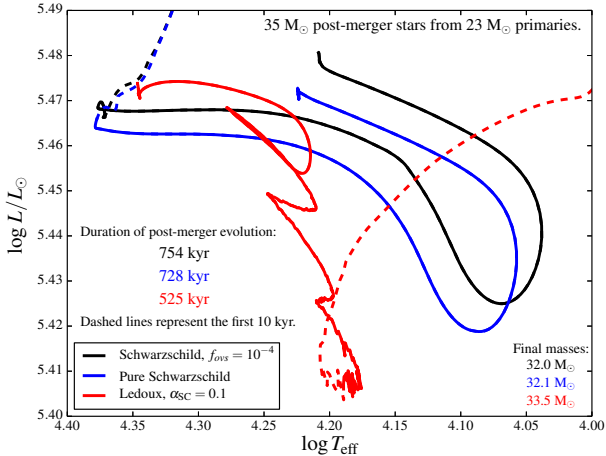, width=9cm}
\caption{\label{fig:mesa_mixing}
We compare three stellar calculations when adopting different assumptions about mixing (as
marked), for the same primary and post-accretion masses (23
$M_{\odot}$ and 35 $M_{\odot}$, respectively).  The post-merger
lifetime and final stellar masses are marked on the plot. The
post-accretion evolution of the models
which assume the Schwarzschild criterion is qualitatively similar,
despite the difference in overshooting. However, adopting
the Ledoux criterion does lead to a qualitatively different evolution
in the HR diagram; most notable is the appearance of sharp
turning points which are characteristic of breathing pulses during core
He burning.}
\end{figure}

\subsection{The effect of assumptions about convective mixing}

In Fig.\,\ref{fig:mesa_mixing}, we present one representative example
in which the same stellar evolution scenario (both pre-merger and
post-merger) was followed using the three different choices of
convective instability physics which were described at the start of this section. The final
location of the models in the HR diagram is surprisingly similar for
all three options, but the shape of the evolutionary track for the
model which adopts the Ledoux
criterion for convection is qualitatively very different from those
which adopt the Schwarzschild criterion. It is well known that the
shape of blue loops is sensitive to which mixing criterion is chosen,
and we speculate that the difference in the shape of the tracks may be a
combination of that effect with breathing pulses in the core during
He burning. The post-merger lifetime of the Ledoux model is also
significantly shorter than for the other calculations, meaning that
the star loses less mass before reaching core collapse. 

Despite those differences, the basic result that early Case B
mergers can lead to BSGs at explosion is robust against changing the
mixing physics in these ways.

\begin{figure}
\centering
\epsfig{file=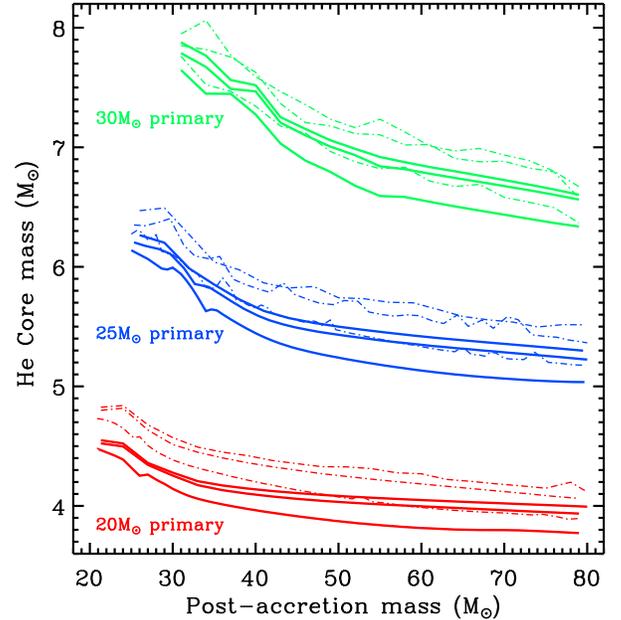, width=9cm}
\caption{\label{fig:he_core_masses}
We show the mass of the He core -- where the boundary of the core is
defined to be where the H mass fraction is lower than 0.01 -- for
the post-accretion calculations described in \S \ref{sec:PPE}. The
solid curves show the core masses at the point when He
core-burning ends, whilst the broken curves are at the end of the
calculation (typically after core carbon burning is complete, which
can sometimes introduce numerical instabilities).
The different models displayed for each primary mass are for different
merger times, as in Figs. \ref{fig:HRmulti}, \ref{fig:lifetimes} \&
\ref{fig:LBVtimes}. Early Case B accretion decreases the He
core masses at the end of core He burning, with more accretion
leading to less massive He cores. }
\end{figure}

\section{The fate of the core: black-hole versus neutron star
  production}
\label{sec:BHvsNS}

In this section we investigate the outcome of core collapse for our models
for LBV progenitors of luminous SNe. 
The particular phase of accretion we have considered -- and the merger scenario
through which that accretion may occur -- enables stars to exist
which are not only BSGs immediately before core collapse but are also
plausibly luminous enough to be potential LBVs.  
However, if those post-accretion stars were not to produce SNe with
standard (or greater) explosion energy, then this model would still fail to
explain the observed SNe which have been inferred to have LBV
progenitors.

It is widely expected that the core-collapse SN engine somehow uses
the gravitational potential energy which is liberated during
core collapse to power the SN explosion \citep[see, e.g.,][]{Colgate+White1966}
However, theorists have so far been unable to
definitively and robustly model the operation of that mechanism
\citep[see,
e.g.,][]{WoosleyWeaver1986,Herant+1994,Fryer1999,Janka+2007,Ugliano+2012,Muller+2012,
  Bruenn+2013,Hanke+2013,Fryer2013,Couch+OConnor2014,Takiwaki+2014,Fernandez+2014}.
Hence we cannot accurately predict
the outcome of core collapse for our stellar models.  Nonetheless, the 
qualitative expectation is that the more tightly bound stellar cores
of more massive stars are those in which the SN shock stalls and fails
to escape, leading to BH formation and a faint SN (or simple direct
collapse with no explosion; see, e.g., \citealt{Fryer1999,Heger+2003}).  
In this respect there are commonly thought to be three regimes for the
outcome of iron-core collapse:
\begin{itemize}
\item{} formation of a NS; 
\item{} formation of a BH `by fallback' (a NS
transiently forms, but the outward-moving shock is not strong enough
to unbind the remainder of the core, and this core material is
accreted by the NS, which then collapses to a BH); 
\item{} formation of a BH by direct collapse.
\end{itemize}
NSs might also form from ``electron capture'' SNe, i.e., not from such iron
core collapse \citep[see, e.g.,][]{PhP+2004ElectronCapture}, which may well also
produce faint SNe.

It is uncertain how luminous a SN should be expected to be  
associated with the formation of a black hole by fallback, although
expectations and observations tend to favour faint fallback SNe \citep[see,
e.g.,][]{Moriya+2010fallback}. 
Despite this, support for strong shocks from fallback SNe would
follow if SN 1987A were demonstrated to have formed a low-mass BH, 
as has occasionally been argued \citep{Brown+Bethe1994,Brown+Lee2004}.  

Perhaps more robustly, expectations exist for the outcome of core collapse for single stars
(or stars which are effectively single) which are mainly based on inferences from
observations. Hence we will compare the core properties of our binary models
to the core properties of single-star models.   We will assume that
the H-rich envelopes are broadly irrelevant to the question of
whether a successful SN shock develops and escapes the star, except
for the influence that they have on the structure of the
core.\footnote{Clearly the dynamical timescale of the H
  envelopes are far longer than the core-collapse timescale.}

The expectation for the upper end of the single-star zero-age main sequence mass which produces a
NS is sometimes quoted as being between 20 and $25~M_{\odot}$
\citep[see, e.g.,][]{Heger+2003, EldridgeTout2004}. This is broadly consistent with
the observationally-inferred upper limit on the progenitor mass for
type IIP SNe, although there is some tension between the data and
simple predictions, with an upper-limit for directly-inferred progenitor masses which
is somewhat below  $20~M_{\odot}$ \citep{Smartt2009}. However, the observational evidence
from type IIP SNe may be compatible with luminous SNe being produced
by ZAMS stars slightly more massive than $20~M_{\odot}$ for reasonable expectations about dust
formation obscuring those more massive SN progenitors \citep[see, e.g.,][]{Walmswell+Eldridge2012}. 

It is trivial to allow an \emph{initially} much more massive star to produce a NS remnant, simply by early
removal of the star's H envelope \citep{Belczynski+Taam2008}. However, the
SNe which we seek to explain are H-rich (as are LBVs); therefore this
mechanism cannot be directly relevant to these events.

Some previous work has suggested that the properties of the CO
core set the properties of the final iron core
which, in turn, controls the outcome of core collapse
\citep{Timmes+1996,WoosleyTimmes1996,Brown+2000,Brown+2001}. 
However, both \citet{Fryer+2002}
and \citet{Sukhbold+Woosley2014} stress that the fate of the
core is not a monotonic function of CO core mass, even if the CO
core mass is broadly a useful indicator of the likely outcome.
There are some significant uncertainties in those results and in our
calculations, especially arising from the the nuclear reaction cross
section for $\alpha$-capture onto $^{12}$C
\citep[see, e.g.,][]{WoosleyWeaver1986,Brown+2001,Brown+Lee2004}, as
well as both physical and numerical issues with the treatment of
convection (see, e.g., \citealt{Sukhbold+Woosley2014} and references
therein, especially \citealt{Rauscher+2002}).
We also note that rapid rotation may well also affect the outcome of core collapse, even
in cases when a ``collapsar'' does not occur \citep[see,
e.g.,][]{FryerHeger2000}. The cores of our
post-accretion stars are not necessarily rapidly rotating, but this
potential effect should also be borne in mind.
We will compare the properties of our stellar merger products to those of
single-star models calculated using exactly the same code, and only
draw conclusions from relative properties rather than absolute values, 
but these uncertainties could potentially still be problematic.

In what follows we compare a range of indicators for the fate of the
core. We will argue that those indicators overall suggest that, despite
their increased mass, the post-accretion objects are no more
likely to collapse to a BH instead of a neutron star (NS) than
the original primary stars, even if the merger product is massive
enough that a BH remnant would normally have been expected.
Whilst the evolution of the post-main-sequence core has
not been completely decoupled from the mass of the merger product, the
pre-merger mass of the primary seems to be much more important to the final
structure of the core than the mass of the merger product.   
Moreover, and surprisingly, several of the indicators suggest that accretion
\emph{increases} the chance that a star can avoid BH formation, and
thereby increases the range of initial stellar masses which produce
luminous core-collapse SNe.

\subsection{He core masses from the Eggleton-code calculations}
\label{sec:PPEcore}

The Eggleton code is unable to follow stellar evolution all the way to the
formation of the iron core. In principle we can calculate the evolution 
to the end of carbon burning, but numerical
instabilities during carbon burning can be troublesome. 

Nonetheless, we will compare the masses of the He cores after the end
of central He burning as a first indication of the eventual fate of the stellar core.
The He core masses are shown in Fig. \ref{fig:he_core_masses}, evaluated both after
core He exhaustion and in the final saved model time-step (i.e., nominally at
the end of C-burning). Early Case B accretion
leads to a \emph{decrease} in the He core masses. The effect is
clearer for 25 and 30 $M_{\odot}$ primaries than for 20 $M_{\odot}$
primaries, and does depend somewhat on the timing of the onset of
accretion (i.e., the temperature or radius of the primary at the onset
of accretion) but seems qualitatively generic. A simplistic interpretation of 
this would be that early Case B accretion may \emph{increase} the range
of initial stellar masses which might produce successful SN explosions
at core collapse. 
  
Since Fig.\ \ref{fig:lifetimes} does not show a strong decrease in
post-merger lifetimes as more mass is
added, the decrease in He core mass shown in Fig.\
\ref{fig:he_core_masses}  does not appear to be a simple lifetime
effect. The effect is at least partly due to dredge-up from the He
core by the convective envelope of the post-merger star.  
This dredge-up may indicate that the details of the thermal relaxation phase after the
merger are important for the final fate of the star, in which case our
assumptions about the entropy of the accreted material should be
examined in more detail. If so, then this uncertainty is related to
the possibility that the merger process itself would directly cause
dredge-up from the core of the primary \citep[see,
e.g.,][]{IvanovaPodsiadlowski2003}, and to the likelihood that the
accreted material will itself be He-enriched (because the the secondary star should have
completed some nuclear burning by the time of the merger).  As
noted earlier, additional He enrichment should be favourable for
the production of BSGs \citep[see,
e.g.,][]{BarkatWheeler1988,HillebrandtMeyer1989}.  Additional
dredge-up from the core may also further increase the range of initial
stellar masses which can produce a successful core-collapse SN after
merger; conversely, it is possible that He-enriched 
envelopes would sufficiently alter the H shell-burning during
the post-merger BSG phase to produce the opposite effect. These combined
uncertainties introduced by the physics of the merger/accretion
process deserve further study, but they add several dimensions to the
parameter space, and so we defer them to future studies.

\begin{figure}
\centering
\epsfig{file=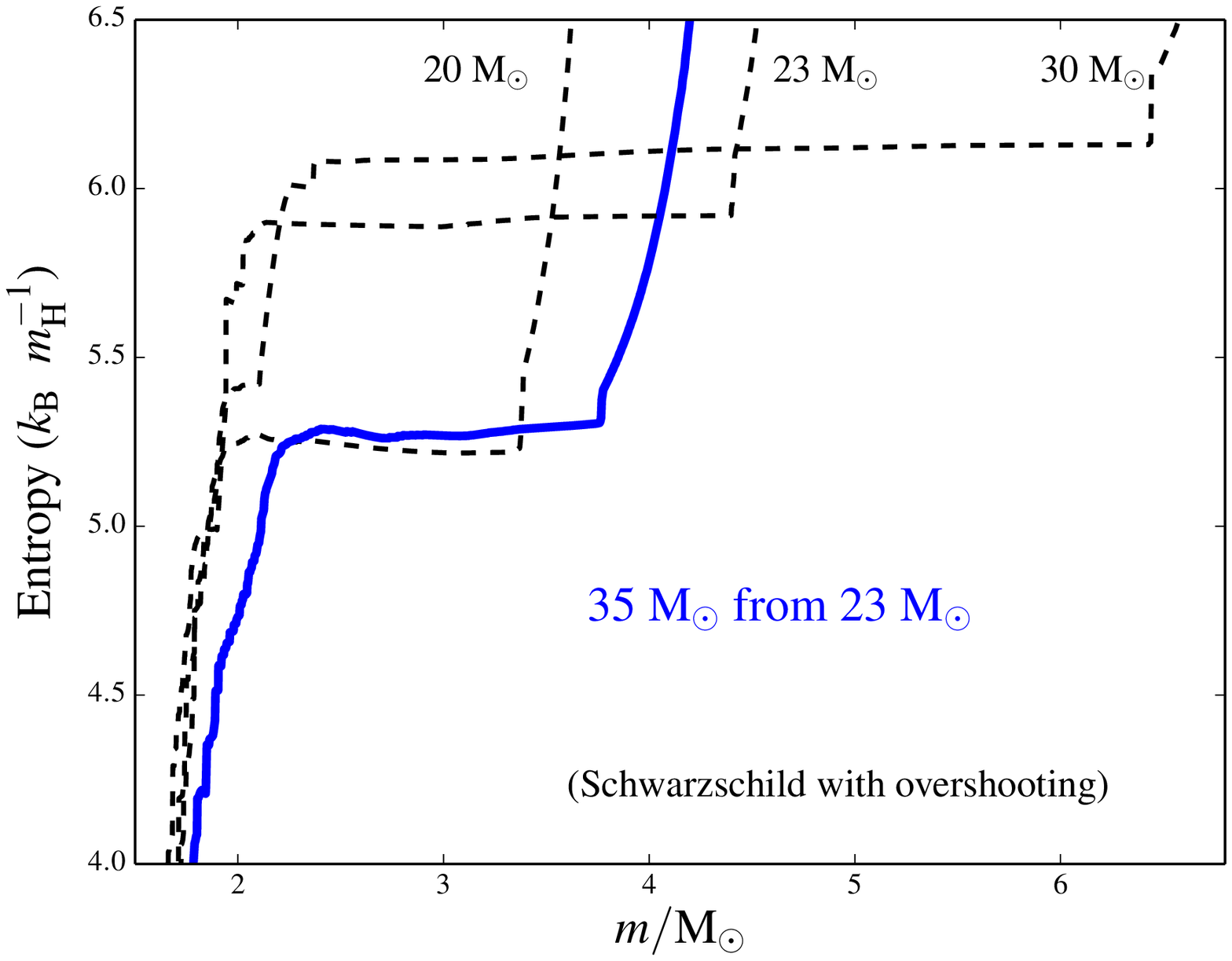,
  width=9cm}
\epsfig{file=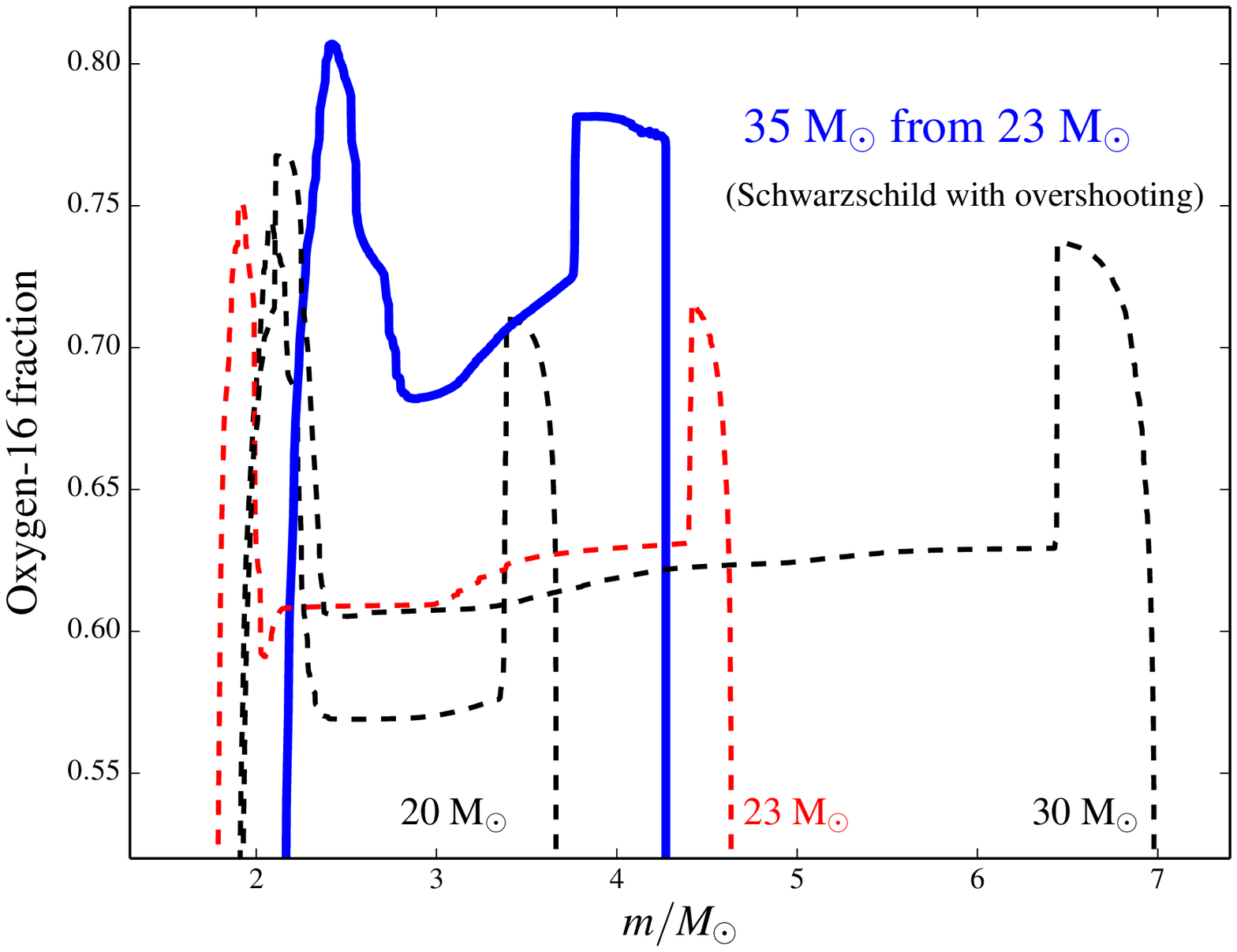,
  width=9cm}
\caption{\label{fig:MESA_infall8_35from23} We compare the details of
  the structure at core collapse for one of our merger
  products/post-accretion stars with that of three single stars (20,
  23 and $30~M_{\odot}$, as marked). In all cases, these models are
  shown when the core infall velocity reaches $10^{8}\,{\rm cm}\,
{\rm s}^{-1}$. The
  $35~M_{\odot}$ merger product was formed by accretion onto a
  $23~M_{\odot}$ primary which had a radius of $100~R_{\odot}$ at the
  onset of accretion and the structure is shown using thick solid curves
    (blue in the online version). 
  Differences between the merger product
  and the $23~M_{\odot}$ single star are therefore a
  result of how the accretion has affected the evolution of the star.
  The upper panel shows a portion of the entropy profiles of the
  stellar cores, and the lower panel the $^{16}$O composition
  profile.  The entropy profile within the CO core of the merger
  product has values closer to those of the $20~M_{\odot}$ single star
  than those of the $23~M_{\odot}$ model.  The mass gained by the
  merger product leads to a reduced overall mass for the final
  O-rich core, and the inner boundary of the O-rich material
  also moves outwards in mass coordinate. }
\end{figure}

\begin{figure}
\centering
\epsfig{file=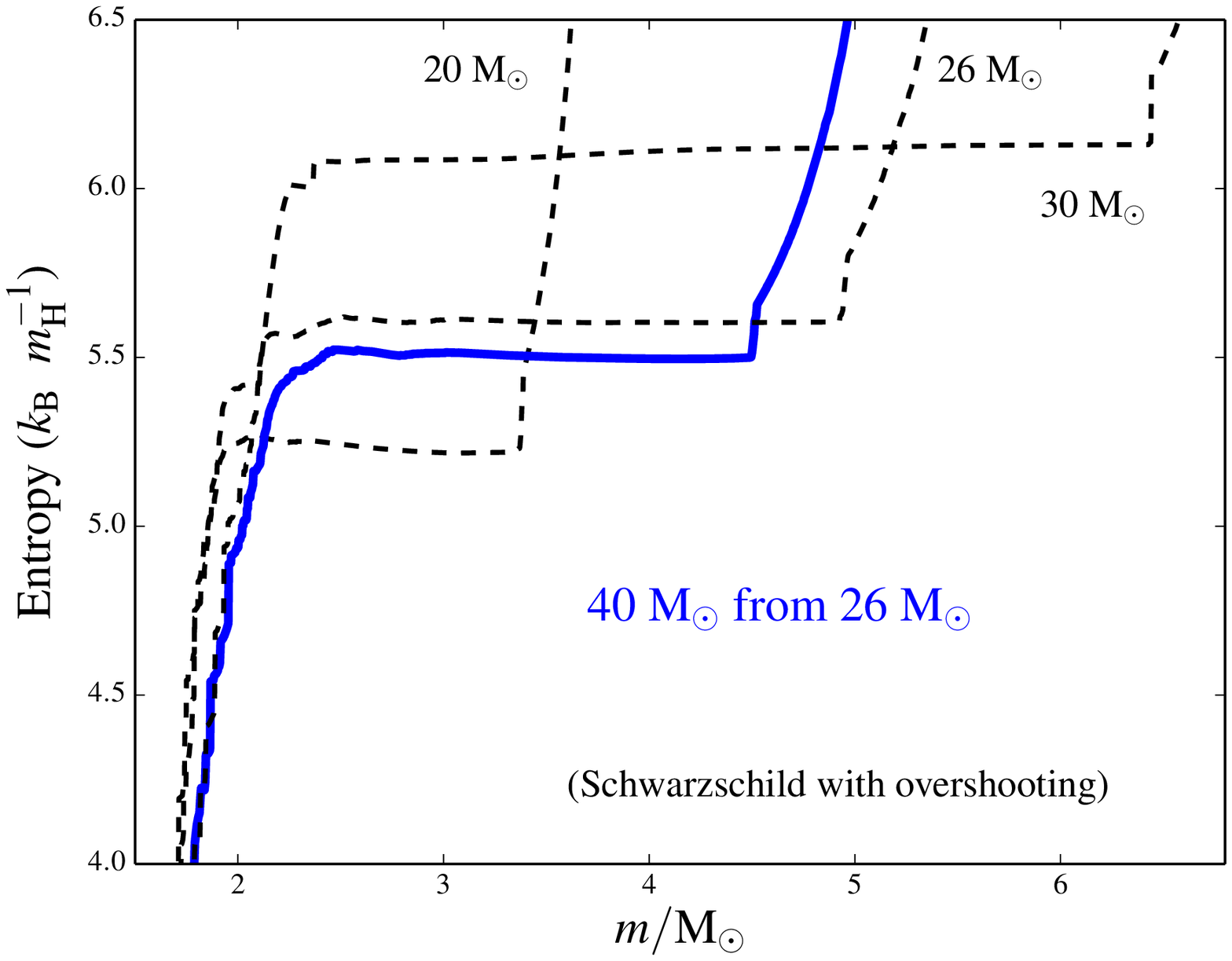,
  width=9cm}
\epsfig{file=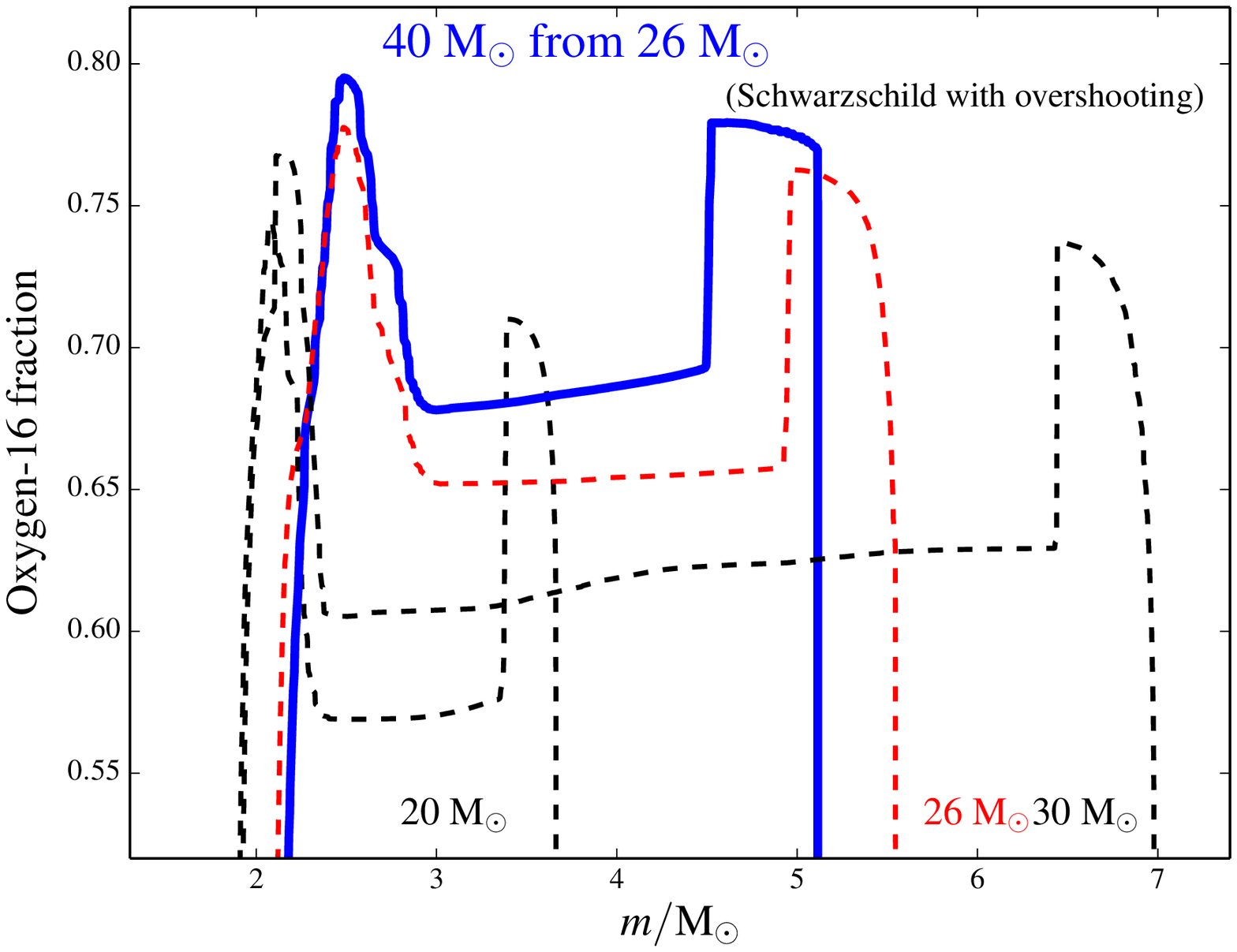,
  width=9cm}
\caption{\label{fig:MESA_infall8_40from26} As in
  Fig.\,\ref{fig:MESA_infall8_35from23}, 
  we compare details of the entropy
  and composition profiles from one post-accretion star to those of
  three single stars. Again, these structures are when the core infall
  velocity reaches $\rm 10^{8}\,cm\,s^{-1}$. In this case, the merger
  product was a $40~M_{\odot}$ star formed by accretion onto a
  $26~M_{\odot}$ primary with a radius of $100~R_{\odot}$. Here we
  also show the structure of a $26~M_{\odot}$ single star for
  comparison.  The qualitative trends are similar to those in Fig.\,
  \ref{fig:MESA_infall8_35from23}, in that both the final CO core mass
  and the specific entropy within the CO core are lower, and the peak
  oxygen abundance is higher.  However, the effects in this case are
  less drastic. In contrast to Fig.\,\ref{fig:MESA_infall8_35from23},
  the location of the inner boundary of the O-rich material in
  the merger product is only marginally further out than the boundary
  in the corresponding single star (i.e., in this case
  $26~M_{\odot}$).}
\end{figure}

\begin{figure}
\centering
\epsfig{file=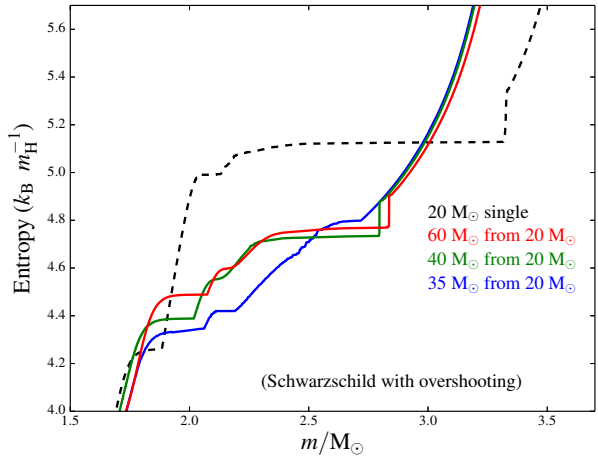,
  width=9cm}
\caption{\label{fig:entropy_profile_threemergers}
We compare details of the entropy profiles at the end of core Si burning for a $20~M_{\odot}$ single star (the black dashed curve) and three stars formed by early Case B accretion onto such
a $20~M_{\odot}$  primary (the solid curves, colored as their
respective labels; in all cases
the accretion began when the primary star had a radius of $30~R_{\odot}$).   The
entropy profiles in this region of the post-accretion stars are more
similar to each other than to that of the single star. Even the
$60~M_{\odot}$  post-accretion star produces a less-massive and
lower-entropy core than the $20~M_{\odot}$  single star.  We also note
that these merger products have far less substantial plateaus in their
entropy profiles, with the corollary that the entropy gradients in this part of
their core are typically shallower; in particular, the $35~M_{\odot}$  post-accretion
star finishes Si burning with no sharp jump in this region of the
entropy profile.}
\end{figure}

\begin{figure}
\centering
\epsfig{file=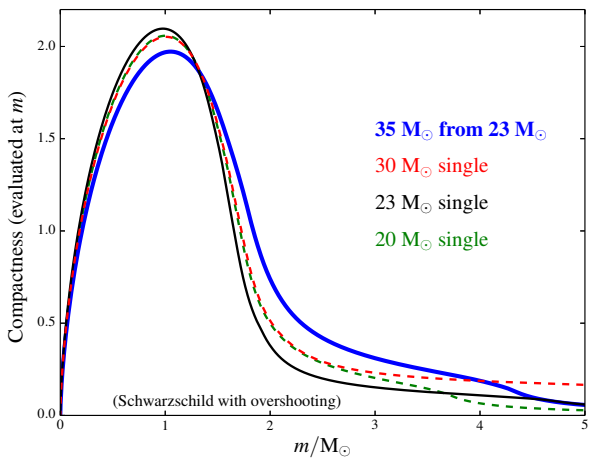,
  width=9cm}
\caption{\label{fig:xi}
At a core infall velocity of $\rm 10^{8} cm~s^{-1}$, we compare the
internal profile of the compactness parameter, $\xi$ for four stellar
models: three single stars with initial masses of $20~M_{\odot}$
(green broken curve), $23~M_{\odot}$ (black curve)
and $30~M_{\odot}$ (red broken curve), and a merger product with a $35~M_{\odot}$
post-merger mass from a $23~M_{\odot}$ primary (the thick blue
curve), i.e., the same models as in Fig.\
\ref{fig:MESA_infall8_35from23}. A color version of this Figure is
available in the online version.}
\end{figure}

\begin{figure*}
\centering
\begin{tabular}{cc}
\epsfig{file=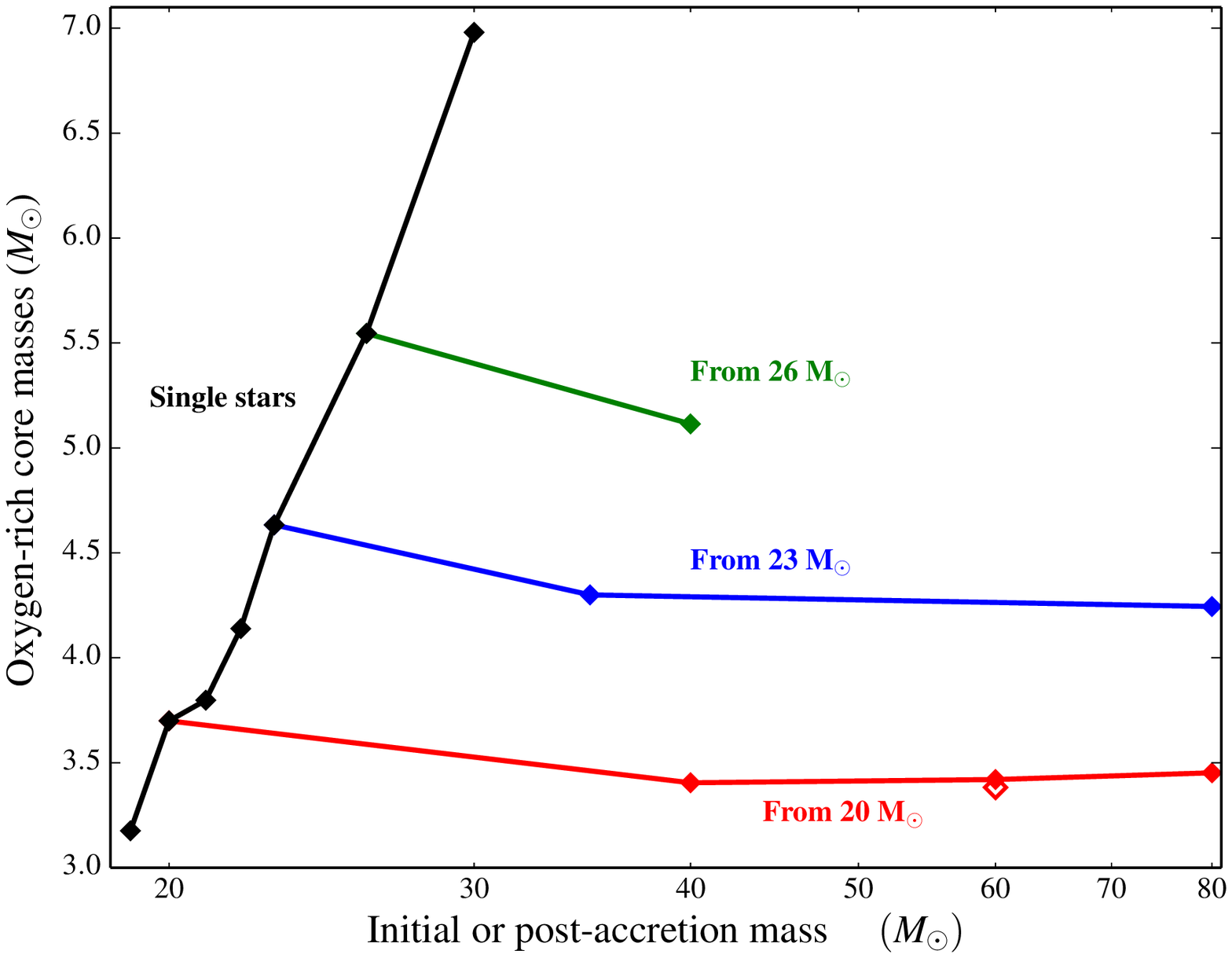,  width=9cm} &
\epsfig{file=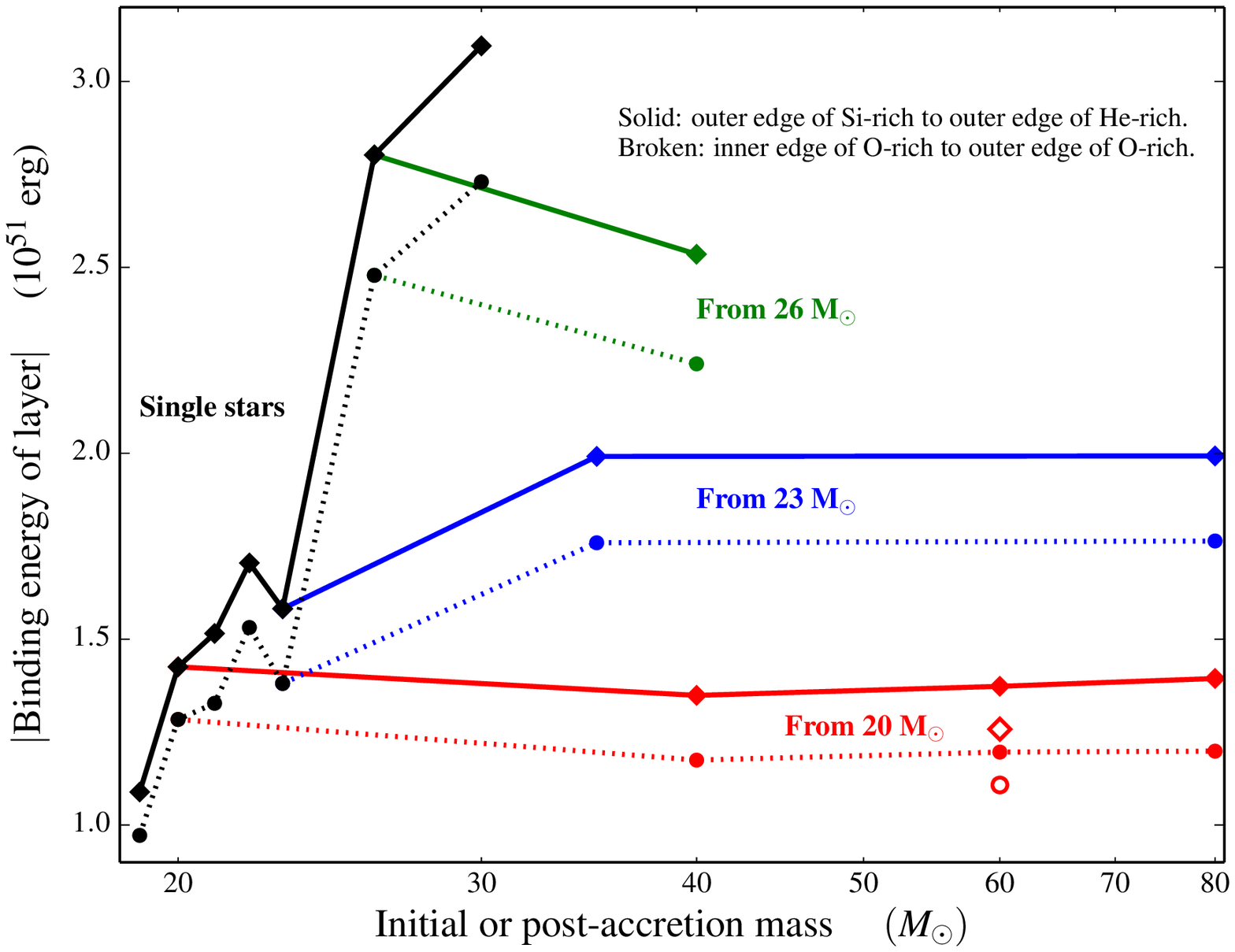,  width=9cm} \\
\epsfig{file=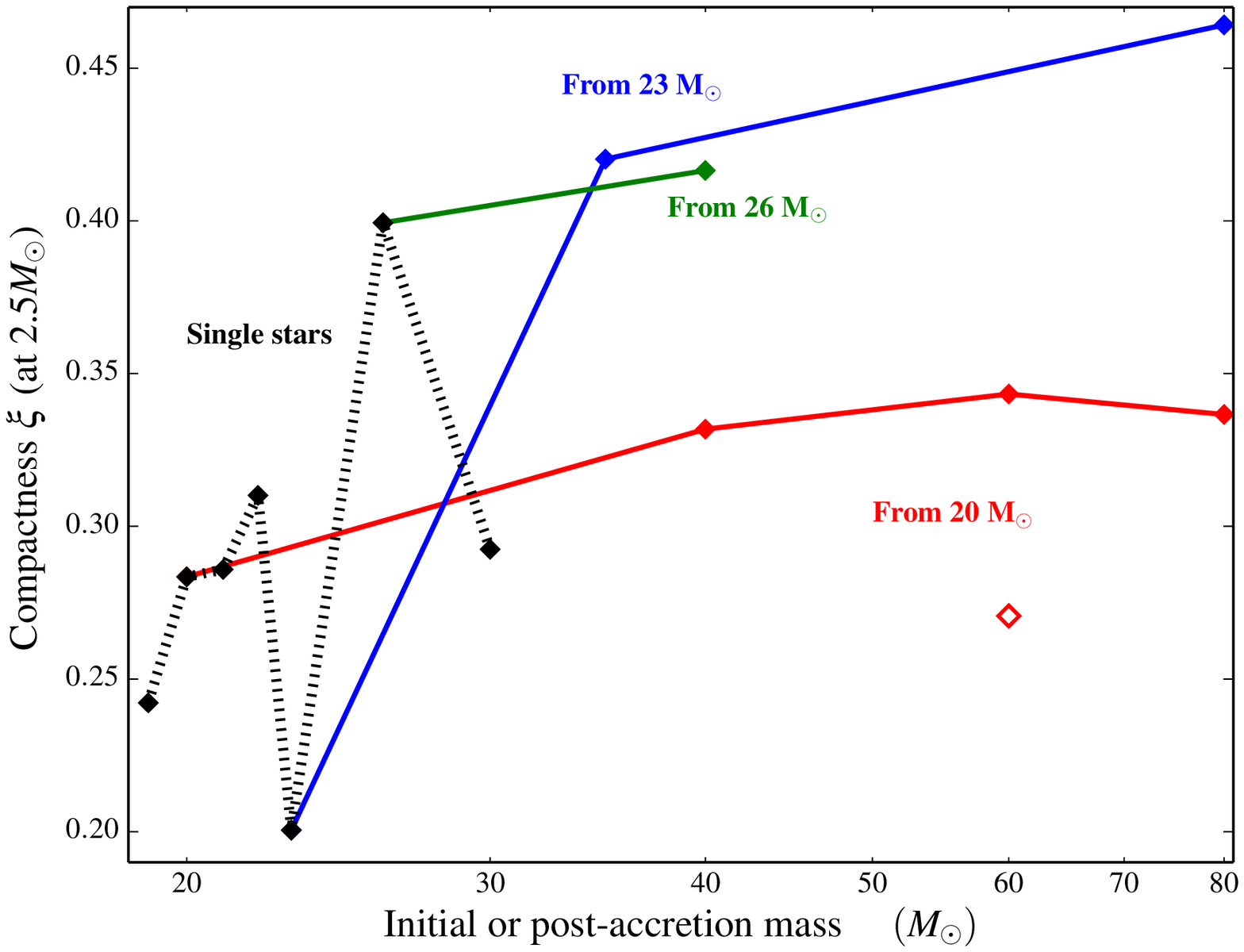,  width=9cm} &
\epsfig{file=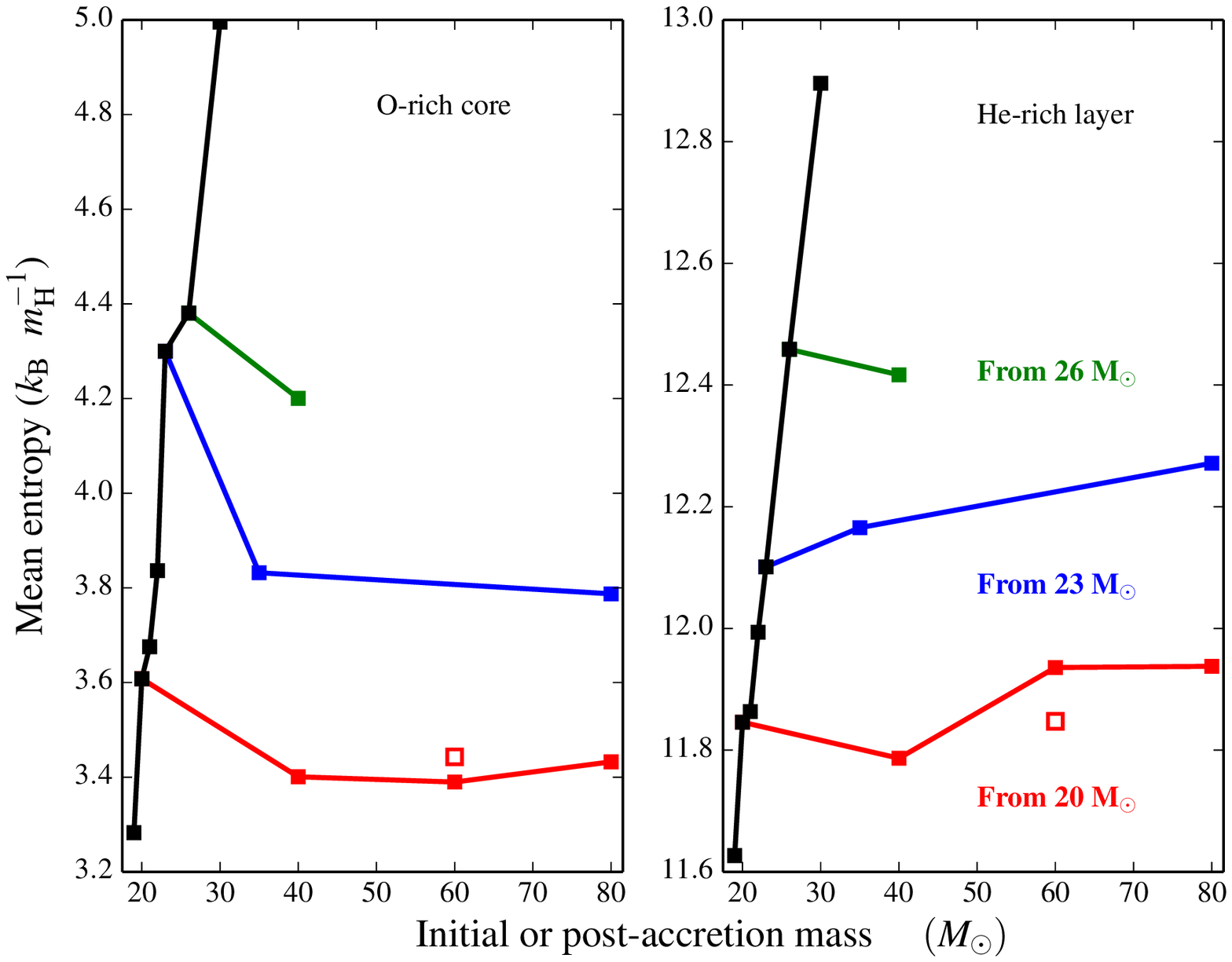,  width=9cm} 
\end{tabular}
\caption{\label{fig:cores_schwovs_infall8}
We compare several quantities evaluated at a core infall velocity of $\rm 10^{8}~cm~s^{-1}$ for 
single-star stellar models (shown in black) and post-accretion stellar
models (plotted using colored lines and symbols). These calculations
assumed no stellar rotation and adopted the Schwarzschild criterion for
convection with overshooting (as described in the text).  For single-star
models the abscissa shows the mass of the initial pre-main-sequence
model, whilst for post-accretion models the abscissa gives the post-accretion
mass (and the initial mass is as labelled). 
For accretion onto $20~M_{\odot}$ primaries, the solid symbols
represent models which began accretion at $30~R_{\odot}$; the larger, hollow symbol
represents a case where the accretion began at a primary radius of $100~R_{\odot}$; for the
models which represent the outcome of accretion onto $23~M_{\odot}$ and
$26~M_{\odot}$ primaries, accretion began at $100~R_{\odot}$. 
\textbf{Upper-left:}
The mass inside the boundary of the O-rich core.
\textbf{Upper-right:} The binding energy of the outer core,
from the outer edge of the Si-rich core to the outer edge of the
He-rich core (shown with symbols joined by solid lines).
This binding energy is dominated by that of the O-rich layer
(given using symbols joined by broken lines).  \textbf{Lower-left:}
The compactness parameter, $\xi_{2.5}$ of \citet{OConnor+Ott2013}. 
\textbf{Lower-right:} The mean entropy per baryon within the
O-rich core (left) or the He-rich layer outside the O-rich
core (right). }
\end{figure*}

\begin{figure}
\centering
\epsfig{file=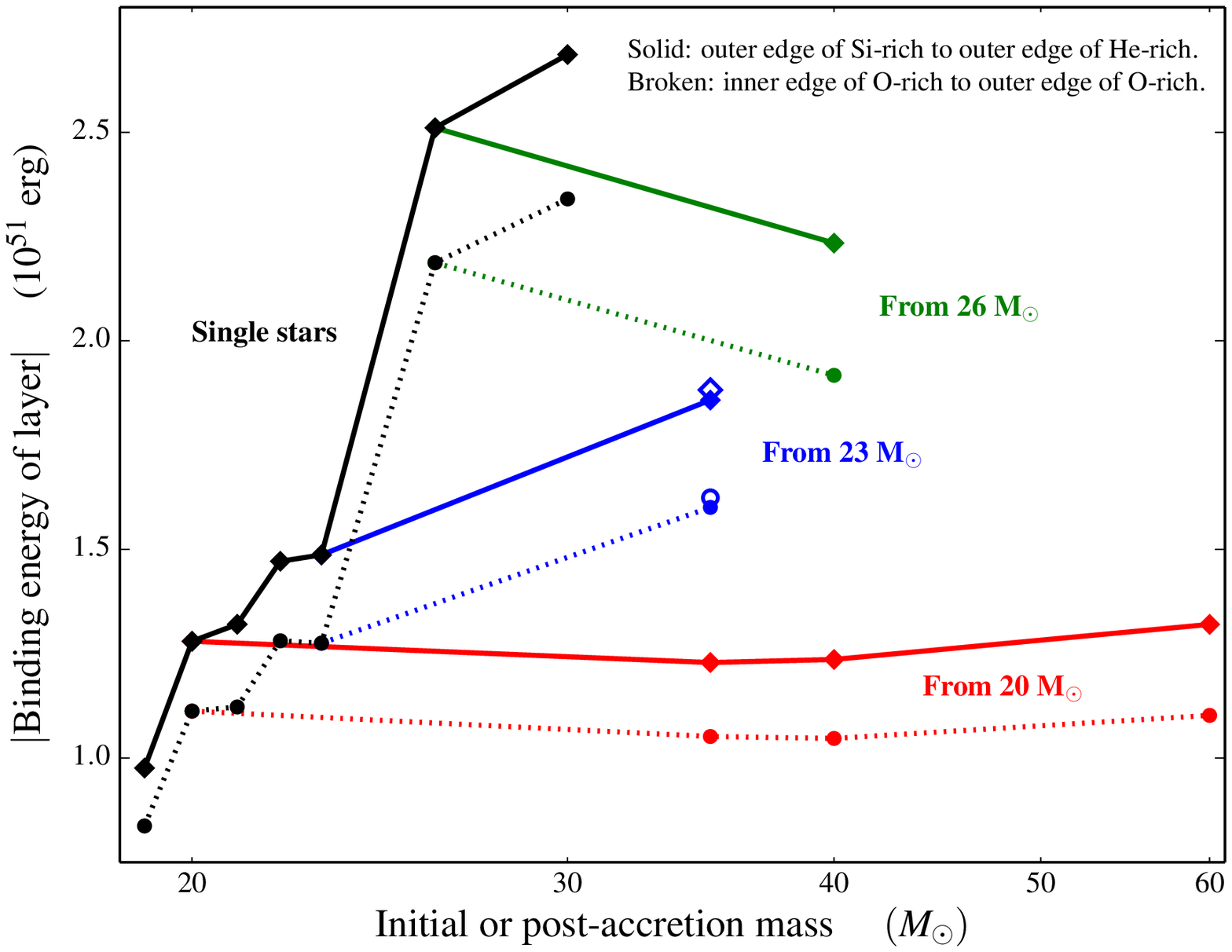, width=9cm}
\epsfig{file=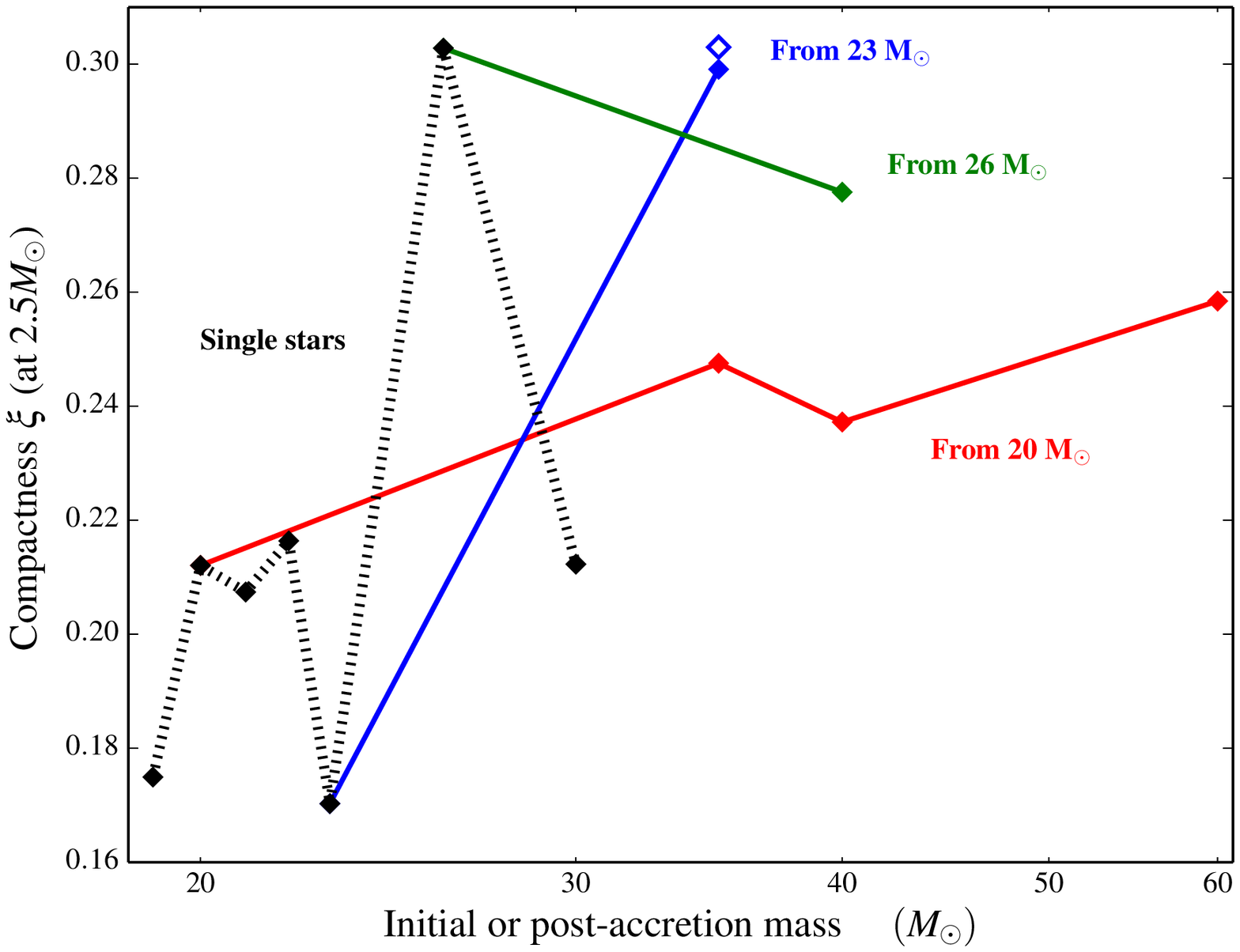, width=9cm}
\epsfig{file=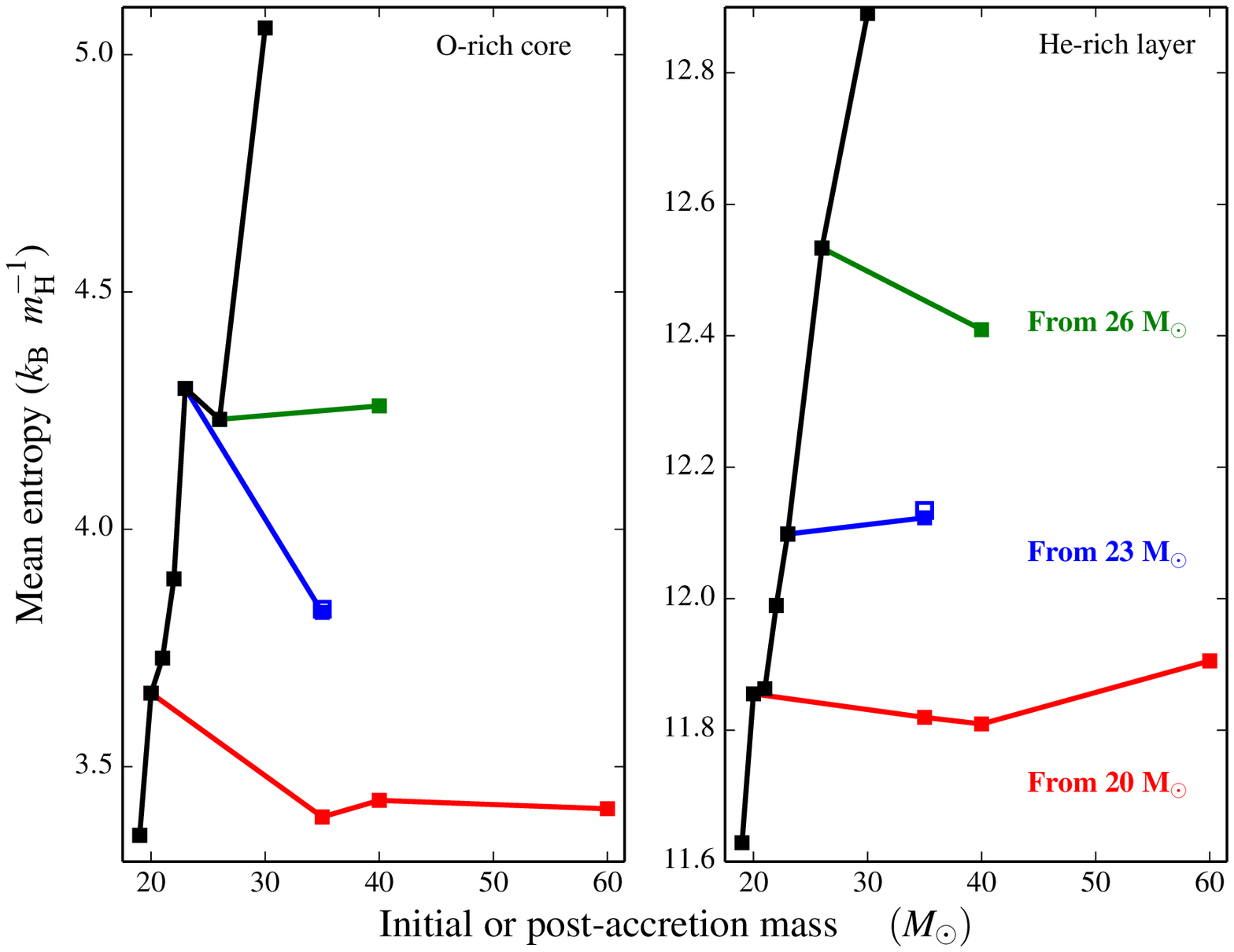, width=9cm} 
\caption{\label{fig:cores_schwovs_Si28}
As Fig. \ref{fig:cores_schwovs_infall8} but for quantities 
at the end of core Si burning, where we do not show the oxygen core
masses again. For almost all of these models, accretion began when the
primary radius was $30~R_{\odot}$; the exception is plotted using hollow blue
symbols, which represent a model for which accretion onto the primary began at $100~R_{\odot}$.}
\end{figure}

\begin{figure*}
\centering
\begin{tabular}{cc}
\epsfig{file=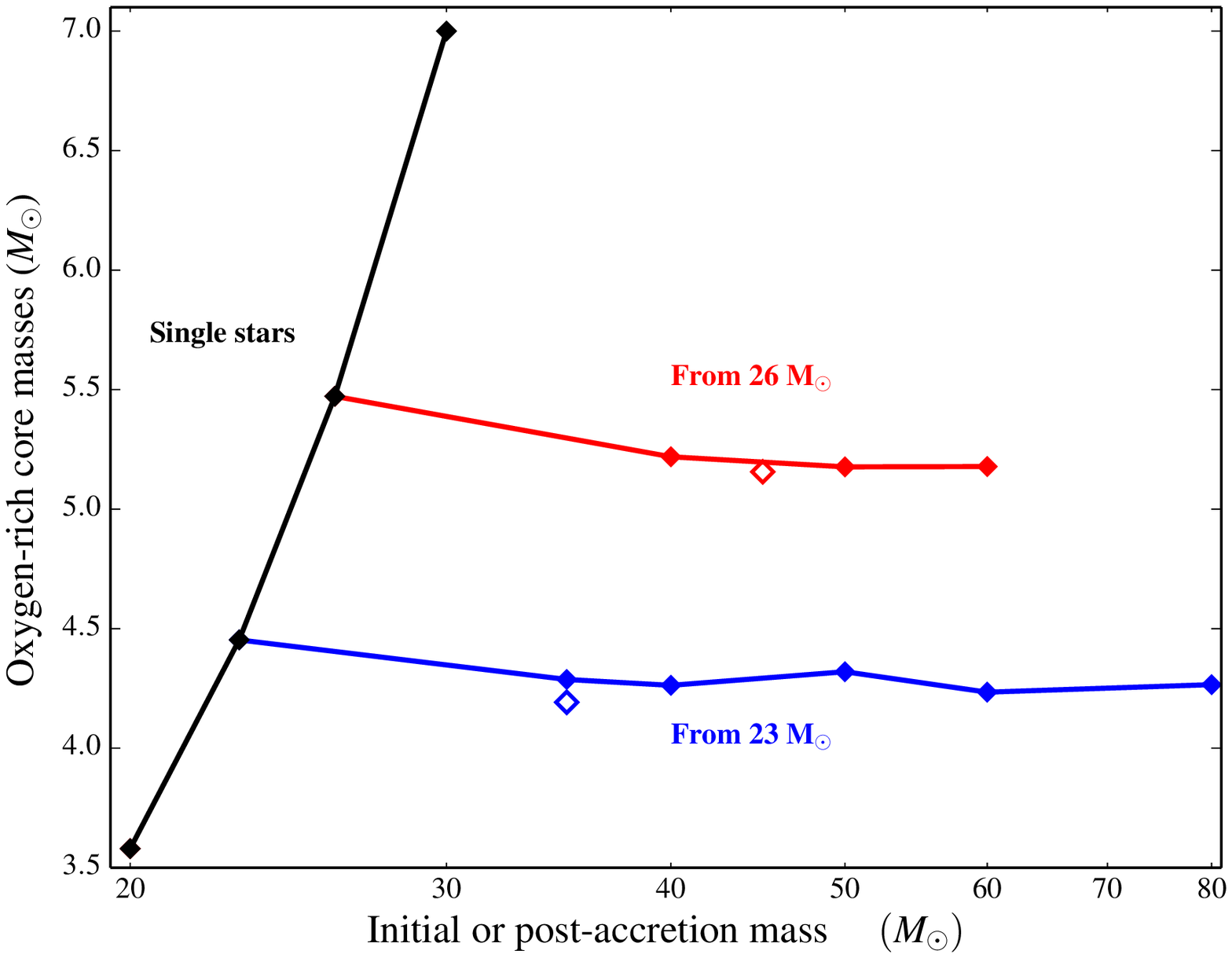, width=9cm} &
\epsfig{file=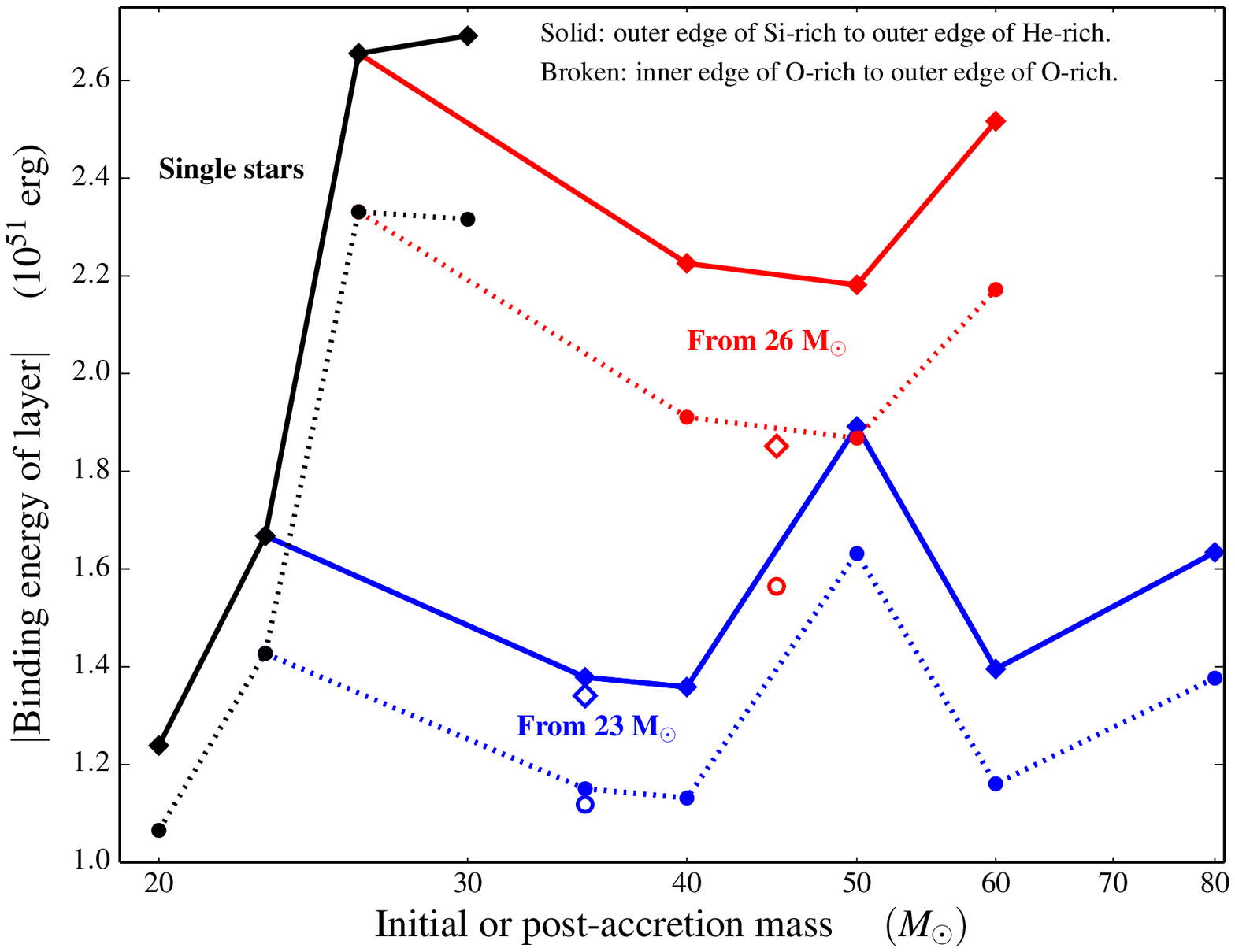, width=9cm} \\
\epsfig{file=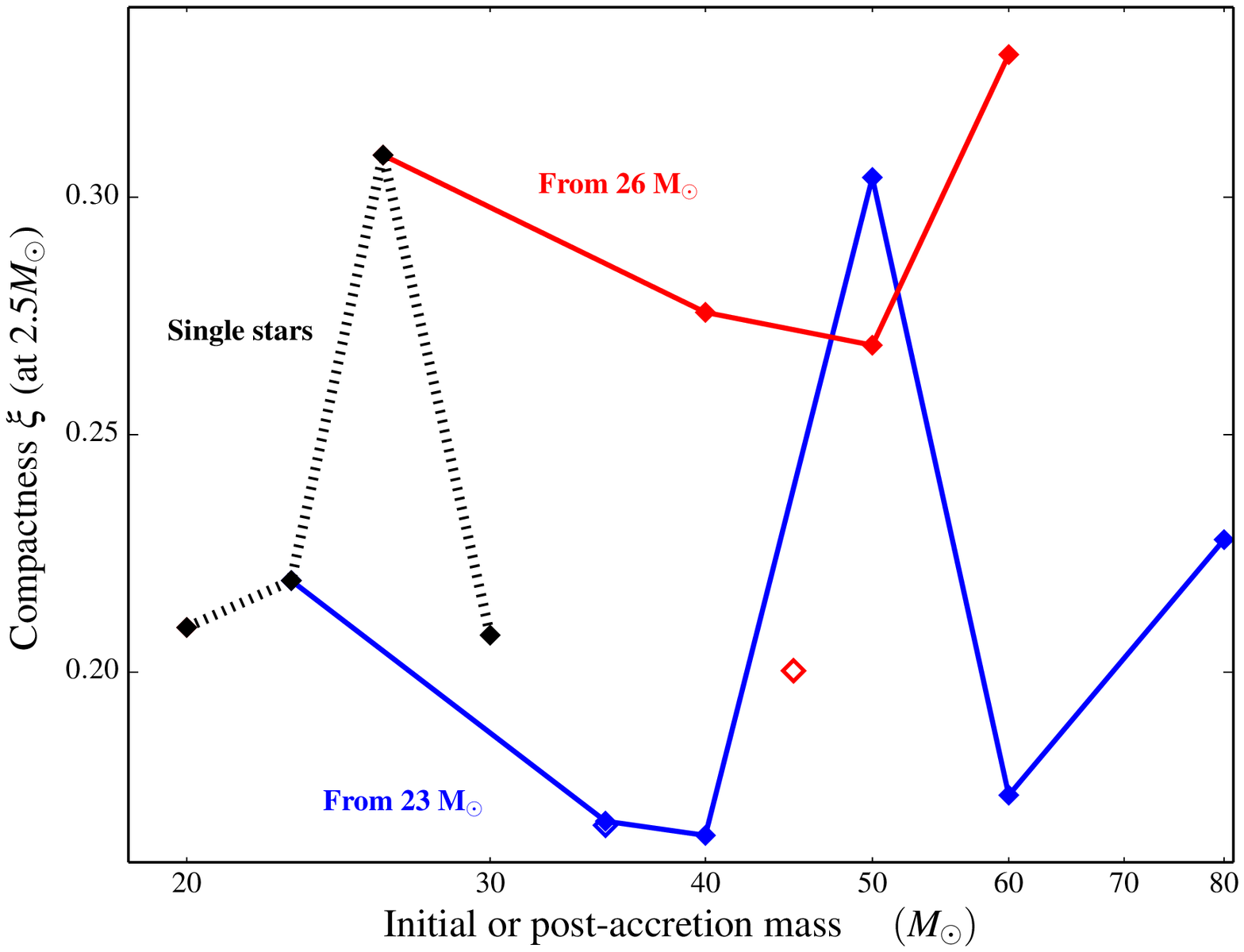,  width=9cm} &
\epsfig{file=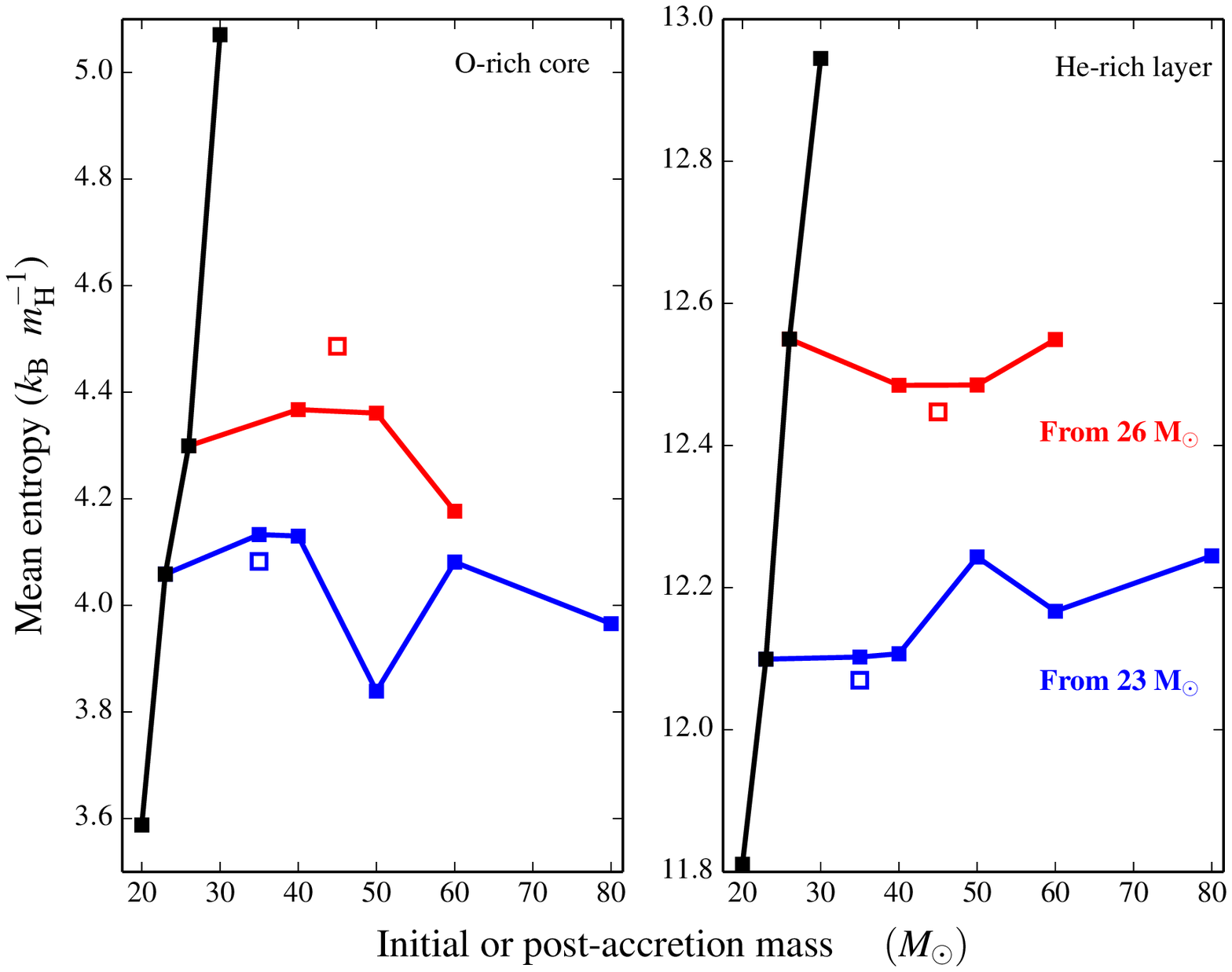, width=9cm} 
\end{tabular}
\caption{\label{fig:cores_pureschw}
As Fig. \ref{fig:cores_schwovs_infall8} but for quantities 
at the end of central Si burning, and for calculations which assume the
Schwarzschild criterion for convection without any overshooting. Here
the solid symbols represent models for which accretion began at
primary radii of $100~R_{\odot}$; the hollow symbols represent models
for which accretion started when the primary had a radius of $30~R_{\odot}$.}
\end{figure*}

\subsection{Indications from the post-Si-burning core based on MESA calculations}

MESA enables us to continue calculating the evolution of our models
until after the end of core silicon burning.\footnote{However, we
  understand that MESA has not been as well validated for silicon burning
  as for earlier phases (L.\ Bildsten [priv.\ comm.]).} Some of the calculations
we present were stopped at the end of central silicon burning, whilst for others we 
followed the collapse of the core until it reached an infall velocity
of $\rm 10^{8}~cm~s^{-1}$; for results where the difference is
important, we will state which termination criterion was used.

\subsubsection{Core profiles of example models}

Figures \ref{fig:MESA_infall8_35from23} and
\ref{fig:MESA_infall8_40from26} compare the final core structures of
example post-accretion models to similar-mass single-star models.  In
both of these examples, the final mass of the O-rich core is smaller
for the post-accretion star than for the single star with the same
initial mass.  However, we also note that the shape of the composition
profile is not intermediate between any of these single-star model
(e.g., the peak mass fraction of $^{16}$O is higher than for any of
these single stars).

The entropy profiles of the post-merger stars
are also altered in the broad direction of lower-mass progenitors, in
particular the value of the entropy plateau in the O-rich core
moves to lower values.\footnote{We
    note that the fact that some of the entropy
profiles slightly decrease outwards is not problematic, since these
stars are undergoing core collapse.}  
However, the entropy profiles do not change in a simple monotonic
way as the single-star mass is increased, since the value at which the
$23~M_{\odot}$ entropy profile plateaus is somewhat above the value at
which the $26~M_{\odot}$ star does so.  This may be related to the 
suggestion that the cores of stars in the region around $23~M_{\odot}$
may be more difficult to explode than those of stars in the region
around $26~M_{\odot}$ (Ilka Petermann,
[priv.\ comm.]; see also \citealt{Sukhbold+Woosley2014}). If it is true
that stars in the range $\approx25$--$30~M_{\odot}$ are relatively
``explodable'', as concluded by \citet{Sukhbold+Woosley2014}, then
this region of initial primary-star masses might be the most
favourable for producing LBV SN progenitors via this binary-accretion scenario.
Further examples of how the
entropy profile is affected by early Case B accretion are given in
Fig.\,\ref{fig:entropy_profile_threemergers}, which shows how 
the core entropy profile of a $20~M_{\odot}$ primary is altered by
increasing amounts of mass gain.  The specific entropy in this region
of the core for post-accretion masses of 35, 40 and $60~M_{\odot}$ are
more similar to each other than to that of the $20~M_{\odot}$ single
star. Moreover, the entropy profile of the $35~M_{\odot}$
post-merger star in Fig.\,\ref{fig:entropy_profile_threemergers} is
distinguished by having almost no plateau in this
region, and no sharp entropy jumps.  If such changes in the shape of
the entropy profile affect how easily these cores explode, then this
effect of binary evolution may be very important. We
note that for this set of models the change in the shape of the entropy profile is
greatest for the \emph{lowest} amount of accreted mass.
We stress that we have not modelled the full parameter space
for which early Case B accretion may be important; we suspect that this
effect may be import for primary stars which are less massive than
those which we have modelled for this work on LBV SN progenitors.

Based on Figs.\,\ref{fig:MESA_infall8_35from23},
\ref{fig:MESA_infall8_40from26}  and
\ref{fig:entropy_profile_threemergers} we might anecdotally conclude
that this supports the idea that early Case B merger products and
post-accretion stars should be \emph{easier} to explode than single
stars with the same initial primary-star mass, as suggested by the
results in \S \ref{sec:PPEcore}. However, an alternative indicator --
the compactness parameter 
\citep[$\xi$, as suggested by][]{OConnor+Ott2011,Ugliano+2012,OConnor+Ott2013} -- is less
favourable.  We do not evaluate $\xi$ at core
  bounce, as was done by \citet{OConnor+Ott2011,OConnor+Ott2013}. However, 
  \citet{Sukhbold+Woosley2014} show that conclusions drawn from
  evaluating $\xi$ at core infall velocities of $10^{8}\,{\rm cm}\,{\rm s}^{-1}$
are roughly equivalent to those at core bounce; they further demonstrate that comparing $\xi$
from models at earlier evolutionary phases  -- even
as far back as O ignition -- produces similar, though less well-developed
trends to those shown by $\xi$ at core-collapse.
The distribution of $\xi(m)$ is shown in Fig.\,\ref{fig:xi} for the same models as shown in
Fig.\,\ref{fig:MESA_infall8_35from23}. Higher values of this
compactness parameter are thought to indicate that cores are more
difficult to explode.  To some extent, conclusions based
on the comparative compactness of these models are sensitive to the location at which $\xi$ is
evaluated. The inner core of the merger product is less compact than
any of the single stars, but between $\approx 1.5 $ and $4~M_{\odot}$,
the merger product possesses a higher value of  $\xi$ than any of the
single stars. (Note also that, over the same range, the 23~$M_{\odot}$
single star is less compact than either of the 20~$M_{\odot}$ or 30
$M_{\odot}$ models).  Outside the O-rich layer (for which, see Fig.\
\ref{fig:MESA_infall8_35from23}), the compactness of the merger
product falls relatively sharply.  If the
relevant mass scale is smaller than 4~$M_{\odot}$, then simple
application of $\xi$ suggests that this merger product
would be \emph{harder} to explode than any of the single stars to
which we compare it.

\subsubsection{Collected indicators for sets of models}

Figs.  \ref{fig:cores_schwovs_infall8}, \ref{fig:cores_schwovs_Si28}
and \ref{fig:cores_pureschw} present potential indicators of the
outcome of core collapse for collections of single and post-accretion
stellar models. There we show four types of characteristics of the cores of
those stars: 
\begin{itemize}
\item{} The mass of the final O-rich core (as shown for
  individual cases in Figs.\
  \ref{fig:MESA_infall8_35from23} \& \ref{fig:MESA_infall8_40from26}).
\item{} The binding energy of the outer core. This is shown for both
  the O-rich layer alone and for all mass from the outside of
  the Si-rich core to the outside of the He-rich core.  More
  precisely, we add the magnitude of the infall kinetic energy to
  the magnitude of the binding energy. (If proto-NS formation in the
  inner core releases a roughly fixed amount of energy, and the
  explosion mechanism also always converts a fixed fraction of that energy
  release, then this quantity would control whether the SN engine is
  energetically capable of ejecting the outer core.)
\item{} The compactness parameter evaluated at a mass coordinate of
  $2.5~M_{\odot}$ (i.e., $\xi_{2.5}$; see \citealt{OConnor+Ott2011,OConnor+Ott2013,Sukhbold+Woosley2014}). 
\item{} The mean specific entropy inside both the O-rich core and
  the remainder of He core (i.e., the He-rich layer outside the
  oxygen core).  
\end{itemize}
For most of the quantities shown in
Figs.\,\ref{fig:cores_schwovs_infall8}, \ref{fig:cores_schwovs_Si28}
and \ref{fig:cores_pureschw}, either the change in the stellar
structure after early Case B accretion is normally weak or the trend
resulting from accretion is in the opposite direction to the trend
produced by increasing single-star mass.  The main exception is the
compactness parameter ($\xi_{2.5}$), and only for the models which
assume overshooting (Figs.\,\ref{fig:cores_schwovs_infall8} and
\ref{fig:cores_schwovs_Si28}).  For those models, if $\xi_{2.5}$ were
a reliable parameter for predicting black-hole formation at core
collapse, then early Case B accretion would broadly increase the
likelihood of black-hole formation (since for those stellar models we
find that $\xi_{2.5}$ increases with accretion).  However, we find
that the trend is mostly reversed for the set of calculations without
overshooting, for which see Fig.\,\ref{fig:cores_pureschw}.  In that
case we find that most of the post-accretion models show a lower value
of $\xi_{2.5}$ than if the primary had been allowed to evolve without
accretion; the $50~M_{\odot}$ star formed from early Case B accretion
onto a $23~M_{\odot}$ primary is a strange and strong exception. We
also note that $\xi_{2.5}$ shows an unclear trend for the single-star
sequence, whether or not we adopt any overshooting when calculating
the single stars; this non-monotonic behaviour has recently been
studied in detail by \citet{Sukhbold+Woosley2014}.

These comparisons also suggest that $\xi_{2.5}$ may be the indicator for which changing the
radius at the start of the accretion phase makes the largest
relative difference (see especially Fig.\,\ref{fig:cores_schwovs_infall8}, but also
Fig.\,\ref{fig:cores_pureschw}),  although we have too few direct comparisons to be sure that this is
generally the case. It is unclear to us how this sensitivity might be interpreted, although
it may be relevant that the other indicators in those plots are
quantities integrated over regions defined by composition criteria, whilst $\xi_{2.5}$
is evaluated at a fixed mass coordinate. 

The binding energy of the outer core also occasionally suggests that
the likelihood of a successful SN explosion could be decreased by
early Case B accretion, though less frequently than would be concluded
from $\xi_{2.5}$.  For the calculations with overshooting, accretion
onto the $23~M_{\odot}$ primary increases the final binding energy
(but not for post-accretion stars created from other primary masses).
For the calculations without overshooting, the $50~M_{\odot}$
star formed from a $23~M_{\odot}$ primary shows an increase in final
core binding energy compared to the $23~M_{\odot}$ single star (this
is the same model which is an outlier to the trend for $\xi_{2.5}$ in
that set of models).

\section{Channels for Early Case B Accretion: Event Rate Calculation and Comparison}
\label{sec:rates}

In this section, we attempt to estimate the rates at which some relevant early
Case B merger or accretion events are likely to occur. 
This section does not consider all possible formation channels, e.g., we make no
attempt to account for dynamical mergers in dense young stellar
clusters.  However, in \S \ref{sec:triples}, we discuss the potential
importance of systems in which the early Case B merger was of the
inner binary in a triple. In that case, the tertiary companion may be able
to transfer even more mass onto the merger product before the SN
explosion.

\subsection{Rates for the early Case B merger channel}
\label{sec:earlyCaseBrates}

We now estimate the rate of luminous SNe that can be accounted for
by our binary-merger model. When trying to explain CCSNe with immediate LBV
progenitors using this scenario, two of the main uncertainties are the
minimum post-merger mass required to produce the LBV phenomenology (since
this governs the post-merger luminosity; see Figs.\,\ref{fig:HR} \& \ref{fig:HRmulti}) and the
maximum pre-merger primary mass which can lead to a CCSN with
canonical explosion energy.  Assuming insignificant mass loss during
the merger, then the maximum post-merger
mass would be $\approx 1.8$ times the primary-star mass (due
to the stability criteria for this early Case B merger process). So a $20~M_{\odot}$
primary may be able to attain a mass of $\approx 36~M_{\odot}$ after
the merger, which is only just consistent with common
expectations for the lower end of the range of initial LBV masses 
(although it has been suggested that even stars with initial masses as low as
$25~M_{\odot}$ might display LBV-like phenomena, for which see
\citealt{Smith+2004,VinkReview2009}).  This estimate ignores any mass ejected during the
merger, although post-merger rapid rotation may well
increase the likelihood of LBV-type outbursts
\citep{Langer1997,Langer1998}.  
Given the uncertainties, we present estimates for a broad range of possible
upper- and lower- limits on the primary-star mass. 

We assume initial population properties 
guided by \cite{Kobulnicky+Fryer2007} and \cite{Sana+2012}. Table
\ref{tab:poprates} 
presents those choices for the binary fraction ($f_{\rm binary}$) and 
initial period and mass-ratio distributions, along with the mass range
of stars which is assumed to produce standard CCSNe (for normalising
the LBV SN rate to the CCSN rate). 
We assume that all orbits are circular, and mostly use a massive-star IMF with a slope of
$-2.5$. This IMF slope is deliberately conservative. Whilst a flatter
mass function has been inferred for the observed population of
massive stars (a slope of $-2.35$ is typically adopted), that mass
function may itself be caused by mergers of massive stars
(i.e., mergers will tend to make the
observed mass function flatter than the true initial mass function),
as discussed by, e.g., \citet{Schneider+2014} and references therein. Table
\ref{tab:poprates} also contains some estimates for an IMF slope of $-2.35$. 

We feel that our adopted normalisation to the CCSN rate
seems likely to be conservative. However, if the apparent upper mass
limit for the progenitors of type IIP
SNe is set by the point at which stars form BHs \citep[see, e.g.,][]{Kochanek2014},
then for any of our population models in which the upper limit for the CCSN
normalisation is at a lower mass than the upper limit on the LBV SN
primary mass, this requires the binary interactions to be able to cause the LBV
progenitors to avoid BH formation at core collapse, or to somehow produce a luminous SN
whilst forming a BH. We have suggested that the former is plausible, but not
proven it.

For most of our rate estimates we have assumed that the distribution
of binary separations $a$ is flat in $\log (a)$, as is
conventional. For these systems we
also adopt a standard normalisation for the separation distribution, chosen under the general
assumption that the range of $a$ runs from $3~R_{\odot}$ to
$10^{4}~R_{\odot}$ \citep[for which see, e.g.,][]{HTP02}. 
This is somewhat conservative, since such massive stars will
not populate the region with $a$ as small as $3~R_{\odot}$;
however this particular choice leads to only a relatively small underestimate. 
Changing the normalisation to one appropriate for  $10<(a/R_{\odot})<10^{4}$
would increase our predicted rates by only $\approx 15$\%, whilst
pessimistically taking $10<(a/R_{\odot})<10^{5}$ would reduce
the predicted rates by a similar amount.

Recent work using data from the VLT-FLAMES survey \citep{Sana+2012,
  Sana+2013} has confirmed the expected high fraction of interacting
binary stars within the massive-star population, but found binary
properties for massive stars somewhat different to those assumed
above. As is conventional, they used single power-law distribution
functions to fit the population parameters (i.e., $f(x) \propto
x^{a}$, where $x$ is the quantity of interest and $a$ is the
exponent to be fitted).  For their Galactic sample they found that
the exponent of the $\log (P)$ distribution function is $–0.55 \pm
0.22$ and that of the $q$ distribution function is $–0.10 \pm
0.58$ \citep{Sana+2012}, i.e., shorter orbital periods and lower-mass
companions were both found to be more common than we adopted. We
note that the constraint on the mass-ratio distribution in
particular is fairly weak; the $f(q) \propto q$ distribution
motivated by \cite{Kobulnicky+Fryer2007} is within $2\sigma$ of
these newer results.  We also provide comparison rates using these
parameters, assuming circular orbits. 

The main source of uncertainty may be our lack of knowledge
about exactly which binaries will merge after reaching contact inside
the HG. It is qualitatively expected that donors with a steep
density gradient in their envelopes (those with radiative envelopes)
are more likely to lead to a merger than those with relatively shallow
density profiles (those with deep convective envelopes). 
The phase during which we assume that the binary will merge is
defined by the post-main-sequence expansion by a factor of 10 in radius
(i.e., from the early- to mid-HG); this might easily be too conservative.
Based on our earlier evolution
calculations, we are confident that -- if the systems merge during
that phase -- they would produce BSG SN progenitors.

In estimating these rates, we have assumed that no
significant amount of mass is ejected during the merger. 
However, loss of material during the merger can be included in the rate
estimates by appropriately increasing the minimum mass-ratio limit for
suitable mergers. For example, for a primary of $20~M_{\odot}$ at
the time of the merger then, if $1~M_{\odot}$ is assumed to be
ejected during the merger, then this corresponds to increasing the
effective minimum mass ratio which
can produce a \emph{suitable} -- i.e., sufficiently luminous -- 
post-merger star from 0.6 to 0.65.

Given all the above, we overall consider the rates presented in Fig.\
\ref{fig:rates} and Table \ref{tab:poprates} likely to be
conservative, although we admit that there are large uncertainties. 
Those estimated rates are typically in excess of one CCSN with an LBV 
progenitors per thousand CCSNe, in some cases approaching one per
hundred CCSNe.

The observationally-derived rates are not precise for this
class of SNe, but they can only be a subset of the SNe which display 
type IIn phenomenology (as discussed in \S \ref{sec:intro}).
Unfortunately, even the absolute rates of SN IIn are not certain. 
Based on our estimates, the rates for LBV SNe following mergers could easily be
large enough to make a substantial contribution to the type IIn SN
class. The rate of type IIn SNe is a small fraction of the overall
CCSN rate \citep[see, e.g.,][]{Kiewe+2012}. 
Pessimistically, we might only explain 1\% of SN IIn (e.g., if SN IIn
constitute 10\% of CCSNe and our mergers produce only one
SN in 1000 CCSNe). Conversely, if SN IIn produce only 2\% of the
volumetric CCSN rate, and only roughly half of those are from true CCSNe, then our
more optimistic estimates for this formation channel could explain all
of the genuinely core-collapse type IIn SNe.

The range of predicted rates for LBV SNe considerably exceeds the
empirical rate for the superluminous SNe (which is estimated to be
between $10^{-3}$ and $10^{-4}$ times the CCSN rate, see
\citealt{Tanaka+2012}). This is as qualitatively expected if special
circumstances are necessary to lead to a radiatively-efficient SNe,
such as the ejection of a particularly massive shell just a few years
before the explosion \citep{Smith+McCray2007}.  However, we cannot be
sure whether appropriate LBV-type mass-ejections would occur
sufficiently often to account for the SLSNe in this way. Clearly it
would help to support this model if such outbursts become more likely
as these stars approach core collapse, perhaps by combining standard
LBV-like instabilities with a driving mechanism similar to that
proposed by \citet{Quataert+Shiode2012}.

\subsubsection{What fraction of LBVs are this type of SN progenitor?}

\begin{figure}
\centering
\epsfig{file=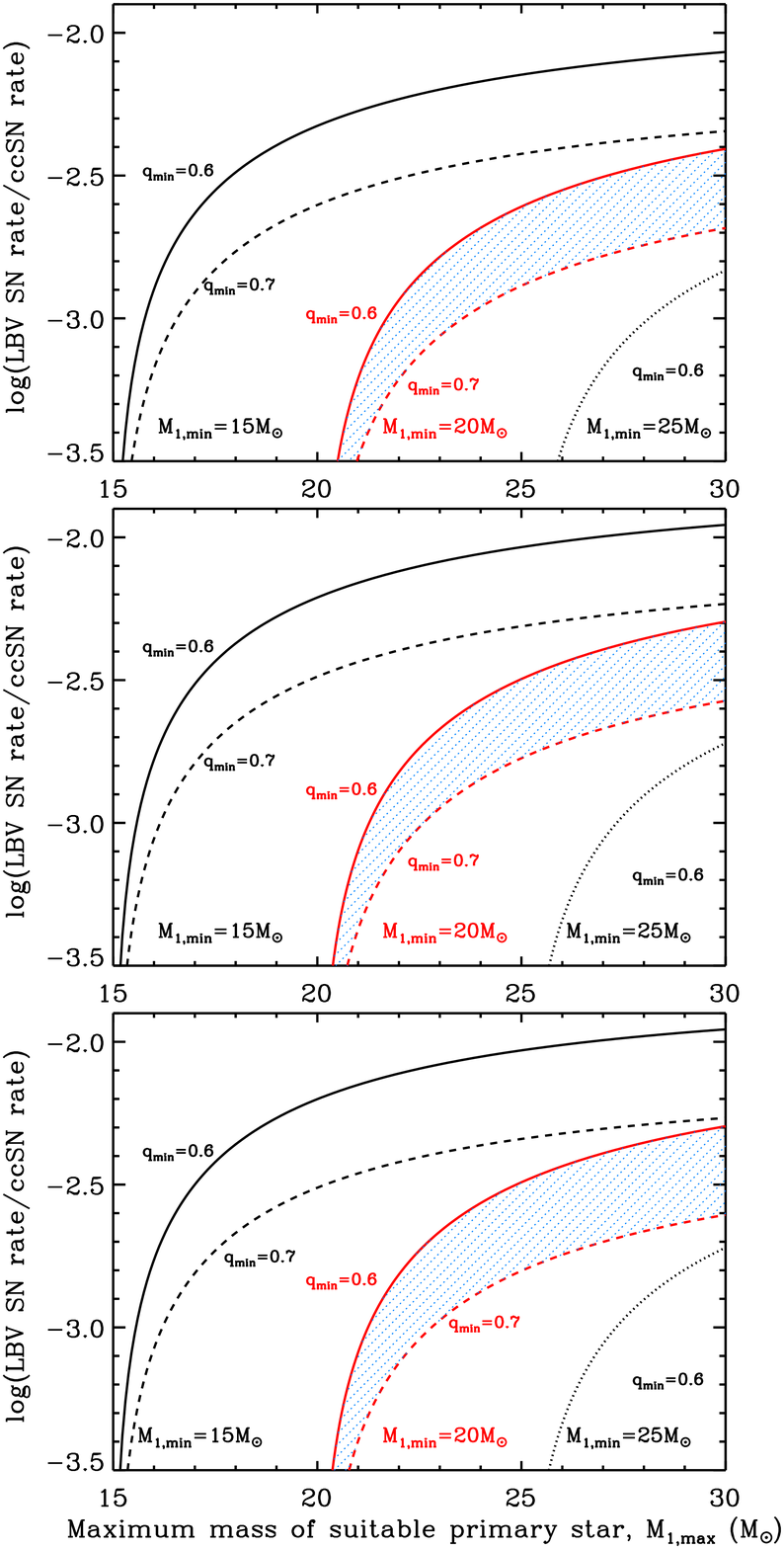, width=8cm}
\caption{\label{fig:rates}
Estimated rates for SNe from our merger scenario, given as a fraction
of the CCSN rate. Initial population assumptions are
given in Table 1 (models KF1, KF2 and S1 are in the top, middle and
bottom panels, respectively). All panels assume that a merger does not
occur if the binary mass ratio ($q$; accretor mass over donor mass) is higher than
0.8. The minimum mass ratio for merger, $q_{\rm min}$, is 0.6 or 0.7, as marked.
The curves which bound the shaded regions assume that the
primary needs to be more massive than $20~M_{\odot}$; those shaded
regions therefore indicate a range of rate estimates for
that assumption. 
More extreme cases are provided by the black curves, which assume that the
primary needs to be more massive than only $15~M_{\odot}$ (solid and
dashed curves) or at least $25~M_{\odot}$ (dotted curves). }
\end{figure}

It would be extremely difficult to give anything like a precise
estimate for the present-day fraction of LBVs which were formed in
this way. This is partly because LBVs may be formed through multiple
binary channels in addition to the portion of the single-star IMF
which produces LBVs. In addition, the duration of the LBV phase may
well be different for LBVs which were formed through different
routes. However, we can make a very rough estimate by comparing our
predicted rates to the formation rate of single massive stars which
are suitable massive to produce LBVs.  To do this, we can compare a
notional single-star LBV birthrate to a single-star CCSN rate (similar
to the normalisation used in Table \ref{tab:poprates} and
Fig.\,\ref{fig:rates}, but here we simply integrate over different
ranges of masses from a single-star IMF for the LBVs and CCSNe).
Thereby we estimate that such single-star LBVs would form at roughly
10\% of the rate at which a notional population of purely single-star
CCSN would occur; this is accurate to within a factor of $\approx 2$
(in either direction) for a range of assumptions.\footnote{For this we
  applied IMF exponents of both $-2.5$ and $-2.35$. For CCSNe, we
  tried combinations of minimum ZAMS masses between 8--10\,$M_{\odot}$
  and maximum ZAMS masses between 20--30\,$M_{\odot}$. For LBVs, we
  adopted minimum ZAMS masses between 35--45\,$M_{\odot}$ and maximum
  ZAMS masses of 100\,$M_{\odot}$ (the integral is not sensitive to the
  upper bound).}  This suggests that LBVs formed from this particular
merger channel -- at $\lesssim$1\% of the CCSN rate -- constitute
$\lesssim$10\% of LBVs. We stress that this estimate neglects several
potentially large factors.  Nonetheless, we would be surprised if many
more than $\sim$10\% of present-day LBVs were to reach core collapse
whilst still in the LBV phase.

\begin{deluxetable*}{cccccccccc||c}
\tablewidth{0.97\textwidth}
\tablecaption{Population assumptions and associated rate estimates
  for the primordial binary early Case B merger channel alone. \label{tab:poprates}}
\tablehead{ \colhead{Name for} &
\multicolumn{5}{c}{Assumptions about initial population \&
    normalisation to CCSN} & 
\multicolumn{4}{c}{Ranges assumed for scenario} & 
\colhead{Rate estimate} \\
\colhead{Fig. \ref{fig:rates}} & 
\colhead{$f_{\rm binary}$} &
\colhead{$M_{1}$ dist.} &
\colhead{$q$ dist.} &
\colhead{$\log P$ dist.}  &
\colhead{CCSN range\tablenotemark{a}} &
\colhead{$M_{\rm 1,min}$}  & 
\colhead{$M_{\rm 1,max}$} & 
\colhead{$q_{\rm min}$ \tablenotemark{b}} & 
\colhead{Radius expansion \tablenotemark{c}} & 
\colhead{$\log \left( \frac{\rm rate}{\rm CC SN rate} \right) $} }  
\startdata
KF1 & 50\%\tablenotemark{d}  & $\propto M_{1}^{-2.5}$ & 
$\propto q$ & flat & 8--40 $M_{\odot}$ &  
15 &    35 &   0.6 &    10 &  $-$2.02 \\ 
\multicolumn{6}{c}{$~$}  &  20 &    35 &   0.6 &    10 &  $-$2.31 \\ 
\multicolumn{6}{c}{$~$}  &  20 &    30 &   0.6 &    10 &  $-$2.41 \\ 
\multicolumn{6}{c}{$~$}  &  20 &    25 &   0.6 &    10 &  $-$2.61 \\ 
\multicolumn{6}{c}{$~$}   &   15 &    35 &   0.7 &    10 &  $-$2.30 \\ 
\multicolumn{6}{c}{$~$}   &   20 &    35 &   0.7 &    10 &  $-$2.59 \\ 
\multicolumn{6}{c}{$~$}   &   20 &    30 &   0.7 &    10 &  $-$2.68 \\ 
\multicolumn{6}{c}{$~$}   &   20 &    25 &   0.7 &    10 &  $-$2.89 \\
\hline 
KF2  & 70\%\tablenotemark{d} &  $\propto M_{1}^{-2.5}$ & 
$\propto q$  & flat  & 8--30 $M_{\odot}$  &
   15 &    35 &   0.6 &    10 &  $-$1.88 \\ 
\multicolumn{6}{c}{$~$} &  20 &    35 &   0.6 &    10 &  $-$2.17 \\
\multicolumn{6}{c}{$~$} &  20 &    30 &   0.6 &    10 &  $-$2.30 \\ 
\multicolumn{6}{c}{$~$} &  20 &    25 &   0.6 &    10 &  $-$2.50 \\ 
\multicolumn{6}{c}{$~$} &   15 &    35 &   0.7 &    10 &  $-$2.15 \\ 
\multicolumn{6}{c}{$~$}  &  20 &    35 &   0.7 &    10 &  $-$2.44 \\ 
\multicolumn{6}{c}{$~$}  &  20 &    30 &   0.7 &    10 &  $-$2.57 \\ 
\multicolumn{6}{c}{$~$}  &  20 &    25 &   0.7 &    10 &  $-$2.77 \\ 
\multicolumn{6}{c}{$~$}  &  15 &    35 &   0.6 &    30 &  $-$1.71 \\ 
\multicolumn{6}{c}{$~$}  &  20 &    35 &   0.6 &    30 &  $-$2.00 \\ 
\multicolumn{6}{c}{$~$}  &  20 &    30 &   0.6 &    30 &  $-$2.13 \\ 
\multicolumn{6}{c}{$~$}  &  20 &    25 &   0.6 &    30 &  $-$2.33 \\
\hline 
{$~$}  & 70\%\tablenotemark{d} &  $\propto M_{1}^{-2.5}$ & 
$\propto q$  & flat  & 8--40 $M_{\odot}$  &
 15 &    35 &   0.6 &    10 &  $-$1.94 \\ 
\multicolumn{6}{c}{$~$}  &    20 &    35 &   0.6 &    10 &  $-$2.23 \\ 
\multicolumn{6}{c}{$~$}  &    20 &    30 &   0.6 &    10 &  $-$2.32 \\ 
\multicolumn{6}{c}{$~$}  &    20 &    25 &   0.6 &    10 &  $-$2.52 \\ 
\multicolumn{6}{c}{$~$}  &    15 &    35 &   0.7 &    10 &  $-$2.21 \\
\multicolumn{6}{c}{$~$}  &    20 &    35 &   0.7 &    10 &  $-$2.50 \\ 
\multicolumn{6}{c}{$~$}  &    20 &    30 &   0.7 &    10 &  $-$2.60 \\
\multicolumn{6}{c}{$~$}  &    20 &    25 &   0.7 &    10 &  $-$2.80 \\ 
\hline 
{$~$}  & 70\%\tablenotemark{d} &  $\propto M_{1}^{-2.35}$ & 
$\propto q$  & flat  & 8--30 $M_{\odot}$  &  15 &    35 &   0.6 &    10 &  $-$1.84 \\ 
\multicolumn{6}{c}{$~$}   &   20 &    35 &   0.6 &    10 &  $-$2.12 \\ 
\multicolumn{6}{c}{$~$}   &   20 &    30 &   0.6 &    10 &  $-$2.25 \\ 
\multicolumn{6}{c}{$~$}   &   20 &    25 &   0.6 &    10 &  $-$2.46 \\
\multicolumn{6}{c}{$~$}   &   15 &    35 &   0.7 &    10 &  $-$2.12 \\ 
\multicolumn{6}{c}{$~$}   &   20 &    35 &   0.7 &    10 &  $-$2.40 \\ 
\multicolumn{6}{c}{$~$}   &   20 &    30 &   0.7 &    10 &  $-$2.53 \\ 
\multicolumn{6}{c}{$~$}   &   20 &    25 &   0.7 &    10 &  $-$2.74 \\
\multicolumn{6}{c}{$~$}   &   15 &    35 &   0.6 &    30 &  $-$1.68 \\ 
\multicolumn{6}{c}{$~$}   &   20 &    35 &   0.6 &    30 &  $-$1.96 \\ 
\multicolumn{6}{c}{$~$}   &   20 &    30 &   0.6 &    30 &  $-$2.09 \\ 
\multicolumn{6}{c}{$~$}   &   20 &    25 &   0.6 &    30 &  $-$2.29 \\
\hline 
S1 & 70\%\tablenotemark{e}  &  $\propto M_{1}^{-2.5}$ & 
$\propto q^{-0.1}$  & $\propto (\log~P)^{-0.55}$ & 8--40 $M_{\odot}$
& 15 &    35 &   0.6 &    10 &  $-$1.92 \\ 
\multicolumn{6}{c}{$~$}   &   20 &    35 &   0.6 &    10 &  $-$2.21 \\ 
\multicolumn{6}{c}{$~$}   &   20 &    30 &   0.6 &    10 &  $-$2.30 \\ 
\multicolumn{6}{c}{$~$}   &   20 &    25 &   0.6 &    10 &  $-$2.50 \\
\multicolumn{6}{c}{$~$}   &   15 &    35 &   0.7 &    10 &  $-$2.23 \\ 
\multicolumn{6}{c}{$~$}   &   20 &    35 &   0.7 &    10 &  $-$2.52 \\ 
\multicolumn{6}{c}{$~$}   &   20 &    30 &   0.7 &    10 &  $-$2.61 \\ 
\multicolumn{6}{c}{$~$}   &   20 &    25 &   0.7 &    10 &  $-$2.81 \\ 
\hline 
{$~$}   & 70\%\tablenotemark{e}  &  $\propto M_{1}^{-2.5}$ & 
$\propto q^{-0.1}$  & $\propto (\log~P)^{-0.55}$ & 8--25 $M_{\odot}$
&   15 &    35 &   0.6 &    10 &  $-$1.83 \\ 
\multicolumn{6}{c}{$~$}   &   20 &    35 &   0.6 &    10 &  $-$2.12 \\ 
\multicolumn{6}{c}{$~$}   &   20 &    30 &   0.6 &    10 &  $-$2.23 \\ 
\multicolumn{6}{c}{$~$}   &   20 &    25 &   0.6 &    10 &  $-$2.45 \\
\multicolumn{6}{c}{$~$}   &   15 &    35 &   0.7 &    10 &  $-$2.14 \\ 
\multicolumn{6}{c}{$~$}   &   20 &    35 &   0.7 &    10 &  $-$2.43 \\ 
\multicolumn{6}{c}{$~$}   &   20 &    30 &   0.7 &    10 &  $-$2.54 \\ 
\multicolumn{6}{c}{$~$}   &   20 &    25 &   0.7 &    10 &  $-$2.76 \\
\multicolumn{6}{c}{$~$}   &   15 &    35 &   0.6 &    30 &  $-$1.67 \\ 
\multicolumn{6}{c}{$~$}   &   20 &    35 &   0.6 &    30 &  $-$1.96 \\ 
\multicolumn{6}{c}{$~$}   &   20 &    30 &   0.6 &    30 &  $-$2.06 \\ 
\multicolumn{6}{c}{$~$}   &   20 &    25 &   0.6 &    30 &  $-$2.28
\enddata
\tablenotetext{a}{For normalisation to the CCSN rate. Secondary stars
  are included in the normalisation using the same
  mass-ratio distribution and $f_{\rm binary}$.}
\tablenotetext{b}{In all cases, $q_{\rm max}=0.8$.}
\tablenotetext{c}{i.e. the range of radius expansion of the primary
  star after the end of the main sequence over which the outcome may be a merger.}
\tablenotetext{d}{With separations $a$ such that 3 $\leq
  (a/R_{\odot}) \leq 10^{4}$.}
\tablenotetext{e}{With separations $a$ such that 3 $\leq
  (a/R_{\odot}) \leq 5 \times 10^{3}$.}
\end{deluxetable*}

\subsection{The stable mass transfer channel}
\label{sec:stableRLOFrates}

The early Case B merger channel is only one of the ways in which a
star might potentially gain
mass at the correct point in its evolution. That merger channel
can naturally be triggered by the expansion of the primary at the
appropriate point in its evolution (i.e., effectively the star
\emph{gains} mass upon expansion, which is the reverse of the normal
expectation), which helps with fine-tuning the timing of the mass
accretion. An alternative way in which a star could gain mass at a suitable time to become
one of the BSG SN progenitors we model is if the 
star is the accreting secondary in a binary system in which the
primary star happens to fill its Roche lobe when the secondary is in
an appropriate phase of evolution. We will call this the stable
mass-transfer formation channel.

If we require that the secondary has already left the main sequence
when the primary transfers mass, then the qualitative conditions for
this formation channel to operate are roughly similar to the
conditions for ``double-core evolution'' to occur \citep[see,
  e.g.,][]{Bethe+Brown1998,Dewi+2006}, but with a more restrictive
limit on the evolutionary phase of the accretor.  For double-core
evolution, this mass transfer leads to an unstable contact phase, and
thence to a special case of common-envelope evolution in which two
cores spiral-in inside the shared envelope.  The Galactic birthrate of
binaries produced from double-core evolution has been variously
estimated to be between $\rm \sim 10^{-6}~yr^{-1}$ and $\rm \sim
10^{-4}~yr^{-1}$ \citep[see, e.g.,][and references
  therein]{Bethe+Brown1998,Dewi+2006}. Only the upper end of those
double-core birthrates are comparable to our estimates for the early
Case B merger channel. Since the constraints on timing for this stable
mass-transfer channel are tighter than for double-core evolution, we
conclude that production of these LBV SN progenitors will probably
occur less frequently via stable mass transfer than through the early
Case B merger channel.

\begin{figure}
\centering
\epsfig{file=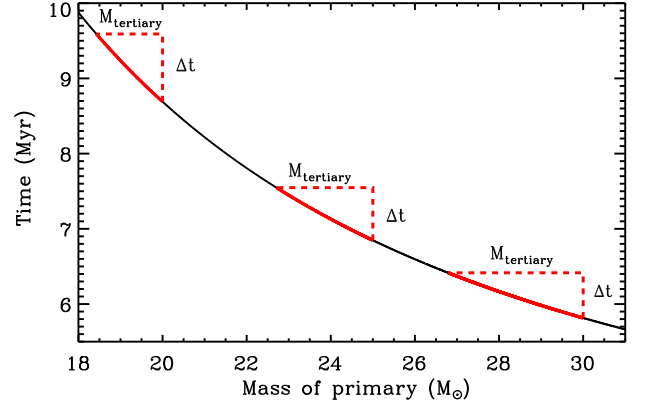,
  width=9cm}
\caption{\label{fig:tertiary_range} We estimate the range of tertiary masses
  for which the outer star in a ``potentially-interacting'' triple
  would evolve to fill its Roche lobe during the lifetime of the
  merger product. The black curve shows an approximate measure of the stellar lifetime before
  expanding to become a RSG ($t_{\rm BGB}$ from the fits of
  \citealt{HPT2000} for a metallicity of 0.02;
 note that for our
  estimates here the absolute value
  is irrelevant, only the slope). We compare this to the rough
  lifetime of the merger product (denoted $\rm \Delta t$) for three
  primary masses (20, 25 and 30 $M_{\odot}$, for which we take post-merger
lifetimes of 0.9, 0.7 and 0.6 Myr respectively, based on
Fig. \ref{fig:lifetimes}).  Even though the post-merger lifetime
decreases for increasing primary mass, the range of potentially-suitable tertiary masses 
  ($\rm M_{\rm tertiary}$) increases at higher primary masses (because
   the gradient of the curve decreases). }
\end{figure}

\begin{deluxetable}{cc|c|ccc}
\tablewidth{0.48\textwidth}
\tablecaption{The estimated fraction of potentially-interacting triples
  which, if the inner binary produces a merger, possess a mass ratio
  which leads to mass transfer onto the merger product. \label{tab:triples}}
\tablehead{ \colhead{~} &
 \colhead{~} &
 \colhead{~} &
\multicolumn{3}{c}{Fraction of interacting triples suitable,}
\\
\colhead{~} &
 \colhead{~} &
 \colhead{~} &
\multicolumn{3}{c}{for a $q$-distribution with exponent:} \\
\colhead{$M_{1}$} &
 \colhead{min($M_{3}$)} &
 \colhead{max($q$)} &
\colhead{0 (i.e., flat)} & 
\colhead{1 (i.e., correlated)} &
\colhead{$-0.1$}}
\startdata
  20 &  18.43 &  1.0 & 0.079 &   0.151 &   0.071 \\ 
 ~ & ~ & 0.99 &  0.069 &   0.131 &   0.062 \\ 
 ~ & ~ & 0.95 & 0.029 &   0.054 &   0.026 \\ 
~&~&~&~&~&~\\
  25 &  22.71 &  1.0 & 0.092 &   0.175 &   0.083 \\ 
 ~ & ~ & 0.99 &  0.082 &   0.155 &   0.074 \\
 ~ & ~ & 0.95 & 0.042 &   0.077 &   0.038 \\
~&~&~&~&~&~\\
  30 &  26.79 &  1.0 & 0.107 &   0.202 &   0.097 \\
 ~ & ~ & 0.99 &  0.097 &   0.183 &   0.088 \\ 
 ~ & ~ & 0.95 &  0.057 &   0.105 &   0.052
\enddata
\end{deluxetable}

\subsection{Post-merger accretion from the tertiary star in a triple}
\label{sec:triples}

Some of the post-accretion models shown earlier in this paper have
been for post-accretion masses which could not be produced if the early
Case B merger mechanism on which was have concentrated (and for which the rates in \S
\ref{sec:earlyCaseBrates} were derived) was the only way for the primary star to gain mass.
One potential route through which unusually massive post-merger stars might be formed involves triple
stars (see, e.g., Fig.\,\ref{fig:HRtriples}).  

As an observational example, the massive binary system R4 in the Small Magellanic Cloud is
presently best explained by a merger in a primordial triple
\citep{Pasquali+2000}.  The current B[e] star in R4 appears younger than the A star
companion, suggesting that it was rejuvenated by a merger; that event could
also lead to the nebula around the system \citep{Pasquali+2000}. 
Some triples -- similar to R4, though with a smaller separation of the remaining
post-merger binary -- might subsequently experience either a second
merger, or mass transfer from the triple onto the merger product. In
either case, such systems could produce a qualitatively different
population of SN progenitors than those which involved only one
merger (or accretion phase).  These stars would also have little time to lose the
angular momentum gained during their second accretion phase. 

Overall rate estimates for triple stars are even more uncertain than for
binary stars. However, we can estimate the fraction of triples in
which the tertiary star would expand away from the main sequence after
the inner binary has merged but before the merger product explodes as
a SN by using the post-merger lifetimes shown in
Fig. \ref{fig:lifetimes}.  We will call triples in which the tertiary
orbit is small enough for the tertiary to fill its Roche lobe at some point in its evolution ``potentially
interacting'' (since the triple would not have the chance to interact if the system has
been disrupted by a SN before the tertiary expands to fill its Roche
lobe). We note that mass transfer from the outer star in a triple onto the inner binary has been
discussed by \citet{Tauris+vdH2014} and \citet{deVries+2014}, although in very different
contexts. \citet{Perets+Fabrycky2009} have also considered how triple
stars may be important in promoting blue-straggler formation.

Fig.\,\ref{fig:tertiary_range} demonstrates how we estimate the
potentially-suitable range of tertiary masses. Given those mass
ranges, we can then estimate the fraction of ``potentially
interacting'' triples which meet the necessary criteria.  These
estimates are shown in Table \ref{tab:triples} for different
assumptions about the distribution of the mass of the tertiary star
relative to the primary star.  Table \ref{tab:triples} also shows how
those estimates change if we limit the mass of the outer
triple to be less than 99\% or 95\% of the mass of the primary. For
all but the more conservative sets of assumptions, these estimates
indicate that $\approx 10$ percent of the ``potentially interacting''
triples in which the inner binary is subject to an early Case B merger
would produce accretion onto the merger product. 

Whilst we do not claim to properly estimate the fraction of O-stars
which are in such ``potentially interacting'' triples, we stress that
O-stars are very commonly found in triple systems.
\citet{Eggleton+Tokovinin2008} found that the combined fraction of
O-star systems with triple or higher
multiplicity was higher than the binary fraction of O-star systems
(the fraction of O-star systems which they detected to have triple multiplicity
was $\approx 2/3$ of the binary fraction; note that this does not
include the systems with higher multiplicity which could contain a
suitable triple-star sub-system).\footnote{It also seems plausible that the fraction of the
``potentially interacting'' triples in which the inner binary
merges is higher than for standard binaries.}
We therefore consider it realistic that as
many as $\sim 10$\% of the early Case-B mergers might potentially
gain further mass from a tertiary star.
Somewhat less than 10\% of our binary merger rate (from \S
\ref{sec:earlyCaseBrates}) would produce a SN rate that matches the
observationally-inferred rate of SLSN-II.  This might well be
coincidence, but nonetheless suggests that further study of this
evolutionary channel is deserved.

\section{Discussion}

\subsection{The relationship between ``normal'' SNe with LBV
  progenitors and ``superluminous'' SNe.}

The calculations presented in this paper have shown that
early Case B accretion is able to
produce BSGs which are plausibly sufficiently luminous to have been
LBVs immediately before core collapse.  Our rate estimates further indicate
that the birthrate from the early Case B binary-merger scenario may well be high
enough to explain the CCSNe of normal luminosity which have been
inferred to have immediate LBV progenitors.  Some particularly luminous LBV
progenitors of CCSNe might be explained by a second phase of mass
accretion from a tertiary companion, or potentially by multiple
mergers in dense stellar systems.

However, this does not guarantee that this class of formation channels can
explain any of the progenitors of SLSNe.  If the hypothesis is
correct that some SLSNe can be explained by greater radiative
efficiency of a SN with otherwise normal energetics, then this might
be explained by CSM produced by an appropriately-timed and
appropriately-massive LBV outburst.  This still appears to be a
reasonable model for at least a subset of the SLSNe, given that the 
amount of CSM required to lead to radiatively-efficient
events is disputed \citep[see, e.g.,][]{Dwarkadas2011}. 
However, it might be that normal LBV outbursts are
incapable of producing the CSM properties which are necessary to
explain SLSNe.

\subsection{Metallicity effects}

Our models have all assumed a metallicity of $\rm Z=0.02$. A priori,
it is unclear what effect changing this assumption would have on our results.
If the initial binary-star population does not vary as a
function of metallicity, then we consider that moderate changes
in metallicity should not lead to a significant change in the rate at
which suitable mergers occur. However, for
metallicities which are low enough that the primary star burns He
as a BSG, then the parameter space for early Case B mergers would be
reduced.

Since it is broadly expected that LBV formation requires higher masses
at lower metallicities (since the Humphreys-Davidson limit moves to higher luminosities),
then the parameter space for our merger model which is capable of forming LBV progenitors
of CCSNe may be reduced.  

Those potential effects on the rates are independent from any effects
on the post-accretion structures. We hope to extend this study of
stellar structures and evolution following early Case B mergers with a
systematic future exploration of potential metallicity effects on the
superficial appearance and the fate of the core,
alongside an investigation of the potential effect of He-enrichment
during the merger.

\subsection{Mergers of $q
  \approx 1$ massive binaries and potential pair-instability SNe}

This paper has mainly concentrated on a scenario which involves a
particular early Case B merger process in which the merger instability
follows a brief contact phase.  Those mergers are only expected to
happen for a limited range of mass ratios, in particular with an upper
limit on the mass ratio ($q<0.8$). We repeat that these mergers are
not a result of the canonical
high-mass-ratio dynamical instabilities. There is an additional
part of parameter space which is expected to lead to an unstable
contact phase and binary mergers: the post-main-sequence merger
of two stars of almost equal mass (i.e., $q \approx 1$). 

Early Case B mergers from this channel would require fine-tuning, in
that the masses would have to be so similar that both stars
are expanding across the Hertzsprung gap at the same time (although massive
stars may prefer similar-mass companions; see, e.g.,
\citealt{Kobulnicky+Fryer2007}). 
In most cases, this `double-core' merger channel
seems likely to lead to merged He cores and thence the formation
of a BH at core collapse, even if the star is then an LBV. 

However, mergers of sufficiently massive post-main-sequence cores should form
massive enough oxygen cores to produce a pair-instability explosion
\citep{Barkat+1967,RakavyShaviv1967,Heger+2003}. This would not
necessarily require an early Case-B merger, as long as both cores were
adequately evolved.
Predicted minimum He-core masses for producing a pair-instability SN are
$\approx$ 64 $M_{\odot}$ \citep{Heger+2003}, which suggests that this
scenario requires a merger of two stars each of initial mass in
excess of $\approx$ 50 $M_{\odot}$ (with some uncertainty coming from
the treatment of
convection in the cores of such stars; note that the fractional core
mass increases with stellar mass). 
Unlike standard pair-instability events, this may well occur even at solar
metallicity.  (See also \citealt{Pan+2012}, who noted that runaway collisions in
stellar clusters could also help to generate pair-instability SNe at solar metallicity.)
Furthermore, in very fine-tuned cases these merger-produced pair-instability SNe
might potentially take place inside a recently-ejected (or
partially-ejected) envelope. If such fine-tuning is possible then this
could lead, in principle, to an unusually
energetic \emph{and} unusually radiatively-efficient SN.

\subsection{Tertiary-star CE ejection}

Another route through which a SN might happen inside
a recently-ejected stellar envelope is a variation of the triple-star
scenario described in \S \ref{sec:triples}. Since we find that the
merger products expand very late in their nuclear evolution, that
post-He-core-burning expansion might trigger the onset of a
CE phase within 10 kyr of core collapse. This may occur if the
expanding merger product is already accreting from a tertiary star (leading
to an unstable contact phase), or if the expansion leads to
dynamically-unstable Roche-lobe overflow onto a tertiary companion
(although this may well require fine-tuning of the separations in
the initial triple system).
In some cases the time to core collapse could be 
comparable to the potential combined duration of the onset and spiral-in of the
CE phase, i.e., a CCSN might occur inside many solar masses of
ejected material. The outcome would be similar to the model of
\citet{Chevalier2012}, although this post-merger scenario may help to
explain why the timing of the ejection was so close to core collapse.

\subsection{Observables and tests of the model}

Our post-merger BSG models have lower core mass fractions
than canonical main-sequence BSGs. The difference is even larger
when compared to LBV SN models which invoke 
rotation-induced mixing, which have cores which account for almost
the entirety of the stellar mass \citep{Groh+2013}.
Unfortunately, it is not clear whether this difference can be practically tested.
The core masses might perhaps be inferred by means of asteroseismology 
if oscillation modes of the stars are sufficiently excited \citep[see,
e.g.,][]{Saio+2013}.  As we expect these BSGs to spend time 
as B[e] supergiants \citep{Podsiadlowski+2006}, we
encourage attempts to determine the structures of sgB[e] stars. 
Alternatively, perhaps reconstruction of the structure of the progenitor star
from a suitable SN would enable discrimination between the
possibilities, either by using time-resolved spectroscopy
\citep[e.g.,][]{Mazzali+2008} or analysis of nebular spectra
\citep[e.g.,][]{Mazzali+2010}.  

Other observables during the BSG/LBV phase might have multiple
interpretations. For example, the bipolar shell surrounding the LBV
candidate G25.5+0.2 might well have been produced by ejecta from a
stellar merger, and the projected peak expansion velocities of the nebula
($\sim 180\,{\rm km}\,{\rm s}^{-1}$) are comparable to surface escape
velocities from HG stars \citep[][]{Clark+2000}. 
However, even if one could prove that the nebula was generated by a
merger, it is unclear how one could unambigously determine whether 
the merger in that particular system was during Case B. Likewise, we would
not be surprised to find surface abundance anomalies in the
envelope of a post-merger BSG/LBV, such as enhancement in helium or nitrogen, 
but conclusions drawn from observations of such enrichment may not be
definitive \citep[for a discussion of how surface nitrogen
abundances and rotational velocities might help to constrain BSG
properties see, e.g,][]{Vink+2010}. 

Clearer surface observables might be provided during the
post-accretion/post-merger contraction phase, but the duration of
this stage of the evolution is relatively brief. 
This phase is also the one most likely to have its appearance affected 
by the details of the merger
physics, so it is the one for which quantitative predictions are the most
uncertain.  Nonethless, the qualitative changes in our models during
this phase have similarities to the rapid temperature increases seen
in Eta Carinae since the Great Eruption
\citep{Rest+2012,Mehner+2014}. We intend to explore this similarity
in the future.

\section[]{Conclusions}

We have explored the results of early Case B accretion, concentrating on
primary-star masses which are at the upper end of the
range which seem likely to produce NSs after core collapse. 
We find that, if massive stars are able to gain sufficient mass soon
after they finish core H burning, they can reach
core collapse with the properties expected of an LBV.  We have demonstrated
that our results are robust against some reasonable variations of the stellar
physics employed. 

The amount of accretion which is required to produce LBV SN
progenitors from such primary stars might be supplied by early Case B mergers following
an unstable contact phase.  Our estimates for the birthrate from this
merger channel are broadly consistent with the inferred rate of
LBV SNe, at more than one per thousand CCSNe and approaching 1 per cent
of the CCSN rate for moderately optimistic assumptions.

Additional contributions to the birthrate may be obtained through
other formation channels, notably from very late Case A contact
binaries which merge after the primary leaves the main sequence.  Or in
some rare cases, a RSG primary may fill its Roche-lobe and transfer its
envelope to the secondary just as the secondary is leaving the MS (see
\S \ref{sec:stableRLOFrates}). In dense stellar environments,
well-timed direct collisions might potentially account for
some similar SN progenitors, including some examples in which even more
mass could be accreted than is possible for co-eval field binary evolution. We have
calculated evolutionary tracks to examine the evolution of stars which
accrete more mass than simple binary evolution should allow,
but we have not attempted to estimate the rate at which such merger
products might be produced.

Surprisingly, stable mass transfer from a tertiary star onto the
product of an early Case B 
merger may be an important channel for the formation of some
extreme SN progenitors. Indeed, rough estimates indicate that this
channel might be more common than early Case B accretion from
stable mass-transfer in a binary. 
This assumes that mass transfer onto the merger product can
occur at any time during the lifetime of the merger product,
which requires less fine-tuning than that which is required to achieve mass
transfer onto a star in the Hertzsprung Gap.

We have also investigated whether the core collapse of these
post-accretion stars is likely to lead to a successful SN explosion. 
To study this we have
compared the core structures of the post-accretion stars to those of
single stars using a range of potential indicators: the mass of the CO
core, the binding energy of the outer core, the compactness parameter
$\xi_{2.5}$ and the mean specific entropy of the core (for both the CO core and the
He-rich layer outside the CO core).
These generally suggest that early Case B accretion onto the
envelope of a star does not significantly increase the
likelihood of BH formation at core collapse. 
However, $\xi_{2.5}$ leads to ambiguous predictions: 
for $\xi_{2.5}$ the predicted effect of accretion differs between our set of
models which adopt significant overshooting and the set which assumes the
Schwarzschild criterion with no overshooting.
Moreover, we suggest that when there are changes in the
final core properties, these are more often in the direction of the core 
becoming more similar to the
core of a \emph{lower-mass} star, not a higher-mass one.  
Despite the uncertainties arising from the assumptions about the
accretion phase (or merger process), the fact that \emph{accretion}
may \emph{increase} the chance of a successful CCSN is striking.   
If this result is confirmed then this effect may be very significant
in understanding the diversity of CCSNe from binary progenitors.

However, the merger products often display a combination of core properties
which are not possessed by any of our single-star models. The fact that
binary interactions can affect observed SN diversity by changing the
distributions of final envelope masses and final core angular momenta
has long been appreciated \citep[see, e.g.,][]{Podsiadlowski+1992}. The effects of binary
interactions on the structure of the final core have been less
widely studied \citep[but see, e.g.,][]{PhP+2004ElectronCapture,Poelarends+2008}.  Our
results add to the evidence that the core collapse of non-rotating stars at fixed
metallicity may be poorly described by a single-parameter family in He
core mass. They also strengthen the idea that binary
interactions are vital for understanding the diversity of
CCSNe. In future we plan to improve the density of our model coverage within the binary
parameter space. We also intend to study the
physics of the merger process, and the potential effects of that merger
physics on the post-merger evolution.

The recently-recognised explosions of LBV stars have sometimes been presented as a
challenge to existing theories of stellar evolution. In contrast, a
class of binary mergers is able to produce events
which naturally match the inferred properties of the relevant SN
progenitors.  This one channel may be able to produce a
diverse range of SN types, ranging from the explosions of yellow
supergiants to superluminous SNe. These SNe and the stellar mergers
which preceded them are extraordinary in their physical and
astrophysical interest and deserve greater theoretical attention.

\section*{Acknowledgements}
 
The final form of this work has been significantly influenced by an
anonymous referee whom we thank for a critical
examination which prompted numerous improvements. 
SJ thanks the Chinese Academy of Sciences and National Science Foundation of 
China for support (grant numbers 11250110055 and 11350110324).
SJ and PhP thank Zhanwen Han for generous
hospitality, Ilka Petermann for useful discussions, and the
attendees at the ``Fireworks 2012'' workshop for helpful comments. 
PhP and JSV thank the Aspen Center for Physics (and
NSF Grant number 1066293) for hospitality during discussions of parts
of this work.
The authors are very grateful to Peter Eggleton, Bill Paxton, and
their respective collaborators for their hard work and for making
their stellar evolution packages available. This
publication has made use of matplotlib \citep[][]{matplotlib}.


\end{document}